\documentclass[12pt]{article}
\usepackage[utf8]{inputenc}

\usepackage{geometry,amsmath,mathtools,blkarray,amssymb,amsthm,setspace,xcolor,tikz-cd,xparse,amsfonts,graphics,comment}

\usepackage{cancel}

\usepackage[shortlabels]{enumitem}
\usepackage[colorlinks,linkcolor=blue,citecolor=blue,urlcolor=blue,bookmarks=false,hypertexnames=true]{hyperref}

\usepackage{thmtools, stackrel, caption,subcaption,graphicx,lscape, makecell, tabularx,bbm}

\usepackage{natbib}

\usetikzlibrary{shapes.geometric,math}
\usepackage{tikz,verbatim}

\allowdisplaybreaks

\newtheorem*{theorem*}{Theorem}
\newtheorem{theorem}{Theorem}
\numberwithin{theorem}{section}
\newtheorem{definition}[theorem]{Definition}
\newtheorem{proposition}[theorem]{Proposition}
\newtheorem{lemma}[theorem]{Lemma}
\newtheorem{corollary}[theorem]{Corollary}
\newtheorem{remark}[theorem]{Remark}
\newtheorem{claim}[theorem]{Claim}

\newcommand{\argmin}{\mathop{\mathrm{argmin}}}

\newenvironment{nospaceflalign*}
 {\setlength{\abovedisplayskip}{0pt}\setlength{\belowdisplayskip}{0pt}%
  \csname flalign*\endcsname}
 {\csname endflalign*\endcsname\ignorespacesafterend}

\def\Re{\mathbf{R}}

\def\N{{\cal{N}}}

\def\D{\mathcal{D}}
\def\E{\mathcal{E}}
\def\F{\mathcal{F}}
\def\P{\mathcal{P}}
\def\M{\mathcal{M}}
\def\C{\mathcal{C}}
\def\ep{\varepsilon}
\def\A{{\cal{A}}}

\def\L{\mathcal{L}}

\newcommand{\bit}{\begin{itemize}}
\newcommand{\eit}{\end{itemize}}
\newcommand{\co}{\text{co.}}
\newcommand{\dis}{\displaystyle}

\declaretheoremstyle[spaceabove=1pt, spacebelow=6pt, headfont=\bfseries, notefont=\bfseries, notebraces={\UTF{0081}i}{\UTF{0081}j}, postheadspace=1em, numbered=no,qed=$\blacksquare$]{myproof}\declaretheorem[title=Proof, style=myproof]{myproof} 

\declaretheoremstyle[spaceabove=1pt, spacebelow=6pt, headfont=\itshape, notefont=\bfseries, notebraces={\UTF{0081}i}{\UTF{0081}j}, postheadspace=1em, numbered=no,qed=$\square$]{mysubproof} \declaretheorem[title=Proof, style=mysubproof]{mysubproof}\renewenvironment{proof}{\begin{mysubproof}}{\end{mysubproof}}

\title{Random Utility with Unobservable Alternatives}

\author{Haruki Kono, \ Kota Saito, \ Alec Sandroni\footnote{
Kono: MIT, \textsf{hkono@mit.edu}. Saito: Caltech, \textsf{saito@caltech.edu}, Sandroni: Caltech, \textsf{asandron@caltech.edu}.  We appreciate the valuable discussions we had with Jean-Paul Doignon, Efe Ok, Faruk Gul, Wolfgang Pesendorfer, Pietro Ortoleva, Peter Klibanoff, Marciano Siniscalchi, and Takashi Ui. In particular, Jean-Paul Doignon as well as three anonymous referees for EC 2023 read an earlier version of the manuscript and offered helpful comments. We also appreciate participants at a presentation at Northwestern University, Princeton University, New York University, University of Tokyo, EC 2023, RUD 2023, and DTEA 2023.  Kono acknowledges financial support from the Funai Overseas Scholarship and the Jerry A. Hausman Fellowship. 
Alec Sandroni acknowledges financial support from the Summer Undergraduate Research Fellowships (SURF) program in 2022, 2023, and 2024 at Caltech.
}}

\begin{document}

\maketitle

\begin{abstract}
The random utility model, a cornerstone in economics, is axiomatized by \cite{falmagne1978representation} and \cite{mcfad91} with the assumption that if a menu is observable, the choice frequencies of all alternatives are also observable. However, in practice, it is common for choice frequencies of some alternatives to remain unobserved. To address this discrepancy, we obtain the testable implications of the random utility model when the choice frequencies of some alternatives are unobservable, which consist of 
nonredundant inequality constraints on observed choice frequencies. 
Our findings indicate that the widespread empirical practice of aggregating unobserved alternatives into a single ``outside option'' fails to capture significant implications of random utility models.
\end{abstract}


\section{Introduction}

\vspace{-0.2cm}

Consider a population of individuals choosing an alternative from various choice sets, where the analyst observes the choice frequencies of each alternative within each set. A foundational framework for interpreting such datasets is the random utility model, which posits a probability distribution over rankings of alternatives, with each ranking representing an individual’s preferences. This model serves as a central tool in economics for linking observed stochastic choice behavior to underlying preference structures.
Existing characterizations of the random utility model by \cite{falmagne1978representation} and \cite{mcfad91} assume that the choice frequencies of {\it all} alternatives in each menu are observable.\footnote{\cite{falmagne1978representation} axiomatized the random utility model under the assumption that choice frequencies for all alternatives in all menus are observable. \cite{mcfad91} do not require that all menus be observed, but when a menu is observed, they assume that the choice frequencies for all alternatives in that menu are observable.}

However, it is often the case that the choice frequencies for some alternatives are missing. For example, consider a set of transportation methods consisting of bus, train, walking, and driving. While it may be possible to estimate the market share of public transportation (bus or train) based on the revenue of bus or train companies, it can be difficult to determine whether a person chooses to drive or walk unless a survey is conducted. As a result, the choice frequencies for walking and driving may not be available. 

There are many other economically significant examples in which the choice frequencies of some alternatives are not observable.\footnote{For instance, consider school choice among private and public schools. Governments can often obtain choice data for public schools but not for private schools. In this case, the choice frequencies for public schools are observable, whereas those for private schools are not. See Section \ref{sec:examples} for further details.} In such situations, empirical researchers often aggregate all unobservable alternatives into a single category and treat it as an outside option, even when they know which specific alternatives are unobservable. We refer to this approach as {\it the outside option approach}. With this approach, all choice frequencies become observable because there is only one outside option, and the choice frequencies of all other alternatives are known.\footnote{The choice frequency of the outside option is then calculated as one minus the sum of all observable choice frequencies.}

The purpose of this paper is twofold. First, we investigate a necessary and sufficient condition for a random utility model to rationalize the observed choice data when the choice frequencies of some alternatives are unobservable. Second, through this investigation, we demonstrate the limitations of the outside option approach. We show that by relying on the outside option approach, researchers may not only mistakenly adopt a random utility model but also overlook valuable information contained in the dataset. One key takeaway from these results is that empirical researchers should use the outside option approach with caution, making a deliberate effort to specify available alternatives in each choice set as much as possible. By doing so, the researchers can exploit all the implications of the random utility model. In Section \ref{subsection:ex}, we  demonstrate these points by providing an example.  In the following, we elaborate on these two objectives in order.

In this paper, we focus on a setup where the choice frequencies of some alternatives are consistently missing, while those for the other alternatives are observable. For such datasets, we derive a finite system of linear inequalities that provides a necessary and sufficient condition for the dataset to be rationalized by a random utility model. Furthermore, the characterization we provide is nonredundant, meaning that none of the inequalities are implied by any others, and removing even one of them would render the condition insufficient.\footnote{\cite{falmagne1978representation}'s characterization is also almost nonredundant. \cite{suck2002binary} and \cite{fiorini2004short} show that omitting just a few inequalities from \cite{falmagne1978representation} results in a nonredundant characterization. The characterization by \cite{mcfad91} involves
infinitely many inequalities and entails redundancy.} As we explain later, the nonredundancy of the conditions facilitates us to show the limitation of the outside option approach.

Our necessary and sufficient condition consists of two key components. The first is the classical nonnegativity of the Block-Marschak polynomials, a condition that appears in Falmagne's characterization. The second is a novel condition that consists of inequalities involving the summation and subtraction of the Block-Marschak polynomials. This novel condition highlights that the nonnegativity of the Block-Marschak polynomials alone is insufficient to rationalize a dataset. Instead, balances across the values of the Block-Marschak polynomials are crucial when dealing with incomplete datasets.

Our findings have significant implications for the widely used outside option approach.  We show that the outside option approach disregards all but the most basic inequalities of the second condition. Importantly, this conclusion is made possible by the nonredundancy of our characterization.\footnote{Without nonredundancy, it could be possible, for instance, that some of inequalities in the second condition are implied by other inequalities. With nonredundancy, each inequality  in the second condition is an independent feature of random utility rationalization. Thus we can identify which implication (i.e., inequality) is  ignored by the outside option approach.} In particular, our results show that even if the dataset is not rationalizable by any random utility model, the outside option approach may erroneously conclude that the population of agents' choices is rationalizable by a random utility model.

To establish our results, we translate the problem into a network flow problem. This approach was originally developed by \cite{fiorini2004short}, who provided a shorter proof of the axiomatization of \cite{falmagne1978representation}. Our methodological innovation is employing a feasibility theorem in network flow theory, which provides a necessary and sufficient condition for the existence of a desirable network flow. This novel tool offers clear insights even in cases where some alternatives are missing. Moreover, we demonstrate that our methodology not only facilitates characterization  but also significantly enhances efficiency when testing our conditions in given datasets.

To further demonstrate the limitation of the outside option approach, we obtain bounds for the missing choice frequencies based on our methods and then compare the bounds with the naive bounds  obtained from the outside option approach.  For this purpose, we provide an efficient algorithm to derive tight bounds for the missing choice frequencies. Our algorithm leverages the network flow structure of the problem, offering a practical tool for further analysis.  Using real datasets, we show that our method yields significantly tighter bounds compared to the naive bounds derived from the outside option approach. Moreover, we show that our methods correctly capture the relative desirability of unobservable alternatives, while the outside option approach misses that information. See Section \ref{sec:app} for details.

\subsection{Motivating Example}\label{subsection:ex}

To demonstrate the importance of analyzing datasets without relying on the outside-option approach, we provide an example, in which the original dataset cannot be rationalized by any random utility model, but under the outside-option approach, the  dataset appears to conform to a random utility model. Moreover, with the outside option approach, we lose critical information about the desirability of alternatives.

Let $\{a,b,c,d\}$ be the set of alternatives. Suppose that we do not observe choice frequencies of alternatives $c$ and $d$.  The left table in Table \ref{tale:rho} shows the observable choice frequencies $\rho(D,x)$ of alternative $x$ from $D$.  

As mentioned, in empirical analyses,  researchers often aggregate all of the unobservable alternatives into a single outside option $x_0$ and then consider a reduced dataset $\hat{\rho}$ in which an outside option $x_0$ represents the set of all unobservable alternatives $\{c,d\}$. Thus, the right table shows the reduced dataset $\hat{\rho}$. (See Definition 3.4 in Subsection 3.1 of the paper for the formal definition of the reduced dataset.)

\begin{table}[h]
    \centering
    \begin{tabular}{c}
        \begin{minipage}{0.52\textwidth}
            \centering
            $\rho$
\[\begin{array}{c|llllll}
\tikz{\node[below left, inner sep=1pt] (def) {$D$};%
      \node[above right,inner sep=1pt] (abc) {$x$};%
      \draw (def.north west|-abc.north west) -- (def.south east-|abc.south east);}
                & a & b & c & d \\
                \hline
           \{a,b\} & 1/2 & 1/2 & - & -\\
            \{a,c\} & 2/3 & - & 1/3 & -\\
            \{a,d\} & 2/3+\ep & - & - & 1/3-\ep\\
            \{b,c\} & - & 1/2 & 1/2 & -\\
           \{ b,d \}& - & 1/2+\ep & - & 1/2-\ep\\
          \{  c,d\} & - & -
             & ? & ?\\
           \{ a,b,c \}& 1/3 & 1/6  & 1/2 & -\\
            \{a,b,d\} & 1/3 +\ep/2 & 1/6 +\ep/2& - & 1/2-\ep\\
            \{a,c,d\} & 1/3 & - & ? & ?\\
         \{b,c,d\} & - & 1/3 & ? & ?\\
      \{a,b,c,d\}& 1/6 & 1/6 & ? & ?
\end{array}\]
            % \caption{$\rho(D,x)$}
            \label{tab:rho-difficult}
        \end{minipage}
        \hfill
        \begin{minipage}{0.52\textwidth}
            \centering
            $\hat{\rho}$
\[\begin{array}{c|ccccc}
\tikz{\node[below left, inner sep=1pt] (def) {$D$};%
      \node[above right,inner sep=1pt] (abc) {$x$};%
      \draw (def.north west|-abc.north west) -- (def.south east-|abc.south east);}
                & a & b & x_0 & \\
                \hline
           \{a,b\} & 1/2 & 1/2 & -\\
            \{a,x_0\} & 1/3 & - & 2/3\\
         \{b,x_0\} & - & 1/3 & 2/3\\
      \{a,b,x_0\}& 1/6 & 1/6 & 2/3
\end{array}\]

% \caption{$\hat{\rho}(D,x_0)$}
            % \label{tab:rho-difficult}
        \end{minipage}
    \end{tabular}
    \caption{The left shows the original choice probabilities $\rho(D,x)$ and the right shows the reduced choice probabilities $\hat{\rho}(D,x)$. The rows indicate choice set $D$ and the columns indicate alternative $x$ and question marks indicate unobservable choice probabilities. We assume $0\le \ep \le 1/3$.}\label{tale:rho}
\end{table}

As the reduced dataset $\hat\rho$ contains coarser information on the choice behavior of consumers than $\rho$, our main motivation is to identify exactly what information is lost in going from $\rho$ to $\hat\rho$ and then to find methods that take advantage of the additional information available in $\rho$ but not $\hat\rho$. We will  conclude that the reduced dataset not only loses information about the rationalizability of $\rho$, but also the relative desirabilities of all alternatives.

Given this motivation, we now 
turn to the example choice data in Table \ref{tale:rho}. Note first that when there is only one unobservable alternative in a choice set, we can calculate the frequency of the unobservable alternative by subtracting the sum of observable choice probabilities from one. (For example, $\rho(\{a,c\},c)=1- \rho(\{a,c\},a)= 1/3$.)

Note also that $\rho$ cannot be rationalized by any random utility model. In fact, monotonicity is violated in $\rho$ (i.e., $\rho(\{a,c\},c) \cancel{\ge} \rho(\{a,b,c\},c)$)\footnote{Since $c$ is an unobservable alternative, we cannot calculate the BM polynomial with respect to the alternative; thus the standard approach using BM polynomials cannot be applied. We constructed this example to violate monotonicity for simplicity and in order to make it obvious that the dataset is not RU-rationalizable. In general, non-RU rationalizability can be more difficult to prove without using our characterizing conditions.}. This violation disappears in the reduced model $\hat{\rho}$ and in fact one can verify that $\hat{\rho}$ is rationalizable by a random utility model, even though the original dataset is not.\footnote{The reduced dataset is rationalizable by a distribution $\mu$ such that $\mu(\succ_1)=1/3$, $\mu(\succ_2)=1/3$, $\mu(\succ_3)=1/6$, and $\mu(\succ_4)=1/6$, where 
$x_0 \succ_1 a \succ_1 b$, $x_0 \succ_2 b \succ_2 a$, $a \succ_3 b \succ_3 x_0$, and $b \succ_4 a\succ_4 x_0$.}

Additionally, in the original dataset, $\rho(D,a) \ge \rho(D,b)$ for all $D$ with $\{a,b\} \subseteq D$ and $\rho(D\cup a,a)\ge \rho (D \cup b, b)$ for any $D$ such that $a \not \in D$ and $b \notin D$.
% \textcolor{red}{Thank you for clarifying this. But then there is no set that satisfies $a \notin D,$ $b \notin D,$ and $\{c, d\} \subsetneq D.$}
Moreover, these inequalities are strict when $D = \{c\}$ or $D = \{d\}$. Thus it can be inferred that $a$ is strictly more desirable than $b$. However, in the reduced dataset, the desirability of $a$ and $b$ is indistinguishable. Consequently, the estimated desirability of $a$ and $b$ may be biased when using the outside-option approach.

Another consequence of the outside option approach is that in the reduced dataset $c$ and $d$ are indistinguishable. In the original dataset, however, we can compare choice sets with different unobservable alternatives to learn about their relative desirability. For example, we have that $\rho(D \cup c) > \rho(D \cup d)$ for all non-empty $D$ such that $c \notin D$ and $d \notin D$. We can also make inferences about the choice frequencies of $c$ and $d$ even when both are in the choice set. For example, suppose we only observe the choice sets $\{b,c,d\}$, $\{b,c\}$, and $\{b,d\}$. Assuming monotonicity of choice frequencies, we may conclude $\rho(\{b,c,d\},c) \le \rho(\{b,c\},c) = 1/2$ and $\rho(\{b,c,d\},d) \le \rho(\{b,d\},d) = 1/2 -\ep$.\footnote{Monotonicity means that $\rho(E,x) \le \rho(D,x)$ for all $x \in D \subseteq E$.}. Then $\rho(\{b,c,d\},c) = 1-\rho(\{b,c,d\},b) - \rho(\{b,c,d\},d) \ge 1-1/3-(1/2 -\ep) = 1/6 + \ep$. Therefore when $\ep > 1/6$, we have that $\rho(\{b,c,d\},d) \le 1/2 - \ep < 1/3 < 1/6 + \ep \le \rho(\{b,c,d\},c)$. We conclude that, under monotonicity (and therefore RUM) the probability that $c$ is chosen from $\{b,c,d\}$ is higher than the probability that $d$ is chosen from $\{b,c,d\}$.\footnote{In the example of Table \ref{tale:rho}, there are only two unobservable alternatives, so we can directly infer the choice frequency of each unobservable alternative when the choice set contains only one unobservable alternative. However, even when there are more than two unobservable alternatives in a choice set, we can still learn about the sum of their choice frequencies, which enables us to compare the desirability of different unobservable alternatives. See Section \ref{sec:app} for further details. In this sense, our conclusion that our methods can extract more information than the outside option approach does not depend on there being only two unobservable alternatives in the dataset.} All of these conclusions suggest that $c$ is more desirable than $d$, which cannot be concluded from the reduced dataset.

This example highlights the importance of analyzing the original dataset without introducing an outside option.
More specifically, by identifying which alternatives are available to the agent as much as possible, we can infer more information about the agents' preferences.
In the previous example, we showed that it is possible to learn more about the relative desirability between the observable alternatives $a$ and $b$ as well as the unobservable alternatives $c$ and $d$ simply by observing the choice probabilities on choice sets containing some but not all unobservable alternatives.

These observations indicate the limitations of the outside option approach. In this paper, we formally show the limitations of the outside option approach by first characterizing the full implications of the random utility model when some choice frequencies are unobservable (Theorem \ref{thm:characterization}); and by showing what implications of the random utility model are lost in the outside-option approach (Proposition \ref{prop:outside}). Moreover, in Section \ref{sec:app}, we formalize our methods to obtain bounds of the unobserved choice frequencies, which gives us information about desirability of the unobservable alternatives. (See Remark \ref{rem:bound_mono} and Proposition \ref{coro:bound2}.) In contrast, the outside option approach yields trivial bounds that prevent us from learning about the relative desirability of these unobservable alternatives.

\subsection{Related Literature}

\vspace{-0.1cm}

We now briefly discuss the related literature. 
It is well known that obtaining a nonredundant characterization of the random utility model with incomplete datasets in general is a challenging problem.\footnote{See \cite{linear_ordering_polytope} for a survey. A more recent paper by \cite{sprumont2022regular} also highlights the difficulty of this problem.}
 For example, when choice frequencies are observed only on binary choice sets, it has been unknown how to obtain a nonredundant characterization of the random utility model since the 1980s; only for the case where the number of alternatives is less than eight, the nonredundant characterizations have been obtained.\footnote{See \cite{reinelt1993note}. This is in contrast with logit model: The logit models  can be axiomatized with binary choice sets.  See \cite{luce2005individual}, \cite{cerreia2023multinomial}, and \cite{cerreia2021canon} for the axiomatization of the logit models.  Recently \cite{petri2023binary} has obtained an axiomatization of a special case of the random utility model, {\it single-crossing random utility models} introduced by \cite{apesteguia2017single} in a binary choice setup.}

\cite{mcfad91} provide a characterization of random utility models without assuming that all menus are observable. However, they do assume that, for each observable menu, the choice frequencies of all alternatives in that menu are observed; thus their approach is not applicable to our setup.\footnote{More recently, \cite{turansickgraaphicalmethods} provides an alternative axiomatization using the network flow approach.} Moreover, our result differs from their results in that their characterization involves infinitely many inequalities and entails redundancy, while ours consists of finite inequalities and  contains no redundant ones.\footnote{\cite{mcfad91} shows that a stochastic choice function is rationalized by a random utility model if and only if all polynomials, which we call the Mcfadden and Ricther poliynomials in section  \ref{sec:mc},  are nonnegative. In section \ref{sec:mc} of the online appendix,  we first show that their characterization fails in our set up where choice frequencies of some alternatives are not observable. Then, we generalize the result by \cite{mcfad91} to make their approach applicable to our setup. However, we will explain that the Mcfadden and Ricther polynomials contain redundancy in an essential way, unlike the Block-Marschak polynomials.}

Other than the papers mentioned so far, only a few papers have studied the characterization of the random utility model with incomplete data. \cite{mcfadden2006revealed} considers a nested structure of choice sets: if choice frequencies are observable in a menu $D,$ then choice frequencies are observable in any larger set $E$ (i.e., $E \supseteq D)$. In this paper, we discuss the random utility characterization under this restriction of available choice sets. \cite{suck2016regular} addresses the truncated complete choice environment, in which only choice sets with at least $k \geq 2$ alternatives are observable. Nevertheless, to the best of our knowledge, our setup, in which the choice frequencies of some alternatives are missing, is novel in the literature. Moreover, these results are special cases of our theorem---cases in which there are no unobservable alternatives.

As mentioned, we use the network-flow theory to prove our results. Since the publication of \cite{fiorini2004short},  some more recent papers have used the network-flow theory to investigate different topics on random utility models. \cite{turansick2022identification} and \cite{chambers_identification} study the identification of random utility models. \cite{chambers2021correlated} provide a new model of random utility with more than one agent.  \cite{doignon2022adjacencies}  characterize the adjacency of vertices and facets of a {\it mulitiple-choice polytope} (i.e., the set of random utility models). None of these papers study incomplete datasets.

Finally, we highlight several studies that develop empirical methods for testing random utility models. 
\cite{kitamura2018nonparametric} propose a nonparametric statistical test for random utility models, focusing on datasets where choices are made from budget sets—a setting distinct from ours. \cite{dean2022better} extend this framework to examine choice overload. 
While their methods are applicable to a variety of choice datasets, they are computationally intensive due to the inefficiency of the model representation they rely on. They rely on the representation of the set of random utility models as the convex hull of vertices (V-representation), but the number of vertices is huge. In contrast, our characterization, which takes advantage of a specific structure of data incompleteness, provides a more computationally efficient foundation for testing based on facet-defining hyperplanes (H-representation) of the random utility model.  See section \ref{sec:online_polytope} of the online appendix for a further explanation.

\vspace{-0.5cm}

\section{Model}

\vspace{-0.2cm}

Let $X$ be a finite set of alternatives. Let $X^* \subseteq X$ be the set of \textit{unobservable alternatives}. We assume that the choice frequencies of the elements of $X^*$ are not observable (even if a choice set includes the alternatives). Let $\tilde{X}\coloneqq  X\setminus X^*$ be the set of \textit{observable alternatives}. 

Let $\D \subseteq 2^X \setminus \emptyset$ be the set of choice sets. Unlike \cite{falmagne1978representation}, we do not assume that $\D = 2^X \setminus \emptyset$. Note that $(\D, \subseteq)$ is a partially ordered set, where $\subseteq$ is the set inclusion. Like \cite{mcfadden2006revealed}, we assume that $\D$ is an {\it upper set} (i.e., $\D$ satisfies the following: $D \in \D,\ E \supseteq D \Longrightarrow E \in \D$). To make our notation simple, let $\M\coloneqq  \{(D,x)\in \D \times \tilde{X} \mid x \in D \}$

Note that for any $(D,x)$, the choice frequency over $(D,x)$ is observable if and only if $(D,x) \in \M$ (i.e., $x \in \tilde{X}$ and $D \in \D$).

\begin{definition}
A nonnegative vector $\rho \in \Re_+^{\M}$ is called an {\it incomplete dataset} if it satisfies the following conditions: for any $D \in \D$, 
\bit
\item[(i)]  if $D \subset  \tilde{X}$, then  $\sum_{x \in D}\rho(D,x)=1$; and 
\item[(ii)] if  $D \not \subseteq \tilde{X}$, then $\sum_{x \in D\cap \tilde{X}}\rho(D,x)\le 1$. 
\eit
\end{definition}

When the context is clear, we will simply call $\rho$  a {\it dataset} instead of an {\it incomplete dataset}. If $\rho$ is an incomplete dataset, then $\rho$ is not defined on $(D,x) \not \in \M$.  This does not mean that we cannot know anything about the choice frequencies of elements in $X^*.$ When $x^* \in X^*$ is the only one unobservable alternative in the choice set $D$ (i.e., when $D\cap X^*=\{x^*\})$, we can calculate $\rho(D,x^*)$ as $
    \rho(D,x^*)
    =
    1
    -
    \sum_{y \in D \setminus x^*}
    \rho(D, y)$
    .
    \footnote{We often omit braces for singletons. Here, $D \setminus x^*$ means $D \setminus \{x^*\}.$}

\begin{definition} A nonnegative vector $\rho^* \in \Re_+^{\{(D,x) \mid x \in D \in 2^X\}}$ is called a {\it complete} dataset if,  for any $D \subseteq X$, $\sum_{x \in D}\rho^*(D,x)=1$.
 \end{definition}

\vspace{-0.4cm}
\subsection{Examples} \label{sec:examples}

\vspace{-0.1cm}

\noindent\textbf{Example 1 (Transportation):}
An analyst is often able to estimate the market share of public transportation methods (i.e., bus or train) based on the revenues of bus or train companies. However, it is sometimes difficult for the analyst to know separately the percentages  of people who drive or walk. In this case, $X=\{ \text{walk, drive, bus, train}\}$, $\tilde{X}=\{ \text{bus, train}\}$ and $X^*=\{ \text{walk, drive}\}$. An example of the set of choice sets is $\D = \big\{\{w,b,t\}, \{w,d, b\}, \{w,d, t\}, \{w,d, b,t\}\big\}$, where $w, d, b,$ and $t$ stand for {\it walk}, {\it drive, bus}, and {\it train}, respectively. This set $\D$ can be obtained from the assumption that depending on the location of homes, some transportation methods are not available.\footnote{Assuming that the analyst believes that the distribution of preferences is independent of the location of homes, it would make sense  to find a single distribution over ranking that describes the choice frequencies across $\D$.}\\
%In this case, the set of choice sets that have at least two unobservable alternatives is $\D^*=\big\{\{w,d, b\}, \{w,d, t\}, \{w,d, b,t\}\big\}$. 

\noindent\textbf{Example 2 (Market Shares of Private Companies):}
One definition of market share is  the percentage of a company's total sales divided by the market's total sales. The market's total sales can be estimated by consumer surveys. However, private companies occasionally  do not disclose their financial information, including their total sales; thus the market shares of private companies are sometimes unobservable. For example, suppose that there are four  companies (i.e., $X = \{a, b, c, d\}$).  If companies $c$ and $d$ are private companies, then we do not know their sales (i.e., $c,d \not \in \tilde{X}$).  Other companies $\{a,b\}$ are public and the information from these companies is disclosed.  In addition, the availability of products may vary across stores, which would give a variation of choice sets (i.e., $\D$). \\

\noindent\textbf{Example 3 (School Choice for Private Schools):}
Applicants submit their choices among public schools so the government knows the percentage of students choosing each public school. However, it might not have access to information on how many students choose each private school. For example, suppose that there are four schools (i.e., $X=\{a,b,c,d\}$). Among them, $c$ and $d$ are private schools for which we do not know the choice frequencies (i.e., $c,d \not \in \tilde{X}$).  The availability of schools may depend on the location of homes, which would give a variation of choice sets (i.e., $\D$).

\begin{comment}
\begin{remark}
In the existing empirical literature on industrial organization (IO), it is common practice for researchers to aggregate all unobservable alternatives into a single category, commonly referred to as the {\it outside option}. For example, in Example 1, the set $\{\text{bus}, \text{train}\}$ would be treated as one outside option, say "public transportations".

As mentioned in the introducdtion, We will demonstrate that the outside option approach overlooks certain implications of the random utility model and may result in biased estimations.  See Section  \ref{sec:imp_outside_option} for the details.   
\end{remark}
\end{comment}

\vspace{-0.4cm}

\subsection{Random-Utility Rationalization}

\vspace{-0.1cm}

Let $\L$ be the set of linear orders on $X$, i.e., binary relations that are irreflexive, asymmetric, transitive, and weakly complete.\footnote{A binary relation is weakly complete if, for any distinct elements $x, y \in X$, either $x \succ y$ or $y \succ x$.} 

\begin{definition}
An incomplete dataset $\rho$ is \it{random-utility (RU) rationalizable} if there exists $\mu \in \Delta(\L)$ such that, for any $(D,x) \in \M$, $
\rho(D,x)=\mu(\ \succ \in \L \mid x \succ y \text{ for all } y \in D\setminus  x)$. We then say that $\mu$ rationalizes $\rho$.
\end{definition}

\begin{definition} Let  $p \in \Re^{\{(D,x) \mid x \in D \in 2^X\}}$. For any $(D,x)$ such that $x \in D \subseteq X$, define $
K(p,D,x)= \sum_{E:E \supseteq D} (-1)^{|E \setminus D|} p(E,x)$. $K(p,D,x)$ is called a {\it Block-Marschak (BM) polynomial}.\footnote{As we will explain below, the BM polynomial is crucial concept to characterize random utility models. The BM polynomial appears in other contexts. For example, \cite{brady2016menu} observe that one of their axioms is equivalent to a multiplicative version of the BM polynomial.}
\end{definition}

%\textcolor{blue}{\fbox{Both $\subseteq$ and $\subset$ are used to represent weak inclusion. Should be unified to $\subset.$}}

%KS Allmost all places we are using $\subseteq$ and $\subsetneq$.  Lets us them instead of \subset. Please change the section you wrote accordingly.

Note that, given an incomplete dataset $\rho \in \Re^{\M}_+$, the BM polynomial $K(\rho,D,x)$ can be calculated if and only if  $(D,x) \in \M$ (i.e.,  $x  \in \tilde{X}$ and $D \in \D$). The next remark provides a meaning of a BM polynomial given a complete dataset $\rho^*$:

\begin{remark}\label{rem:rhostark}
Assuming that a complete dataset $\rho^*$ is RU-rationalizable, we can provide a meaning of a BM polynomial. Suppose that  there exists  $\mu \in \Delta(\L)$ such that, for any $x \in D \subseteq X$, $
\rho^*(D,x)=\mu(\ \succ \in \L \mid x \succ y \text{ for all } y \in D\setminus  x)$. By the M\"{o}bius inversion formula, 
it follows that 
\begin{equation}\label{eq:meaning_BM}
K(\rho^*,D,x)=\mu(\succ\in \L| D^c\succ x \succ D\setminus x),
\end{equation}
where $D^c \succ x$ means that $y \succ x$ for all $x \in D^c$ and $x \succ D\setminus x$ means that $x \succ y$ for all $x \in D \setminus x$.\footnote{To see this notice that $\mu(\succ\in \L| D^c\succ x \succ D\setminus x)= \mu(\bigcup_{E \supseteq D}\{\succ\in \L| E^c\succ x \succ E\setminus x\})=\sum_{E \supseteq D} \mu(\succ\in \L| E^c\succ x \succ E\setminus x)$. Thus, we have $\rho^*(D,x)=\sum_{E \supseteq D} \mu(\succ\in \L| E^c\succ x \succ E\setminus x)$. By applying the M\"{o}bius inversion to this equation, we obtain $\mu(\succ\in \L| D^c\succ x \succ D\setminus x)=K(\rho^*,D,x)$.} Thus, $K(\rho^*, D, x)$ can be interpreted as a measure of population of agents whose preference satisfies $D^c\succ x \succ D\setminus x$.
\end{remark}

%\textcolor{red}{
%\textbf{HK wants to define this:}
%Let $\M^{**} = \{(D, x) \in 2^X \times X \mid x \in D, x \in X^* \ \text{or} \ D \notin \D\}.$
%Note that given an incomplete dataset $\rho \in \Re^{\M}$ and $x \in D,$ the BM polynomial $K(\rho,D,x)$ can be calculated if and only if $(D, x) \notin \M^{*}.$

%KS Lets define in section 4 because we will not use the notation until the section and the reader will forget anyway

%\textcolor{blue}{\fbox{for $D \in \D?$}}

\vspace{-0.5cm}

\section{Main Results} \label{sec:theorem}

\vspace{-0.2cm}

Remember that $X^*\coloneqq X \setminus \tilde{X}$ is the set of unobservable alternatives. Recall that $(2^{X^*}, \subseteq)$ is a partially ordered set with the set inclusion $\subseteq.$ 
Consider a collection $\E$ of subsets of $X^*$; we assume that $\E$ is an upper set.\footnote{Recall the property of an upper set: if $D \in \E, D\subseteq E    \Longrightarrow E \in \E$. In the example in which $X^*=\{d,e\}$, all upper sets in $2^{X^*}$ are $
    \emptyset,
    \
    \{\{d, e\}\},
    \
    \{\{d, e\}, \{d\}\}, 
    \
    \{\{d, e\}, \{e\}\}$, 
    $\{\{d, e\}, \{d\}, \{e\}\}$, and    $\{\{d, e\}, \{d\}, \{e\}, \emptyset\}
    .$ The complement $\E^c$ is a {\it lower set} (i.e., $\E^c$ satisfies the following: $E \in {\mathcal E}^c, D\subseteq  E\implies D \in \E^c$). We use the concept of lower set in the proof.}

To characterize the RU-rationalizability of incomplete data, the following collection of choice sets is  fundamental.

\begin{definition}
A nonempty collection $\C$ of subsets of $X$ is called a {\it test collection} if  there exists a set $A \subseteq \tilde{X}$ of observable alternatives and a nonempty  upper set $\E \subseteq 2^{X^*}$ of unobservable alternatives such that $\C = \{A \cup E \mid E \in \E\}.$   Moreover, the test collection is said to be {\it essential} if $\emptyset \neq A \neq \tilde{X}$ and $\E \neq 2^{X^*}.$
\end{definition}

The following is our main theorem.

\begin{theorem} \label{thm:characterization}
(a) An incomplete dataset $\rho \in \Re_+^{\M}$ is RU-rationalizable if and only if the following two conditions hold:
\bit
\item (i) for any $(D,x) \in \M$ such that  $1<|D|<|X|$, the polynomial $K(\rho, D, x)$ is nonnegative; and
\item (ii)
for any essential test collection $\C \subseteq \D$,
\eit
\begin{equation}\label{eq:th1}
\Bigg(\sum_{(D,x): D \in {\C},  D\cup x \not\in {\C}}  K(\rho, D \cup x, x)- \sum_{(F,y): F \not\in {\C}, F\cup y \in {\C}, y \in \tilde{X}}  K(\rho, F \cup y, y)\Bigg) \ge 0. 
\end{equation}
(b) Moreover, the inequality conditions in (i) and (ii) are independent: for any inequality condition in (i) or (ii), there exists an incomplete dataset $\rho \in \Re_+^{\M}$ that violates the inequality but  satisfies all the other conditions in (i) and (ii).\footnote{Note that by statement (i), such $\rho$ is not RU-rationalizable.}
\end{theorem}

%\textcolor{red}{HK: For the second term of (1), $F \cup y \in \C$ and $y \notin X^*$ imply $F \notin \C$? Oh ok, $y$ can be in $F.$}

We first make comments on  statement (a) of the theorem. Recall that for any $(D, x)$, the BM polynomial $K(\rho, D, x)$ is computable based on the observable data if and only if $(D,x) \in \M$. Thus, condition (i) is testable. Also, when $\C$ is a test  collection, $D \in \C$ and $D \cup x \notin \C$ imply that $x \in \tilde{X}$.\footnote{Since $\C$ is a test collection, $\C=\{A \cup E| E \in \E\}$ for some $A\subseteq \tilde{X}$ and an upper set $\E \subseteq 2^{X^*}$.  If $x \in X^*$, then $D \in \C$ implies $D \cup x \in \C$ by the definition of test collections (especially by the fact that $\E$ is an upper set).} Thus, the first term as well as the second term in condition (ii) can be calculated based on the available data; so the condition (ii) is also testable.

The necessity of condition (i) in Theorem \ref{thm:characterization} follows from \cite{falmagne1978representation}, who shows that a {\it complete} dataset is RU-rationalizable if and only if all BM polynomials are nonnegative.\footnote{In fact, when $\tilde{X}=X$ (i.e., $X^*=\emptyset$), our theorem reduces to the statement of \cite{falmagne1978representation} and \cite{mcfadden2006revealed}, although our proof does not rely on their proofs.} Novel conditions appear in (ii), which  means that the nonnegativity of the BM polynomials is insufficient for the dataset to be RU-rationalizable because balances across the values of BM polynomials are essential for RU-rationalizability when the dataset is incomplete. For example, one BM polynomial being too large may not be a good sign for RU-rationalizability. In Remark \ref{rem:mean} and Subsection \ref{sec:sketch}, we provide a further explanation of condition (ii).

%the empirical literature on industrial organization

As mentioned, in empirical analyses, researchers often consolidate all unobservable alternatives into a single alternative, often referred to as the {\it outside option}. 
%This practice of aggregation persists even when the components of $X^*$ are clearly identified.
In Proposition \ref{prop:outside}  of the subsequent section, we demonstrate that the outside option approach fails to account for substantial implications of RU-rationalizability. Specifically, it disregards all inequalities specified in condition (ii), except in instances where the essential test collection is a singleton.

Statement (b) is a crucial element of Theorem \ref{thm:characterization}; it not only provides a necessary and sufficient condition but also ensures nonredundancy.\footnote{Geometrically, our theorem delineates all facet-defining inequalities of the {\it random utility polytope}. Refer to Section \ref{sec:online_polytope} in the online appendix for the polytope's definition.} This nonredundancy is essential in elucidating the limitations of the outside option approach. See the discussion after Proposition \ref{prop:outside}. 

The nonredundancy of our conditions is in contrast to the result obtained by \cite{mcfad91}. As we will explain in Section \ref{sec:mc} of the online appendix, the conditions in \cite{mcfad91} are redundant in an essential way. Statement (b) in Theorem \ref{thm:characterization} may be surprising given the known difficulty of obtaining a nonredundant characterization of the random utility model when datasets are incomplete.

In the next remark, we provide a meaning of condition (ii) as we explained the meaning of the BM polynomials in Remark \ref{rem:rhostark}.

\begin{remark}\label{rem:mean}
Suppose that an incomplete dataset $\rho$ is RU-rationalizable by $\mu$. To understand the meaning of condition (ii) in Theorem \ref{thm:characterization}, fix  a test collection $\C$ and assume that $\C$ is a singleton set containing a set $D$. Since $\C$ is a test collection, we have $X^* \subseteq D$. Then,  the left hand side of  (\ref{eq:th1}) simplifies to 
\begin{equation}\label{eq:mean_new}
\sum_{x^* \in X^* \cap D}\mu(\succ \in \L|  D^c \succ x^* \succ D \setminus x^*).
\end{equation}
Thus the meaning of the condition (ii) for this case is the non-negativity of the measure on population of agents whose preference satisfies $D^c \succ x^* \succ D \setminus x^*$ for some $x^* \in D \cap X^*$.\footnote{We show this in the proof of Proposition \ref{prop:outside}} 

Remember that when the dataset $\rho^*$ is complete, we have  $K(\rho^*, D, x) = \mu(\succ \in \mathcal{L} $ $\mid D^c \succ x \succ D \setminus \{x\})$, as explained in Remark~\ref{rem:rhostark}. This expression closely resembles equation~(\ref{eq:mean_new}). The key difference is that the alternative $x^*$ is unobservable in our setting, and we are summing over measures. This distinction arises because we cannot compute each individual BM polynomial $K(\rho, D,x^*)$. Nevertheless, we still obtain a testable implication on the sum of the unobservable BM polynomials $\sum_{x^* \in D \cap X^*} K(\rho,  D,x^*)$, which  must be nonnegative. For the general expression of condition~(\ref{eq:th1}), see Section~\ref{section:mean_new} in the online appendix.
\end{remark}

\vspace{-0.4cm}

\subsection{Implication for the outside option approach}\label{sec:imp_outside_option}

In this section,  we first formalize the outside option approach. Subsequently, we demonstrate that the outside option approach overlooks all inequalities in condition (ii), except when the essential test collection is a singleton.

We first formalize the outside option approach as follows. In the outside option approach, we represent the set $X^*$ of all unobservable alternatives as a single alternative $x_0$. 
Thus, the set of all alternatives in the outside  option approach is defined as $\hat{X}= \tilde{X} \cup x_0$, where  $\tilde X$ is the set of observable alternatives. Consequently, we consider the {\it reduced} choice sets, denoted by $\hat{\D},$ which aggregates all elements of $X^*$ into the outside option $x_0$.\footnote{We ignore the data on choice sets that contain only some (but not all) element(s) of $X^*$. Formally, $\hat \D
    \coloneqq
    \left\{
        D \in \D
        \mid
        D \cap X^*
        =
        \emptyset
    \right\}
    \cup
    \left\{
        (D \setminus X^*) \cup x_0
        \mid
        D \in \D
        ,
        D \supseteq X^\ast
    \right\}
    \subseteq
    2^{\hat X}
    .$}

\begin{definition}\label{rem:def_outside}\
 Let $\hat{\rho} \in \Re_+^{\{(D,x)| x \in D \in \hat{\D}\}}$ be the reduced dataset on the reduced choice set defined as follows. We define   the choice frequencies of observable alternatives remain the same, i.e., $\hat{\rho}(D,x)=\rho(D,x)$ for all $x\in D\cap \tilde{X}$, where $\rho \in \Re^{\M}_+$ is the original (non-reduced) incomplete dataset.  Then, the choice frequency of the outside option can be defined as $
\hat{\rho}(D,x_0)=1-\sum_{x \in D \cap \tilde{X}}\rho(D,x)$.
\end{definition}

See Table \ref{tale:rho} in Subsection \ref{subsection:ex} for the example of $\hat{\rho}$.

Our definition assumes that the composition of the outside option is constant across choice sets—that is, $x_0$ always represents the same set $X^*$ of unobservable alternatives. Although this corresponds to the idealized setup often assumed in the empirical literature, it may not hold in real datasets: in practice, the interpretation of $x_0$ can vary across choice sets, and this variation is typically unobservable to the analyst. \citet{saito2025aggregate} examines such cases by taking the reduced dataset as the primitive of the model. That paper shows that the implications of random utility are further weakened. The paper also found that the case we study, where the composition of the outside option is fixed across choice sets, represents an idealized benchmark in which the model retains relatively strong implications. As such, our current definition serves as a natural starting point for analyzing the limitations of the outside option approach. We show that even under this idealized scenario, key implications of the random utility model may still be lost.

%Our definition of the outside option does not encompass all scenarios. For instance, some propose introducing an outside option $x_0$ even when only a subset of the unobservable alternatives appears in a menu. \citet{saito2025aggregate} examine such cases and show that the implications of the random utility model become severely limited. By contrast, as \citet{saito2025aggregate} also demonstrate, our definition represents one of  best-case scenarios—where the model retains relatively strong implications and estimation bias is minimal. This makes it a natural starting point for exploring the broader limitations of the outside option approach.

We say that the reduced dataset $\hat{\rho}\in \Re_+^{\{(D,x) \mid x \in D \in \hat{\D}\}}$ is {\it RU-rationalizable} if there exists a probability distribution $\hat{\mu}$ on the set $\hat{\L}$ of linear orders on $\hat{X}$ such that for all $(D,x)$ such that $x \in D \in \D$, $\hat{\rho}(D,x)=\hat{\mu}(\hat{\succ} \in \hat{\L} \mid x\ \hat{\succ}\  y \text{ for all } y \in D \setminus x )$.\footnote{In \cite{saito2025aggregate}, we refer to this as aggregated RU-rationalizability because the distribution $\hat{\mu}$ is defined over linear orders on $\hat{X}$, in which all unobservable alternatives in $X^*$ are aggregated into a single alternative $x_0$.}

%\textcolor{red}{HK: I rewrote the remark a bit. Do we need to make this a remark?}
%KS. I think having remark makes it easier for us to refer the definition of the outside option approach.  Alec and I think definning it verbally makes it more accessible. 

% \begin{lemma}
% \textcolor{red}{HK: The standard definition of signed measures doesn't contain $\sum_{\succ \in \L} \mu(\succ) = 1.$ Also, additivity is necessary.}
% Let ${\cal M}\subseteq \{(D,x) \mid x \in D \subseteq X \}$

% For any stochastic choice function $\rho \in \P$, there exists a signed measure $\mu$ on $\L$ (i.e., $\mu$ is a real-valued function defined on $\L$ such that $\sum_{\succ \in \L} \mu(\succ)=1$) such that for all $(D,x)$ such that $x \in D \in \D$
% \[
% \rho(D,x)= \mu(\succ \in \L \mid x \succ y \text{ for all } y \in D \setminus x). 
% \]
% \end{lemma}

% \begin{proof}
% The result follows from 
% \end{proof}

\begin{proposition}\label{prop:outside}
Let $\rho\in \Re^{\M}_+$ be an incomplete dataset. Suppose that $\rho$ satisfies 
\begin{itemize}
\item condition (i) of Theorem \ref{thm:characterization} and; 
\item condition (ii) of Theorem \ref{thm:characterization} for any  singleton essential test collection $\C \subseteq \D$.
\end{itemize} 
Then the reduced dataset $\hat{\rho}\in \Re_+^{\{(D,x) \mid x \in D \in \hat{\D}\}}$ is RU-rationalizable.
\end{proposition}

To see the implication of the proposition, suppose that an original incomplete dataset 
$\rho$ violates condition (ii) of Theorem \ref{thm:characterization} for some non-singleton essential test collections but satisfies all of the other conditions of Theorem \ref{thm:characterization}. Such $\rho$ exists because of the independence of each inequality as stated in statement (b) of Theorem \ref{thm:characterization}. Statement (a) of Theorem \ref{thm:characterization} implies that $\rho$ is not RU-rationalizable. Nevertheless, Proposition \ref{prop:outside} implies that the reduced dataset $\hat{\rho}$ is RU-rationalizable, thus 
researchers may erroneously conclude that the true data-generating process follows the random utility model. 
 In this way, the reduced dataset $\hat\rho$ loses critical implications of the original dataset $\rho$. In particular, Proposition \ref{prop:outside} shows that the outside option approach discards 
substantial implications of random utility model---all but the most basic inequalities in condition (ii). Note that in this interpretation, we need statement (b) of Theorem \ref{thm:characterization}, which states that each inequality condition in the theorem is independent. This highlights the significance of our nonredundancy result, particularly when contrasted with \cite{mcfad91}, which includes redundant conditions that preclude interpretations like this.

\begin{remark}\label{rmk:examplecondition}   To illustrate the implication of Proposition \ref{prop:outside},  remember the incomplete dataset $\rho$ in Table \ref{tale:rho} in Subsection \ref{subsection:ex}.  Let $\ep=0$. 
\begin{itemize}
\item  One can observe that $\rho$ violates condition (ii) with ${\cal C}=\{\{a,c,d\}, \{a,c\}\}$: 
\[
\Bigg(\sum_{(D,x): D \in {\C},  D\cup x \not\in {\C}}  K(\rho, D \cup x, x)- \sum_{(F,y): F \not\in {\C}, F\cup y \in {\C}, y \in \tilde{X}}  K(\rho, F \cup y, y)\Bigg)  = -\frac{1}{6}
\]
\item Therefore, by Theorem \ref{thm:characterization}, the original dataset $\rho$ is not RU-rationalizable. However, since $\rho$ satisfies condition (i) and condition (ii) when ${\cal C}$ is a singleton, Proposition \ref{prop:outside} implies that the reduced dataset $\hat{\rho}$ is RU-rationalizable, despite the fact that the true dataset $\rho$ is not.
\end{itemize}
\end{remark}

In the forthcoming section (Section \ref{sec:sketch}), we provide an intuition of our proof of statement (a). Moving forward to Section \ref{sec:app}, we apply the theorem to derive insights into unobserved choice frequencies; we identify the sets of possible values for these unobservable frequencies. This section further illustrates the limitations of the outside option approach. In  Section \ref{sec:conclusion}, we provide concluding remarks on the comparison between the outside option approach and our approach.

%\subsection{Tentative: If $\hat{\rho}$ is RU rationalizable, then it satisfies condition (i) and condition (ii) for singleton collection.}

\vspace{-0.4cm}

\subsection{Intuition of the proof }\label{sec:sketch}

\vspace{-0.1cm}

In this subsection, we outline the proof of statement (a) of Theorem \ref{thm:characterization}.  All formal proofs and the proof of statement (b) are in the appendix.

\subsubsection{Network Flow for complete dataset }

\usetikzlibrary{math}

\tikzmath{
    \abcd=0; 
    \abc=-3;
    \abd=-1;
    \acd=1;
    \bcd=3;
    \ab=-5;
    \ac=-3;
    \ad=-1;
    \bc=1;
    \bd=3;
    \cd=5;
    \a=-3;
    \b=-1;
    \c=1;
    \d=3;
    \emp=0;
    \scale=1.5;
} 

\begin{figure}
    \centering
  \begin{tikzpicture}[scale=.5, transform shape]
    \tikzset{vertex/.style = {shape=circle,draw,thick,minimum size=4em}}
    \tikzset{-to/.style = {thick,arrows={-to}}} 
    
    \node[vertex] (abcd) at (\abcd*\scale,5) {\footnotesize $a,b,c,d$};
    
    \node[vertex] (abc) at (\abc*\scale,2.5) {$a,b,c$};
    \node[vertex] (abd) at (\abd*\scale,2.5) {$a,b,d$};
    \node[vertex] (acd) at (\acd*\scale,2.5) {$a,c,d$};
    \node[vertex] (bcd) at (\bcd*\scale,2.5) {$b,c,d$};
    
    \node[vertex] (ab) at (\ab*\scale,0) {$a,b$};
    \node[vertex] (ac) at (\ac*\scale,0) {$a,c$};
    \node[vertex] (ad) at (\ad*\scale,0) {$a,d$};
    \node[vertex] (bc) at (\bc*\scale,0) {$b,c$};
    \node[vertex] (bd) at (\bd*\scale,0) {$b,d$};
    \node[vertex] (cd) at (\cd*\scale,0) {$c,d$};
    
    \node[vertex] (a) at (\a*\scale,-2.5) {$a$};
    \node[vertex] (b) at (\b*\scale,-2.5) {$b$};
    \node[vertex] (c) at (\c*\scale,-2.5) {$c$};
    \node[vertex] (d) at (\d*\scale,-2.5) {$d$};
    \node[vertex] (emptyset) at (\emp*\scale,-5) {$\emptyset$};

    \draw[-to] (abc) -- (abcd);
    
    \draw[-to] (abd) -- (abcd);
    \draw[-to] (acd) -- (abcd);
    \draw[-to] (bcd) -- (abcd);
    
    \draw[-to] (ab) -- (abc);
    \draw[-to] (ac) -- (abc);
    \draw[-to] (bc) -- (abc);
    \draw[-to] (ab) -- (abd);
    \draw[-to] (ad) -- (abd);
    \draw[-to] (bd) -- (abd);
    \draw[-to] (ac) -- (acd);
    \draw[-to] (ad) -- (acd);
    \draw[-to] (cd) -- (acd);
    \draw[-to] (bc) -- (bcd);
    \draw[-to] (bd) -- (bcd);
    \draw[-to] (cd) -- (bcd);
    
    \draw[-to] (a) -- (ab);
    \draw[-to] (b) -- (ab);
    \draw[-to] (a) -- (ac);
    \draw[-to] (c) -- (ac);
    \draw[-to] (a) -- (ad);
    \draw[-to] (d) -- (ad);
    \draw[-to] (b) -- (bc);
    \draw[-to] (c) -- (bc);
    \draw[-to] (b) -- (bd);
    \draw[-to] (d) -- (bd);
    \draw[-to] (c) -- (cd);
    \draw[-to] (d) -- (cd);
    
    \draw[-to] (emptyset) -- (a);
    \draw[-to] (emptyset) -- (b);
    \draw[-to] (emptyset) -- (c);
    \draw[-to] (emptyset) -- (d);
    \end{tikzpicture}
\caption{Network Flow for the case in which $X=\{a,b,c,d\}=\tilde{X}$ and $X^*=\emptyset$ and $\D=2^X \setminus \emptyset$.}\label{fig:0}
\end{figure}

Before providing the sketch of our proof, we provide an overview of the network flow approach used by \cite{fiorini2004short} for a complete dataset $\rho^*$. Recall that a {\it network} is a pair of a node set $\cal N$ and a set of directed arcs (i.e., edges) $\cal A \subseteq \cal N \times \cal N$.  Two nodes $s$ (source) and $t$ (terminal) play special roles as explained below. We set ${\cal N}= 2^X, {\cal A}=\{(D, D \cup x) \mid D \subseteq X, x \not \in D\}, s=\emptyset$, and $t=X$. See Figure \ref{fig:0} for an example. In the setup, each $\emptyset-X$ directed path corresponds to a unique ranking  $\succ$.  For example in Figure \ref{fig:0},  the directed path $\emptyset-\{a\}-\{a,b\}-\{a,b,c\}-X$ corresponds to the ranking: $d\succ c \succ b\succ a$.   

We now construct  a vector (i.e., {\it flow}) $r \in \Re_+^{\{(D,D\cup x) \mid x \in D \in 2^X\}}$ on the network assuming that a complete dataset $\rho^*$ is RU-rationalizable by some $\mu \in \Delta(\L)$, or
\begin{equation}\label{eq:rational}
\rho^*(D,x)=\mu(\ \succ \in \L \mid x \succ y \text{ for all } y \in D\setminus  x) \text{ for all }(D,x) \text{ s.t. }x \in D \subseteq X.
\end{equation}
To each arc of the network, we assign the sum of the values of $\mu(\succ)$ over linear orders $\succ$ such that the directed path corresponding to $\succ$ goes through the arc. Given the construction, the value at an arc $(D \setminus x, D)$ is $\mu(\{\succ \in \L \mid D^c \succ x \succ  D\setminus x\})$.\footnote{To see this, note that $\succ$ passes through the arc $(D \setminus x ,D )$ if and only if $x$ is ranked just above the elements of $D\setminus x$. That is, $\succ$ passes through the arc $(D \setminus x,D )$ if and only if $D^c \succ x \succ D \setminus x$. For example, the the path corresponding to $\succ$ passes through the arc $(\emptyset, \{a\})$ if and only if $a$ is the worst element under $\succ$. Therefore the value assigned to the arc is equal to a measure over rankings whose worst elements are $a$.} By the M\"{o}bius inversion formula, as explained in Remark \ref{rem:rhostark}, this value equals to $K(\rho, D,x)$. Note the constructed flow $r$ satisfies all the following constraints: 
\begin{align}
&\sum_{x \in X} r(X \setminus x,X)=1, \label{eq:0sum1}\\
&\sum_{x \in D} r(D\setminus x, D)= \sum_{y \not \in D} r(D, D \cup y) \text{ for any } D \in 2^X \text{ s.t. } 1\le |D|\le |X|-1,\label{eq:0in=out}\\
&r(D \setminus x, D)=K(\rho^*, D, x)\text{ for all }(D, x) \text{ such that } x \in D \in 2^X.\label{eq:0con} 
\end{align}

Condition (\ref{eq:0sum1}) means that the sum of {\it inflows} to $X$ (i.e. flows going into $X$) must be $1$; the condition (\ref{eq:0in=out}) means that for each node $D$, the sum of inflows to the node $D$ equals to  the sum of {\it outflows} from $D$ (i.e. flows coming out from $D$); finally condition (\ref{eq:0con}) means that the value of flow at an arc equals to the corresponding value of the BM polynomial. 

The above observation shows that the three conditions are necessary for $\rho^*$ to be RU-rationalizable. \cite{fiorini2004short} proved that the conditions are also sufficient.

\begin{lemma}[\cite{fiorini2004short}]\label{lem:fio}
Given a complete dataset $\rho^* \in \Re^{\{(D,x) \mid x \in D \in 2^X\}}_+$, there exists $\mu \in \Delta(\L)$ satisfying (\ref{eq:rational})  if and only if there exists $r \in \Re_+^{\{(D\setminus x,D) \mid x \in D \in 2^X\}}$ satisfying (\ref{eq:0sum1}), (\ref{eq:0in=out}) and (\ref{eq:0con}).
\end{lemma}

It can be shown from the definition of the BM polynomial that (\ref{eq:0sum1}) and (\ref{eq:0in=out}) are satisfied automatically under the assumption of (\ref{eq:0con}). Thus this implies the following:

\begin{theorem*}   
[\cite{falmagne1978representation}]\label{lem:fio}
Given a complete dataset $\rho^* \in \Re^{\{(D,x) \mid x \in D \in 2^X\}}_+$, there exists $\mu \in \Delta(\L)$ satisfying (\ref{eq:rational}) if and only if $K(\rho^*, D, x) \ge 0$ for all $(D, x)$ such that $x \in D \in 2^X$.
\end{theorem*}

\subsubsection{Network flow for incomplete dataset}

Now let's consider the case in which the dataset is incomplete.  Remember in this case that $K(\rho, D,x)$ can be calculated if and only if $(D,x) \in \M$.  We say an arc $(D \setminus x,D)$ is {\it observable} if $(D,x) \in  \M$  (and hence we can calculate the value $K(\rho, D  ,x)$ of a flow at the arc); $(D\setminus x,D)$ is {\it unobservable} if $(D,x) \not\in \M$ (and hence we {\it cannot} calculate the value $K(\rho, D ,x)$ of a flow at the arc). See Figure \ref{fig:lem3}, for the illustration.

We extend the result by \cite{fiorini2004short} to incorporate the case of incomplete datasets as follows:

\begin{lemma}\label{lem:p1-is-p2}
Given an incomplete dataset $\rho\in \Re^{\M}_+$,\\
(P1)\quad there exits  $\mu \in \Delta(\L)$ such that 
\begin{equation}
\rho(D,x)=\mu(\ \succ \in \L \mid x \succ y \text{ for all } y \in D\setminus  x) \text{ for any }(D,x) \in \M
\end{equation}
if and only if \\
(P2)\quad  there exits  $r \in \Re_+^{\{(D\setminus x,D)| x\in D \in 2^X\}}$ satisfying (\ref{eq:0sum1}), (\ref{eq:0in=out}), and 
\begin{equation}
r(D \setminus x, D)=K(\rho, D, x)\text{ for all }(D, x) \in \M.
\end{equation}
\end{lemma}

\usetikzlibrary{math}

\tikzmath{
    \abcd=0; 
    \abc=-3;
    \abd=-1;
    \acd=1;
    \bcd=3;
    \ab=-5;
    \ac=-3;
    \ad=-1;
    \bc=1;
    \bd=3;
    \cd=5;
    \a=-3;
    \b=-1;
    \c=1;
    \d=3;
    \emp=0;
    \scale=1.5;
} 

\begin{figure}
    \centering
  \begin{tikzpicture}[scale=.5, transform shape]
    \tikzset{vertex/.style = {shape=circle,draw,thick,minimum size=4em}}
    \tikzset{-to/.style = {thick,arrows={-to}}} 
    
    \node[vertex] (abcd) at (\abcd*\scale,5) {\footnotesize $a,b,c,d$};
    
    \node[vertex] (abc) at (\abc*\scale,2.5) {$a,b,c$};
    \node[vertex] (abd) at (\abd*\scale,2.5) {$a,b,d$};
    \node[vertex] (acd) at (\acd*\scale,2.5) {$a,c,d$};
    \node[vertex] (bcd) at (\bcd*\scale,2.5) {$b,c,d$};
    
    \node[vertex] (ab) at (\ab*\scale,0) {$a,b$};
    \node[vertex] (ac) at (\ac*\scale,0) {$a,c$};
    \node[vertex] (ad) at (\ad*\scale,0) {$a,d$};
    \node[vertex] (bc) at (\bc*\scale,0) {$b,c$};
    \node[vertex] (bd) at (\bd*\scale,0) {$b,d$};
    \node[vertex] (cd) at (\cd*\scale,0) {$c,d$};
    
    \node[vertex] (a) at (\a*\scale,-2.5) {$a$};
    \node[vertex] (b) at (\b*\scale,-2.5) {$b$};
    \node[vertex] (c) at (\c*\scale,-2.5) {$c$};
    \node[vertex] (d) at (\d*\scale,-2.5) {$d$};
    \node[vertex] (emptyset) at (\emp*\scale,-5) {$\emptyset$};

    \draw[-to, dashed, orange] (abc) -- (abcd);
    
    \draw[-to, dashed, orange] (abd) -- (abcd);
    \draw[-to] (acd) -- (abcd);
    \draw[-to] (bcd) -- (abcd);
    
    \draw[-to, dashed, orange] (ab) -- (abc);
    \draw[-to] (ac) -- (abc);
    \draw[-to] (bc) -- (abc);
    \draw[-to, dashed, orange] (ab) -- (abd);
    \draw[-to] (ad) -- (abd);
    \draw[-to] (bd) -- (abd);
    \draw[-to, dashed, orange] (ac) -- (acd);
    \draw[-to, dashed, orange] (ad) -- (acd);
    \draw[-to] (cd) -- (acd);
    \draw[-to, dashed, orange] (bc) -- (bcd);
    \draw[-to, dashed, orange] (bd) -- (bcd);
    \draw[-to] (cd) -- (bcd);
    
    \draw[-to] (a) -- (ab);
    \draw[-to] (b) -- (ab);
    \draw[-to, dashed, orange] (a) -- (ac);
    \draw[-to] (c) -- (ac);
    \draw[-to, dashed, orange] (a) -- (ad);
    \draw[-to] (d) -- (ad);
    \draw[-to, dashed, orange] (b) -- (bc);
    \draw[-to] (c) -- (bc);
    \draw[-to, dashed, orange] (b) -- (bd);
    \draw[-to] (d) -- (bd);
    \draw[-to, dashed, orange] (c) -- (cd);
    \draw[-to, dashed, orange] (d) -- (cd);
    
    \draw[-to] (emptyset) -- (a);
    \draw[-to] (emptyset) -- (b);
    \draw[-to, dashed, orange] (emptyset) -- (c);
    \draw[-to, dashed, orange] (emptyset) -- (d);
    \end{tikzpicture}
\caption{Network Flow\label{fig:lem3} for the case in which $X=\{a,b,c,d\}$, $\tilde{X}=\{a,b\}$, and $X^*=\{ c,d\}$ and $\D=2^X \setminus \emptyset$.  The solid arrows correspond to observable arcs: the dotted arrows correspond to unobservable arcs.}
 
% \begin{scriptsize}
% Note: The figure shows the  Boolean lattice of degree four, which corresponds to the network defined by (\ref{eq:network_def}) for the case in which $X=\{a,b,c,d\}$ and $X^*=\{ c,d\}$ and $\D=2^X \setminus \emptyset$.  The solid arrows correspond to observable arcs: the dotted arrows correspond to unobservable arcs.
% \end{scriptsize}
\end{figure}

Note that unlike the case of complete data, Fiorini's approach (i.e.,  mapping the RU-rationalizability problem into a network flow) does not provide direct testable conditions without existential quantifiers. 
%as (P2) requires the existence of a flow that fills in the unobservable arcs to satisfy all of the conditions in Lemma \ref{lem:fio}.  

In the following,  we translate $(P2)$ to conditions without existential quantifiers. To provide an intuition on how to do so,  consider Figure \ref{fig:4} in which $X=\{a,b,c,d\}$, $\tilde{X}=\{a,b\}$, $X^*=\{ c,d\}$, and $\D=2^X \setminus \emptyset$. Let $\C=\{\{a,c\}, \{a,d\}, \{a,c,d\}\}$.  In the figure, red flows are observable outflows from $\C$; yellow flows are unobservable inflows to $\C$; blue flows are observable inflows to $\C$. Note that there are no unobservable outflows. 
%This is because $\C$ is complete.

By the equality between inflows and outflows with respect to $\C$, we have that the sum of red outflows equals to the sum of yellow inflows and the sum of the blue inflows.\footnote{Note also that the values of red outflows and blue inflows can be calculated based on observable dataset as follows: (Red outflows)$=\sum_{(D,x): D \in {\C},  D\cup x \not\in {\C}, (D\cup x,x) \in \M}$ $K(\rho, D \cup x, x)$; (Blue inflows)$=\sum_{(E,y): E \not\in {\C}, E\cup y \in {\C}, (E\cup y, y)\in \M}  
        K(\rho, E \cup y, y)$.} 
Although each yellow inflow is unobservable, we can calculate the sum of yellow inflows as the sum of red outflows minus the sum of blue inflows. If $\rho$ is RU-rationalizable then the sum of yellow inflows is nonnegative, and hence the sum of red outflows minus the sum of blue inflows must be  nonnegative. This is an implication that is directly testable based on the observable dataset since we can calculate the value of blue inflows and  red outflows.

%\textcolor{blue}{\fbox{there is another blue arrow joining cd to acd.}}

To generalize this idea, we introduce one definition:

\begin{figure}
\begin{center} 
  \begin{tikzpicture}[scale=.5, transform shape]
    \tikzset{vertex/.style = {shape=circle,draw,thick,minimum size=4em}}
    \tikzset{-to/.style = {thick,arrows={-to}}} 

    \node[vertex] (abcd) at (\abcd*\scale,5) {\footnotesize $a,b,c,d$};
    
    \node[vertex] (abc) at (\abc*\scale,2.5) {$a,b,c$};
    \node[vertex] (abd) at (\abd*\scale,2.5) {$a,b,d$};
    \node[vertex, fill=green!10] (acd) at (\acd*\scale,2.5) {$a,c,d$};
    \node[vertex] (bcd) at (\bcd*\scale,2.5) {$b,c,d$};
    
    \node[vertex] (ab) at (\ab*\scale,0) {$a,b$};
    \node[vertex, fill=green!10] (ac) at (\ac*\scale,0) {$a,c$};
    \node[vertex, fill=green!10] (ad) at (\ad*\scale,0) {$a,d$};
    \node[vertex] (bc) at (\bc*\scale,0) {$b,c$};
    \node[vertex] (bd) at (\bd*\scale,0) {$b,d$};
    \node[vertex] (cd) at (\cd*\scale,0) {$c,d$};
    
    \node[vertex] (a) at (\a*\scale,-2.5) {$a$};
    \node[vertex] (b) at (\b*\scale,-2.5) {$b$};
    \node[vertex] (c) at (\c*\scale,-2.5) {$c$};
    \node[vertex] (d) at (\d*\scale,-2.5) {$d$};
    \node[vertex] (emptyset) at (\emp*\scale,-5) {$\emptyset$};
    
    \draw[-to, dotted, thin] (abc) -- (abcd);
    \draw[-to,dotted, thin] (abd) -- (abcd);
    \draw[-to,dotted, thin] (acd) -- (abcd);
    \draw[-to,dotted, thin] (bcd) -- (abcd);
    \draw[-to,dotted, thin] (ab) -- (abc);
    \draw[-to,dotted, thin] (ac) -- (abc);
    \draw[-to,dotted, thin] (bc) -- (abc);
    \draw[-to,dotted, thin] (ab) -- (abd);
    \draw[-to,dotted, thin] (ad) -- (abd);
    \draw[-to,dotted, thin] (bd) -- (abd);
    \draw[-to,dotted, thin] (ac) -- (acd);
    \draw[-to,dotted, thin] (ad) -- (acd);
    \draw[-to,dotted, thin] (cd) -- (acd);
    \draw[-to,dotted, thin] (bc) -- (bcd);
    \draw[-to,dotted, thin] (bd) -- (bcd);
    \draw[-to,dotted, thin] (cd) -- (bcd);
    
    \draw[-to,dotted, thin] (a) -- (ab);
    \draw[-to,dotted, thin] (b) -- (ab);
    % \draw[-to,dotted, thin] (a) -- (ac);
    \draw[-to,dotted, thin] (c) -- (ac);
    % \draw[-to,dotted, thin] (a) -- (ad);
    \draw[-to,dotted, thin] (d) -- (ad);
    \draw[-to,dotted, thin] (b) -- (bc);
    \draw[-to,dotted, thin] (c) -- (bc);
    \draw[-to,dotted, thin] (b) -- (bd);
    \draw[-to,dotted, thin] (d) -- (bd);
    \draw[-to,dotted, thin] (c) -- (cd);
    \draw[-to,dotted, thin] (d) -- (cd);
    
    \draw[-to,dotted, thin] (emptyset) -- (a);
    \draw[-to,dotted, thin] (emptyset) -- (b);
    \draw[-to,dotted, thin] (emptyset) -- (c);
    \draw[-to,dotted,thin] (emptyset) -- (d);

    \draw[-to, red] (acd) -- (abcd);
    \draw[-to, red] (ac) -- (abc);
    \draw[-to, red] (ad) -- (abd);
    \draw[-to, green] (ac) -- (acd);
    \draw[-to, green] (ad) -- (acd);
    \draw[-to, orange , dashed] (a) -- (ac);
    \draw[-to, blue] (cd) -- (acd);
    \draw[-to, blue] (c) -- (ac);
    \draw[-to, orange, dashed] (a) -- (ad);
    \draw[-to, blue] (d) -- (ad);
    \end{tikzpicture}
\caption{Outflows from $\C\equiv \{\{a,c\}, \{a,d\}, \{a,c,d\}\}$ and inflows to $\C$.}\label{fig:4}
\end{center}
\vspace{-0.3cm}
\begin{footnotesize}
Note: Red flows are observable outflows from $\C$; yellow flows are unobservable inflows to $\C$; blue flows are  observable inflows to $\C$. Note that green flows are flows contained in $\C$ and are not relevant to the value of $\delta_{\rho}(\C)$, which is the net observable outflow from $\C$.
\end{footnotesize}
\end{figure}

\begin{definition} 
A collection ${\C} \subseteq 2^{X}$ is said to be complete if $D \in \C \implies \forall x \in X^*, D \cup x \in \C$.
\end{definition}

\noindent Note that a collection $\C$ of subsets is complete if and only if there are no unobservable outflows from $\C$.  In the example above, $\C=\{\{a,c\}, \{a,d\}, \{a,c,d\}\}$ is complete, and hence there are no unobservable outflows from $\C$.

We need one more definition: for any ${\C} \subseteq 2^X$, define 
\begin{align}\label{eq:delta_o}
    \delta_{\rho} (\C)
    =
    \left(
        \sum_{\substack{(D,x): D \in {\C},  D\cup x \not\in {\C},\\ (D\cup x,x)  \in \M}}  
        K(\rho, D \cup x, x)
        - 
        \sum_{\substack{(E,y): E \not\in {\C}, E\cup y \in {\C},\\ (E\cup y, y)  \in \M}}  
        K(\rho, E \cup y, y)
    \right)
    \nonumber
    \\
    +
    1\{X \in {\C}, \emptyset \not \in {\C}\}
    -
    1\{\emptyset \in {\C}, X \not \in {\C}\}.
\end{align}

\noindent For any $\C \subseteq 2^X$, $\delta_{\rho}(\C)$ is the net observable outflows from $\C$. To see this notice that the first term is the values of the observable outflows from $\C$ and the second term is the values of the observable inflows to $\C$.\footnote{The function
$\delta$ can also be defined with $\D=2^X\setminus \emptyset$ and $X^*\neq \emptyset$ as follows: for any complete dataset $\rho^*$, define $    \delta_{\rho^*} (\C)
    =
   \left(
        \sum_{\substack{(D,x) \\ D \in {\C},  D\cup x \not\in {\C}}}  
        K(\rho^*, D \cup x, x)
        -
        \sum_{\substack{(E,y) \\ E \not\in {\C}, E\cup y \in {\C}}}  
        K(\rho^*, E \cup y, y)
    \right)\\
    +1\{X \in {\C}, \emptyset \not \in {\C}\}-1\{\emptyset \in {\C}, X \not \in {\C}\}.
$ We will use this definition later.}  In the example above, $\delta_{\rho}(\C)$ equals to the sum of red outflows minus the sum of blue inflows when $\C =\{\{a,c\}, \{a,d\}, \{a,c,d\}\}$. 

Based on the discussion above, if a  solution to $(P2)$ exists, then $\delta_{\rho}(\C)\ge 0$ for any complete collection $\C$. The next proposition shows the converse also holds:

% $
%     \delta_{\rho^*} (\C)
%     =
%    \big(
%         \sum_{(D,x): D \in {\C},  D\cup x \not\in {\C}}  
%         K(\rho^*, D \cup x, x)
%         -$
%         $\sum_{(E,y): E \not\in {\C}, E\cup y \in {\C}}  
%         K(\rho^*, E \cup y, y)
%     \big)$
%     $+1\{X \in {\C}, \emptyset \not \in {\C}\}-1\{\emptyset \in {\C}, X \not \in {\C}\}$. 
% \eit  

%\textcolor{blue}{\fbox{the last bullet is not easy to understand.}}

\normalsize

\begin{proposition} \label{prop:characterization-complete}
Given an incomplete dataset $\rho \in \Re^{\M}_+$, a solution to (P2) exists if and only if $\delta_{\rho} (\C) \geq 0$ for any complete collection ${\C}$ such that $\emptyset \not \in \C$.
\end{proposition}

%\textcolor{blue}{\fbox{I think the proof below should be in appendix.}}

We provide a proof of proposition in the appendix by using the maximum-flow theorem from \cite{ford2015flows}. The condition given in Proposition \ref{prop:characterization-complete} uses only the observable choice data to characterize a solution to (P2) since the value (\ref{eq:delta_o}) of $\delta_{\rho}$ depend only on the values of $\rho$ on $\M$; moreover the condition does not contain existential quantifiers.

Although Proposition \ref{prop:characterization-complete} together with Lemma 3.5 successfully characterizes RU-rationalizability based only on the available data, the condition has some redundancy. The next corollary provides  a nonredundant characterization, which proves statement (a) of Theorem \ref{thm:characterization}.

\begin{corollary} \label{cor:essential-is-enough}
Given an incomplete dataset $\rho \in \Re^{\M}_+$, a solution to (P2) exists if and only if (i) $K(\rho, D, x) \geq 0$ for all $(D,x) \in \M$ such that $1<|D|<|X|$ and (ii) $\delta_{\rho} (\C) \geq 0$ for any essential test collection $\C \subseteq \D$.
\end{corollary}

In the appendix, we provide a proof of statement (b), which claims the nonredundancy of conditions appear in (i) and (ii). This proof is the most intricate part of our proofs.

%\textcolor{blue}{\fbox{move proof to appendix.}}

\vspace{-0.5cm}

\section{Bounds for Unobservable Choice Probabilities} \label{sec:app}

\vspace{-0.2cm}

%In the previous section, we established a necessary and sufficient condition for an incomplete dataset to be RU-rationalizable. Given this result, 

In this section we obtain bounds for unobservable choice probabilities. In practice,  predicting unobservable choice probabilities is important. Recall, for instance, the transportation example (Example 1) in Section \ref{sec:examples}.  In this example, how people commute is not observable unless they use public transportation. That is, $X=\{ \text{bus, train, walk},$ $\text{drive}\}$, $\tilde{X}=\{\text{bus, train}\}$, and $X^*=\{ \text{walk, drive}\}$. (Remember that $\tilde{X}$ is the set of observable alternatives and $X^*$ is the set of unobservable alternatives.) Suppose that the government is considering introducing a new tax on gasoline to encourage people to commute by public transportation.
To assess the potential impact of the new policy, it is crucial for the government to know the percentage of people who commute by private cars.

The most naive approach is merely to bound the fraction below by zero and above by the percentage of people who did {\it not} use public transportation. In Remark \ref{rem:bound_naive2}, we will observe that this naive approach corresponds to the  outside option approach.  The main purpose of this section is to show that specifying the available alternatives as precisely as possible in each choice set allows us to improve the bounds, assuming the random utility model. We first demonstrate this with only the monotonicity condition in Section \ref{subsec:mono_bounds}. Then in Section \ref{subsec:tightbounds}, we further improve the bounds by using the full implication of the random utility model.

%\footnote{These observations apply generally to our setup and do not require that we ever observe choice sets containing only one unobservable alternative as is the case in our simplified motivating example and the lottery dataset.}

%A more careful way, which is the main object of interest here, is to characterize the upper and lower  bounds of missing choice frequencies, assuming the choices are consistent with the random utility model.  

Finally, in Section \ref{section:data}, we apply both the naive approach and our proposed approach to a dataset from \cite{mccausland2020testing}.\footnote{The dataset is available on the journal's website.} We compare the resulting bounds to highlight the differences between the outside option approach and our method, thereby demonstrating the practical significance of our approach.

We first obtain the implication of the outside option approach.

%While the outside option approach is almost uninformative, the more careful approach that we are proposing in this section is likely to produce a finer prediction and to be helpful in policy making.
% Although the fraction is not identified even under the random utility assumption in the situation where the government does not know how people commute unless they use public transportation, i.e., $X=\{ \text{bus, train, walk, drive}\}$ and $X^*=\{ \text{walk, drive}\}$, it allows the government to compute upper and lower bounds of the proportion of drivers as we will show below, and it could be helpful in the policy making.

%In this section, we will propose a method to compute the upper and lower bounds of missing choice frequencies assuming that the dataset is RU-rationalizable. 

\begin{remark}\label{rem:bound_naive2}
Let $\rho \in \Re_{+}^{\M}$ be a given incomplete dataset. Let $\hat{\rho}\in \Re_+^{\{(D,x)| x \in D \in \hat{\D}\}}$ be the reduced dataset in the outside option approach defined in Definition \ref{rem:def_outside}. 
Choose any $(D,x^*)$ such that $D \in \D$, $ X^* \subseteq D$, and  $x^* \in X^*$. The bounds for the  unobservable choice frequency $x^*$ from $D$ derived from the restriction that $\hat\rho$ is RU-rationalizable are  
\begin{equation}\label{eq:bound_naive2}
\Big[0, 1- \sum_{a \in  D \cap \tilde{X}}\rho(D,a)\Big].   \end{equation}

To see this, notice that every menu in the reduced dataset $\hat{\D}$ either contains all or none of the unobservable alternatives. Therefore, there is no way to distinguish any unobservable alternative from another. Thus, with the outside option approach,  RU-rationalizability does not have any implications beyond the fact that the probabilities must sum up to $1$. 
%When $\hat{\rho}$ is not RU rationalizable, independent of the value of unobservable choice frequencies, the dataset is not RU-rationalizable. Therefore the identification interval is empty.

% Let $\rho \in \Re_{+}^{\M}$ be a given incomplete dataset such that the reduced dataset $\hat\rho$ is RU-rationalizable. Let $\rho^*$ be a complete dataset that coincides with $\rho$ on $\M$.
% Choose any $(D,x^*)$ such that $D \in \D$, $ X^* \subseteq D$, and  $x^* \in X^*$. The identification bound for the  unobservable choice frequency $\rho^*(D, x^*)$ derived from the restriction that the reduced dataset $\hat{\rho}\in \Re_+^{\{(D,x)| x \in D \in \hat{\D}\}}$ defined in Definition \ref{rem:def_outside} is RU-rationalizable is  
% \begin{equation}\label{eq:bound_naive2}
% \Big[0, 1- \sum_{a \in  D \cap \tilde{X}}\rho(D,a)\Big].   \end{equation}

% To see this, notice that every menu in the reduced dataset $\hat{\D}$ either contains all or none of the unobservable alternatives. Therefore, $\hat\rho$ depends entirely on $\rho$. Thus every complete dataset $\rho^*$ that agrees with $\rho$ produces the same rationalizable reduced dataset. The bounds obtained in \ref{eq:bound_naive2} are therefore just the bounds implied by $\rho^*$ coincides with $\rho$ on $\M$. Thus, with the outside option approach,  RU-rationalizability does not have any implications beyond the fact that the probabilities must sum up to $1$. 
\end{remark}

\vspace{-0.4cm}

\subsection{Bounds derived from Monotonicity}\label{subsec:mono_bounds}

\vspace{-0.1cm}

In this section, we focus exclusively on the monotonicity condition, which requires that the choice frequency of an alternative from a given choice set does not increase when an additional alternative is added to the set.  The purpose of this section is to provide an intuitive explanation on how specifying the available alternatives as precisely as possible in each choice set allows us to improve the bounds.

Fix $D \in \D$ that contains some unobservable alternative $x^* \in X^*$. We will obtain the bound for the choice frequency of $x^*$ from choice set $D$. The basic idea is that by comparing the choice probabilities on the choice sets $D$ and $D \setminus y^*$ for each $y^* \in D \cap X^*$, we can learn about the relative desirability of each alternative $y^* \in D \cap X^*$.

\begin{remark}\label{rem:bound_mono} Let $\rho^*$ be a complete dataset that coincides with $\rho$ on $\M$ and satisfies monotonicity. Define $F\equiv \{y ^* \in D \cap X^*| D \setminus y^* \in \D\}$.  For any $(D,x^*)$ such that $x^* \in D \cap X^*$ and $D\in \D$, the bounds for the  unobservable choice frequency $\rho^*(D, x^*)$ become
%\Big[ L(x^*), 1-\sum_{a \in D \cap \tilde{X}} \rho(D, a) -\sum_{y^* \in  D \cap (X^* \setminus x^*)} L(y^*) \Big],
\begin{equation}\label{eq:bound_mono}
\Big[ L(x^*), 1-\sum_{a \in D \cap \tilde{X}} \rho(D, a) -\sum_{y^* \in (D \cap X^*) \setminus x^*} L(y^*) \Big],
\end{equation}
where 
\begin{align}
L(y^*) = 
\begin{cases}
\sum_{a \in D \cap \tilde{X} }\big(\rho(D \setminus y^* , a)-\rho(D, a)\big) &\text{ for all }y^* \in F,\\
0 &\text{ otherwise.\footnotemark}
\end{cases}
\end{align}
\footnotetext{Note that if $\rho$ is not RU-rationalizable, it is possible that $L(x^*) > 1-\sum_{a \in D \cap \tilde{X}} \rho(D, a) -\sum_{y^* \in (D \cap X^*) \setminus x^*} L(y^*) $ which we may interpret as an empty  set. \label{footnote:emptybound}}
The lower bound $L(x^*)$ shows that the larger the difference between $\rho(D ,\cdot )$ and $\rho(D\setminus x^* ,\cdot)$ is, the more desirable $x^*$ is. These bounds (\ref{eq:bound_mono}) are useful as they allow us to compare the relative desirability of unobservable alternatives.\footnote{When $\D$ only contains menus of the form $D$ and $D \setminus y*$ for $y^* \in X^*$ for some $D \subseteq X$, these bounds are actually tight. That is, they coincide with the RUM bounds derived in the next section. However, when the choice frequencies for other menus are available, monotonicity and RUM have further implications. \label{footnote:nottight}}

By comparing the bounds  (\ref{eq:bound_mono}) and the bounds  (\ref{eq:bound_naive2}) from the outside option approach obtained in Remark \ref{rem:bound_naive2}, the improvement of the bounds can be simply summarized as 
\[
\sum_{y^* \in D \cap X^*} L(y^*).
\]
Since $L(y^*)$ is the difference between $\rho(D ,\cdot )$ and $\rho(D\setminus y^* ,\cdot)$, we may interpret $\sum_{y^* \in D \cap X^*} L(y^*)$ as the total difference between the outside option approach and our approach.
\end{remark}

%To see how to obtain bounds, let $\rho\^*$ a complete stochastic choice function that (i) coincides on the given $\rho$ on $\D$ 

To see how to obtain the result, first notice that the upper bounds can be obtained easily given the lower bounds.\footnote{Suppose that we obtain the lower bound $L(y^*)$ for all $y^* \in D \cap X^*$. To get the upper bound, simply subtract the lower bounds from 1 to obtain $\rho^*
(D,y^*) \le 1-\sum_{a \in D \cap \tilde{X}} \rho(D, a) -\sum_{y^* \in D \cap (X^* \setminus x^*)} L(y^*)$.} To get the lower bounds assume that $x^* \in F$. Then, by monotonicity, we have $\rho^*(D , y^*) \le  \rho^*(D \setminus x^* , y^*)$ for all $y^* \in X^* \setminus x^*$.  Thus we have 
$\rho^*(D, x^*)+ \sum_{a \in D \cap \tilde{X} }\rho(D , a)
=1-\sum_{y^* \in (D \cap X^*) \setminus x^*} \rho^*(D , y^*)
\ge  1- \sum_{y^* \in (D \cap X^*) \setminus x^*} \rho^*(D \setminus x^* , y^*) 
=\sum_{a \in D \cap \tilde{X} }\rho(D \setminus x^* , a)$. It follows that 
$\rho^*(D, x^*)\ge \sum_{a \in D \cap \tilde{X} }\Big[\rho(D \setminus x^* , a)-\rho(D , a)\Big] \equiv L(x^*)$.

In the next remark, we demonstrate how to calculate the bounds by using example in Table \ref{tale:rho} in Section \ref{subsection:ex}

\begin{remark}\label{rem:mono_bounds_app}  
We will obtain bounds of unobservable choice frequencies from the choice set $\{b,c,d\}$. In the dataset $\rho$ in Table \ref{tale:rho}, we observe 
$\rho(\{b,d\},b)=1/2+\ep, \rho(\{b,c\},b)=1/2$, and $ \rho(\{b,c,d\},b)=1/3$ as alternative $b$ is observable. From these we can calculate the lower bounds $L(c)$ and $L(d)$ as follows:
\begin{align*}
&L(c) = \rho(\{b,d\},b) - \rho(\{b,c,d\},b) = 1/2+\ep - 1/3 = 1/6+\ep,\\
&L(d) = \rho(\{b,c\},b) - \rho(\{b,c,d\},b) = 1/2 - 1/3 = 1/6.
\end{align*} 
Thus, we also can obtain upper bounds for alternative $c
$ and $d$ as  $1-\rho(\{b,c,d\},b) -L(d)  = 1/2$ and $1-\rho(bcd,b) -L(d) = 1/2 -\ep$, respectively. Thus, as observed in the introduction, the bounds for alternative $c$ are $[1/6+\ep, 1/2]$; the bounds for alternative $d$ are $[1/6, 1/2-\ep]$.

When $\ep \ge 1/6$, observe that the bounds for  alternative $c$ locate higher than the bounds for  alternative $d$, which suggests that alternative $c$ is more desirable to alternative $d$. Thus remark \ref{rem:bound_mono} generalizes the analysis done in Section \ref{subsection:ex}.
\end{remark}

%We directly show that the choice frequency of alternative 1 exceeds the choice frequency of alternative 0 using only the monotonicity condition. See Section \ref{sec:application_mono} of the appendix. It should be noted that this technique does not give the tight bounds from equation \ref{eq:coro:bound2}.

\vspace{-0.4cm}

\subsection{Bounds derived from RU rationality}\label{subsec:tightbounds}

\vspace{-0.1cm}

In this section, we further improve the bounds by assuming full RU rationality based on our characterization.

Let $\rho \in  \Re_+^{\M}$ be a given incomplete dataset. 

\begin{definition}\label{def:gamma_0}
  Assume that $\rho$ is RU-rationalizable. A complete dataset  $\rho^* \in \Re_+^{\{(D,x) \mid x \in D \in 2^X\}}$ is said to be RU-consistent with $\rho$ if it satisfies the followings
 \begin{itemize}
\item[(i)] $\rho=\rho^*$ on $\M$; and  
\item[(ii)] there exists $\mu \in \Delta(\L)$ such that for any $(D, x)$ with $x \in D \in 2^X$, $\rho^* (D, x)=
    \mu (\succ \in \L \mid x \succ y \ \text{for all} \ y \in D \setminus x)$.
\end{itemize}
\end{definition}

Let $\Gamma$ be the set of complete datasets $\rho^* \in \Re_+^{\{(D,x)|x \in D \in 2^X\}}$ that are RU-consistent with the given incomplete dataset $\rho$.  Remember that $\rho(D,x)$ is defined (i.e., observable) if and only if $(D,x) \in \M$.
With $(D, x) \not \in \M$ fixed, the goal in this section is to obtain bounds of $\rho^* (D, x)$ for some $\rho^* \in \Gamma$. As is pointed out in page 1399 of \cite{manski2007partial}, since $\Gamma$ is convex and conditions (i) and (ii) are linear, the identified set is an interval with
the upper and lower bounds given by $\overline \rho (D, x) \equiv \max_{\rho^* \in \Gamma} \rho^* (D, x)$ and $\underline \rho (D, x) \equiv \min_{\rho^* \in \Gamma} \rho^* (D, x),$ respectively.

Based on the idea of (P1) in Section \ref{sec:sketch}, $\Gamma$ can be written as
\vspace{-0.1cm}
\begin{align}\label{eq:boundP1}
    \left\{
        \rho^*
        \in 
        \Re_+^{\{(D,x) \mid x \in D \in 2^X\}}
        \ \bigg | \ 
            \begin{array}{l}
                 \text{There exists $\mu \in \Delta(\L)$ satisfying } \\
                 \text{the conditions (i) and (ii) in Definition \ref{def:gamma_0}}
            \end{array}
    \right\}
    \normalsize
\end{align}
The formulation (\ref{eq:boundP1}) is closely related to (P1): it can be shown that if $\mu$ satisfies the conditions (i) and (ii), it is a solution to (P1). 
To obtain the next proposition, we rewrite $\Gamma$ in  the spirit of  (P2) exploiting the network flow structure.    

\begin{proposition}\label{coro:bound2}
For any $(D, x) \not \in \M$,  the  upper bound is obtained by 
\begin{align}\label{eq:coro:bound2} 
   \overline \rho (D, x)=  \max_{r\in \Re_+^{\{(D,x) \mid x \in D \in 2^X\}}} 
    \sum_{E: E \supseteq D}
    r (E \setminus x, E)
 \end{align}
 subject to the following constraint: for all $D \subseteq X$,
 \begin{align} \label{eq:dmnd=sply}
    \sum_{(D, y) \not\in \M: y \in D} r (D \setminus y, D)
    -
    \sum_{(D \cup y, y) \not\in \M: y \notin D} r (D, D \cup y)
    =
    \delta_{\rho}(D),
   \end{align}
where $r(E\setminus x, E)=K(\rho, E,x)$ for all $(E,x) \in \M,$ and $\delta_{\rho} (\cdot)$ is defined by (\ref{eq:delta_o}). The lower bound $\underline \rho (D, x) $ solves a similar problem with a min replacing the max.
\end{proposition}

See subsection \ref{sec:bounds_new} in the appendix for the proof. Compared with the original formulation (\ref{eq:boundP1}) based on (P1), our  bounds in Proposition \ref{coro:bound2} is computationally more efficient. 
This is because this problem can be seen as a minimum-cost transshipment problem, which is well known in the network-flow theory literature.\footnote{See, for example, \cite{ahuja1988network} and \cite{ford2015flows}.} 
One of the key properties of this problem is that it is a linear program with a constraint that has {\it an incidence matrix} as its coefficient.\footnote{An incidence matrix is a matrix representation of network structure, which is defined as a matrix consisting only of $0, 1$ and $-1$ with each column having exactly one element of $1$ and $-1$.} For this specific problem, a practical polynomial time algorithm, called the {\it network simplex algorithm}, can be applied.\footnote{We refer readers to \cite{orlin1993polynomial} and \cite{orlin1997polynomial} for further computational aspects of the algorithm.}  
When $\D = 2^X\setminus \emptyset$, the bound (\ref{eq:coro:bound2}) can be further simplified as shown in online appendix \ref{sec:online_bounds}.

\cite{kitamura2019nonparametric} also consider bounds of counterfactual choice probabilities based on the formulation of \cite{kitamura2018nonparametric}.
Theoretically, their bounds coincide with (\ref{eq:coro:bound2}) in our setup, but it is possible that the bounds are much harder to compute than our bounds.
This is because they compute the bounds directly from (P1), or equivalently (\ref{eq:boundP1}), ignoring the network structure.
Since the network simplex algorithm relies heavily on the fact that the constraint of the linear program is written with an incidence matrix, the original form (P1) does not have its benefit in terms of computational efficiency.

\vspace{-0.4cm}

\subsection{Application to Lottery Data}\label{section:data}

\vspace{-0.1cm}

We now apply the methods developed in the the previous sections to a stochastic-choice dataset from the experiment conducted by \cite{mccausland2020testing}. In the experiment, the authors fixed a set $X = \{0, 1, 2, 3, 4\}$ of five lotteries and asked 141 participants to choose one from each subset of $X$. Each participant made decision six times for each choice set. See \cite{mccausland2020testing} for further details. We aggregate these choice frequencies to construct a complete dataset denoted by $\rho$.

The lottery dataset is nearly RU-rationalizable but not exactly. 
As our method can be applied only to RU-rationalizable datasets, we first fit a random utility model to the dataset to get a calibrated dataset that is close to the original one and is RU-rationalizable.\footnote{The monotonicity bounds can be applied to the original dataset, although as mentioned in footnote \ref{footnote:emptybound} if the data deviates too much from RUM the bounds may become empty.} 
The detail of this procedure is in subsection \ref{sec:calib} of the online appendix.
We mask the choice probabilities of lotteries $0$ and $1$ and pretend not to observe them; in other words, we set $X^* = \{0, 1\}$, $\tilde{X}=\{2,3,4\}$, and $\D = 2^X \setminus \emptyset$.

Under this setup, we will compute three types of bounds on the probabilities of lotteries $0$ and $1$ being chosen in a given choice set $D$ that contains both lotteries. 

%For a given choice set that contains both lotteries $0$ and $1,$ we compute the identified sets of the choice frequencies of lotteries $0$ and $1$ by applying our method to the calibrated RU-rationalizable dataset. 

\begin{figure}[h]
\begin{center}
    \includegraphics[width=1\textwidth]{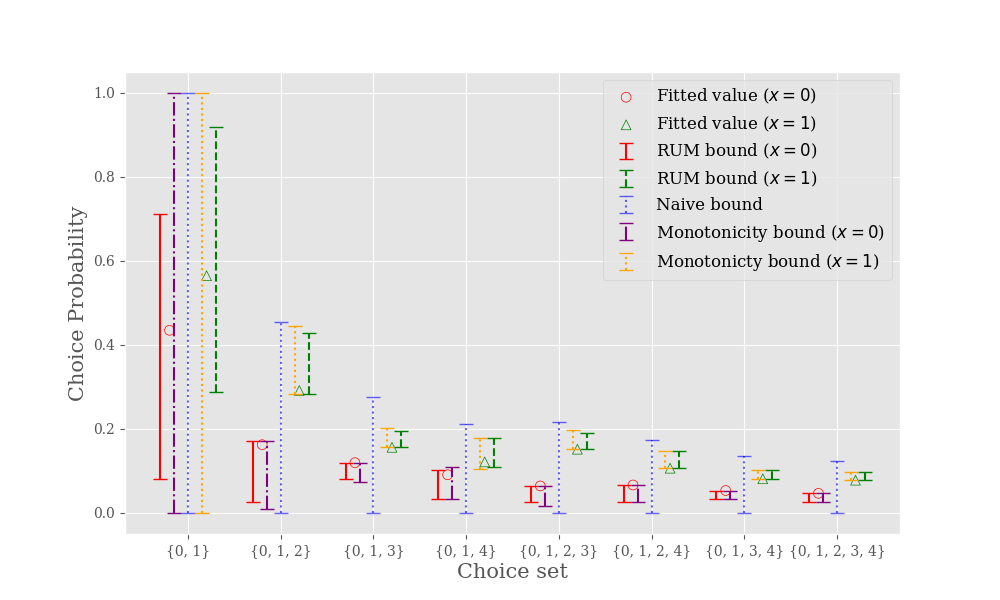}
    \caption{Comparison between the bounds of the choice probabilities of lotteries $0$ and $1$.  The RUM bounds for $0$ and $1$ are shown in red and green, respectively. The monotonicity bounds for $0$ and $1$ are show in purple and yellow, respectively. The naive bounds for both are shown in blue.}
    \label{fig:id_comparison}
   \end{center}
\end{figure}

The first type of bounds is (\ref{eq:bound_naive2})
based on the outside option approach: 
$\Big[
        0
        , 
        1 - \sum_{x \in D \cap \{2, 3, 4\}} \rho (D, x)
    \Big]$.
Note that the bounds for lotteries 0 and 1 are identical, as the outside option approach provides no information to distinguish among unobservable alternatives. The bounds are shown in blue dotted interval in Figure \ref{fig:id_comparison}. We call these bounds {\it naive bounds}.

The second type of bounds is derived from the monotonicity condition obtained in Remark \ref{rem:bound_mono}. As demonstrated in Remark \ref{rem:mono_bounds_app}, the bounds can be calculated easily given the dataset $\rho$.  

The third and the last type of bounds is the one that exploits full implication of RUM and is computed by the linear program (\ref{eq:coro:bound2}).\footnote{In the dataset, we have $\D=2^X \setminus \emptyset$. Under this setup, we can further simplify the bound (\ref{eq:coro:bound2}) into (\ref{eq:lp-full}), as shown in the online appendix \ref{sec:online_bounds}. We use (\ref{eq:lp-full}) to calculate the values.}  These  bounds are shown in red for alternative $0$ and in green for alternative $1$ in Figure \ref{fig:id_comparison}. We call these bounds {\it RUM bounds.}

% \begin{figure}[h]
% \begin{center}
%     \includegraphics[width=1\textwidth]{images/id_comparison_nonpara_fitted.png}
%     \caption{Comparison between the identified sets of the choice probabilities of lotteries $0$ and $1.$ The identified sets for the probability of choosing $0$ are shown in red and the identified sets for the probability of choosing $1$ are shown in green. The naive bounds for both are shown in blue.}
%     \label{fig:id_comparison}
%    \end{center}
% \end{figure}

% \begin{figure}[h]
% \begin{center}
%     \includegraphics[width=1\textwidth]{images/id_comparison_monotonicity_no01.png}
%     \caption{Comparison between the RUM identified sets and the approximate bounds derived from monotonicity in section \ref{subsec:mono_bounds}. The identified sets for the probability of choosing $0$ are shown in red and the identified sets for the probability of choosing $1$ are shown in green. The approximate bounds for $0$ are purple and yellow for $1$. The approximate bounds are very close to the bounds derived from RUM.}
%     \label{fig:id_comparison_mono}
%    \end{center}
% \end{figure}

There are two key takeaways from the figure. First, our bounds (i.e., the RUM bounds and the monotonicity bounds) are substantially tighter than the native bounds, particularly when the choice set is large. While the RUM bounds are slightly narrower than the monotonicity bounds, the two are quite similar except for the choice set $\{0,1\}$.\footnote{While the given monotonicity bounds for $\{0,1\}$ are not close to the RUM bounds, this is only because the bounds derived in remark \ref{rem:bound_mono} use monotonicity in a simple way to aid the exposition. Indeed as mentioned in footnote \ref{footnote:nottight}, since there are additional menus beyond $\{0,1\}$, $\{0\}$, and $\{1\}$, monotonicity has further implications beyond the given bounds. If these are taken into account, the monotonicity bound becomes very close to the RUM bound even on $\{0,1\}$.} This suggests that our methods effectively leverage additional information about unobserved choice frequencies. In particular, for the choice set $\{0,1,2,3,4\}$, the RUM bounds and the monotonicity bounds are extremely tight and closely approximate the true values.

Second, and more importantly, in the RUM bounds for the choice frequency of the alternative 1 are always higher than those of alternative 0. Therefore, we can conclude that in any RU-rationalization, the probability that alternative $1$ is chosen is higher than the probability that alternative $0$ is chosen in all menus but  $D = \{0,1\}$. A similar conclusion can be made for the monotonicity bounds. On the other hand, in the outside option approach, the bounds for alternatives 1 and 0 are exactly identical. This difference indicates that our method exploits the information that alternative 1 is better than alternative 0, while the outside option approach loses this information completely. 
This observation suggests that the estimation of the desirability of unobserved alternatives would be biased in the outside option approach, though it is beyond the scope of the current paper.

\vspace{-0.5cm}

\section{Concluding Remark}\label{sec:conclusion}

\vspace{-0.2cm}

In this paper, we formally demonstrate the limitations of the outside option approach. We begin by characterizing the full implications of the random utility model when some choice frequencies are unobservable (Theorem \ref{thm:characterization}), and then identify which implications are lost under the outside option approach (Proposition \ref{prop:outside}). Furthermore, in Remark \ref{rem:bound_mono} and Proposition \ref{coro:bound2}, we develop methods to bound the unobserved choice frequencies, revealing information about the desirability of unobservable alternatives. This information is entirely lost in the outside option approach, as all unobservable alternatives are aggregated into a single alternative in that approach. Taken together, these results underscore the value of accurately identifying the set of alternatives available to each agent. In a transportation context, for example, measuring access to train or bus stations can help determine which options are  available to a commuter, enabling a more reliable inference of their preferences.

\vspace{-0.5cm}

\appendix

\section*{Appendix}

\vspace{-0.5cm}

\section{Proof of Statement (a) of Theorem \ref{thm:characterization}}

\vspace{-0.4cm}

\subsection{Proof of Lemma \ref{lem:p1-is-p2}}

We first prove that (P1) implies (P2). Fix a solution $\mu$ to (P1). Define a complete dataset $\rho^* \in \Re^{\{(D,x) \mid x \in D \in 2^X\}}_+$ by $\rho^*(D,x)=\mu(\{\succ \in {\cal L} \mid  x \succ y \text{ for all  } y \in D \setminus x\})$ for all $ x \in D \in 2^X$. Then $\rho^*=\rho$ on $\M$. Moreover, by Lemma \ref{lem:fio}, we have $r \in \Re_+^{\{(D\setminus x,D) \mid x \in D \in 2^X\}}$ that satisfies (\ref{eq:0sum1}), (\ref{eq:0in=out}) and $r(D \setminus x, D)=K(\rho^*, D, x)$ for all $(D, x)$  such that $x \in D \in 2^X$. Since $\rho^*=\rho$ on $\M$, thus we have $r(D \setminus x, D)=K(\rho, D, x)$ for all $(D, x) \in \M$. Thus, $r$ is a solution to (P2).

%\textcolor{blue}{\fbox{Terminlogy is not consistent in the proof too.}}

%\textcolor{blue}{\fbox{is the def of $\rho^*$ correct?}}

Next we prove that (P2) implies (P1).  Fix a solution $r$ to (P2). For any $(D,x) \not\in \M$, $\rho^*(D,x) \equiv \sum_{E: E\supseteq D}r(E\setminus x,E)$. Then  $\rho^*=\rho$ on $\M$, where $\rho$ is the given incomplete dataset. Thus, we obtain a complete dataset $\rho^* \in \Re^{\{(D,x)\mid  x \in D \in 2^X\}}_+$.  Then by the M\"{o}bius inversion, we have $r(D \setminus x, D)= K(\rho^*, D,x)$ for all $(D,x)$ such that $x \in D \in 2^X$. Then $r$ satisfies (\ref{eq:0sum1}), (\ref{eq:0in=out}), and (\ref{eq:0con}). Then by Lemma \ref{lem:fio}, there exists $\mu \in \Delta(\L)$ such that $\rho^*(D,x)=\mu(\{\succ \in {\cal L} \mid  x \succ y \text{ for all } y \in D\setminus x\})$ for all $(D,x)$ such that $x \in D \in 2^X$. Since $\rho=\rho^*$ on $\M$,  (\ref{eq:rational}) holds for any $(D,x)\in \M$. Thus,
$\mu$ is a solution to (P1).

\vspace{-0.4cm}

\subsection{Proof of Proposition \ref{prop:characterization-complete}}

\vspace{-0.1cm}

\subsubsection{Lemma}

\vspace{-0.1cm}

Fix a network $(\N,\A)$. Remember that $\N$ is the set of nodes; $\A$ is the set of arcs.  Consider a function $r: {\cal A} \to \Re$. For any node $D \in {\cal N}$, let $r(D, {\cal N})\equiv \sum_{E \in {\cal N}} r(D, E)$; $r({\cal N},D)\equiv \sum_{E \in {\cal N}} r(E, D)$\footnote{We define $r(D, {\cal N})=0$ if $(D,E) \not \in \A$ for any $E\in \N$; Similarly, $r({\cal N},D)=0$ if $(E,D) \not \in \A$ for any $E\in \N$.}.  A function $r : \cal A \to \Re$ is called a {\it flow} on a network $(\cal N, \cal A)$ if it satisfies the following conditions:
\begin{align}
&\dis  r(s, {\cal N})-r({\cal N},s)=1, \label{eq:sum1}\\
&\dis  r(D, {\cal N})-r({\cal N},D)=0 \quad  \forall D\in \N \setminus \{s,t\},\label{eq:in=out}\\
&\dis  r({\cal N},t)-r(t, {\cal N})=1. \label{eq:sum1t}
\end{align}
$r(D, \N)$ is the sum of {\it outflows} from $D$; $r(\N, D)$ is the sum of {\it inflows} to $D$. Thus   (\ref{eq:sum1}) means the net outflow from $s$ is one; (\ref{eq:in=out}) means the inflows equal to the outflows at each node $D\not \in \{s,t\}$; (\ref{eq:sum1t}) means the net inflows to $t$ is one.

%\textcolor{blue}{\fbox{A flow is not just a function but must satisfy some additional constraints.}}

%To introduce the next lemma,  we define a basic concept on network flow theory. A {\it  network flow } is an acyclic directed graph $({\cal N}, {\cal A})$, where   ${\cal A}$ is the set of arcs and ${\cal N}$ is the set of nodes includes the source node $s$ and the terminal node $t$. 

The following lemma provides a necessary and sufficient condition for the existence of a nonnegative flow satisfying some capacity constraints.  For each arc $(D,E)$, let $l(D,E)$ and $u(D,E)$ be exogenously given lower and upper bounds of the flow $r(D,E)$ of the arc. We prove the result using the maximum-flow theorem from \cite{ford2015flows}. We provide the proof  in section \ref{lem:network-flow-feasibility} of the online appendix.\footnote{We appreciate prof. Ui who pointed out  a similar result appears in \cite{NetworkFlows}.}

\begin{lemma} \label{lem:network-flow-feasibility}
Let $l,u: {\cal A} \to \Re_+$ be such that $l(D,E) \le u(D,E)$ for $(D,E) \in {\cal A}.$
There exists a flow $r:{\cal A} \to \Re_+$ such that 
\begin{align}
&\dis   l(D, E) \le r(D, E) \le u(D, E)\quad \forall (D, E) \in {\cal A}
\end{align}
if and only if the following condition holds for all $\C\subseteq {\cal N}$
\begin{eqnarray}\label{eq:1}
\sum_{(D,E) \in \C \times \C^c} u(D,E)-\sum_{(D,E) \in \C^c \times \C }l(D,E)\ge
\left\{
\begin{array}{lll}
1 &\text{  if }t \not\in \C,  s \in \C,\\
-1 &\text{ if }t \in \C,  s \not\in \C,\\
0 &\text{ otherwise}. 
\end{array}
\right.
\end{eqnarray}
\end{lemma}

%\textcolor{blue}{\fbox{does this lemma assume $\cal A = \N \times \N?$ If so, it should be mentioned; if not, why does $\cal A$ below have $(D \cup x, D)?$}}

To interpret condition (\ref{eq:1}), remember that for any collection $\C \subseteq  \N$, $r(D,E)$ is called an {\it outflow} from $\C$ if $D \in C$ and $E \not \in \C$; $r(D,E)$ is called an {\it inflow} to $\C$ if $D \not\in C$ and $E  \in \C$. Thus the left-hand side is the sum of the upper bounds of outflows from $\C$ minus the sum of the lower bounds of inflows to $\C$.\footnote{In network-flow theory, this value is called residual capacity of a cut $(\C, \C^c)$.} On the other hand, the right-hand side is the net outflow from  $\C.$

%By applying Lemma \ref{lem:network-flow-feasibility} to the network flow defined by (P2), we can obtain a necessary and sufficient condition that is testable for the existence of a solution $r$ to (P2). Note that (\ref{eq:0sum1}) corresponds to (\ref{eq:sum1}) ; (\ref{eq:0in=out}) corresponds to  (\ref{eq:in=out}); and  (\ref{eq:0sum1})  and (\ref{eq:0in=out}) implies $\sum_{x \in X} r(\emptyset, x)=1$, which corresponds to (\ref{eq:sum1t}). 

%See Figure \ref{fig:lem3} for the case $|X|=4$. 

\subsubsection{Main proof of Proposition \ref{prop:characterization-complete}}

To apply the lemma to our setup, let
\begin{eqnarray}\label{eq:network_def}
\begin{array}{lll}
&{\cal N}= 2^X%\text{ where } X \text{ is the set of alternatives},
, \ \ {\cal A}=\{(D, D \cup x) \mid D \subseteq X, x \not \in D\},\  s=\emptyset, \ \ t=X,\\
&l(D, D\cup x)=u(D, D\cup x)= K(\rho, D\cup x,x) \text{ if }(D\cup x, x) \in \M,\\
&l(D, D\cup x)= 0  \text{ and } u(D, D\cup x)= +\infty  \ 
\text{ if }(D\cup x, x) \not \in \M.
\end{array}
\end{eqnarray}

Under the setup (\ref{eq:network_def}), there exists a solution $r$ to (P2) $\Leftrightarrow$ there exists a flow $r$ that satisfies the conditions in Lemma \ref{lem:network-flow-feasibility} $\Leftrightarrow$ the condition (\ref{eq:1}) holds for any $\C \subseteq {\mathcal N}$, where the first equivalence holds by the setup and the second equivalence holds by Lemma \ref{lem:network-flow-feasibility}. Thus, to show the proposition, it suffices to prove that the condition (\ref{eq:1}) holds for any $\C \subseteq {\mathcal N}$ if and only if  $\delta_{\rho} (\C) \geq 0$ for any complete collection ${\C}$ such that $\emptyset \not \in \C$.

\noindent\textbf{Step 1:} Suppose that (\ref{eq:1}) holds for any $\hat{\C} \subseteq {\mathcal N}$. For any $\C \subseteq {\mathcal N}$ such that all outflow from $\C$ is observable, then $\delta_{\rho} (\C) \geq 0$.
\begin{myproof}
Fix any $\C \subseteq {\mathcal N}$.  Note that 
\[
\sum_{(D,E) \in {\C}^c \times {\C}}l(D,E)=\sum_{(E,y): E \not\in {\C}, E\cup y \in {\C}, (E\cup y,y)\in \M}  
        K(\rho, E \cup y, y).
        \]
        Assume that any outflow from $\C $ is observable.  Then $u$ does not take the value of $+\infty$. Thus we have \[\sum_{(D,E) \in {\C} \times {\C}^c}u(D,E)=\sum_{(D,x): D \in {\C},  D\cup x \not\in {\C}, (D\cup x,x) \in \M} K(\rho, D \cup x, x).
        \]
        Thus the left-hand side minus the right-hand side of (\ref{eq:1}) equals to the value of $\delta_{\rho}(\C)$. By the suppostion of the statement that  (\ref{eq:1}) holds for any $\hat{C} \subset {\mathcal N}$, we have $\delta_{\rho}(\C)\ge 0$. 
\end{myproof}

We use Step 1 to prove the following:

\noindent\textbf{Step 2:} If the condition (\ref{eq:1}) holds for any $\hat{\C} \subseteq {\mathcal N}$, then $\delta_{\rho} (\C) \geq 0$ for any complete collection $\C \subseteq {\mathcal N}$ such that $\emptyset \not \in \C$.

\begin{myproof}
Fix any complete collection $\C \subseteq {\mathcal N}$ such that $\emptyset \not \in \C$. By the disjoint additivity, we have $\delta_{\rho}(\C) = \delta_{\rho}(\C \cap \D) + \delta_{\rho}(\C \setminus \D)$. 

Since $\D$ is an upper set and $\C$ is complete, there are no unobservable outflows from $\C \cap \D$. Thus by the supposition of this step,  it follows from Step 1 that $\delta_{\rho}(\C \cap \D) \ge 0$.  

Since $\emptyset \notin \C \setminus \D $ and $\C \setminus \D$ has no observable inflows, by the definition of $\delta_{\rho}$, there are no negative terms in $\delta_{\rho}(\C \setminus \D)$. Thus, we have $\delta_{\rho}(\C \setminus \D) \ge 0$. It follows that $\delta_{\rho}(\C) \ge 0$.
\end{myproof}

Step 2 proves one way of the proposition. In the following, we show the other way.

\noindent\textbf{Step 3}: If $\delta_{\rho} (\hat{\C}) \geq 0$ for any complete collection $\hat{\C} \subseteq {\mathcal N}$ such that $\emptyset \not \in \hat{\C}$, then the condition (\ref{eq:1}) holds for any $\C \subseteq {\mathcal N}$.

\begin{myproof} Fix any  $\C \subseteq {\mathcal N}$. If $\C$ has an unobservable outflow, then the left-hand side of (\ref{eq:1}) becomes infinite and (\ref{eq:1}) holds as desired.  In the following consider the case where $\C$ has no unobservable outflows (i.e.,  all outflows are observable), which implies $\C$ is complete. As in the proof of Step 1, this implies that 
 the left-hand side minus the right-hand side  of (\ref{eq:1}) equals to $\delta_{\rho}(\C)$.  
 
 If $\emptyset \not \in \C$, by the supposition of this step, $\delta_{\rho}(\C)\ge 0$ because $\C$ is complete. Thus, (\ref{eq:1}) holds.

Assume $\emptyset \in \C$ in the following.  Since $\C$ is complete,  $2^{X^*} \subseteq \C$.  Remember, by the assumption of the case, there exist no unobservable arcs $(D, D \cup x)$ coming out from $\C$. This means that any arc $(D, D \cup x)$ coming out from $\C$ is observable. Thus, we have $D\cup x \in \D$.  Since $D\cup x  \not \in \C$, by the completeness of $\C$, we have $x\not \in X^*$.

Now we build a certain subset of $\C$ inductively: initialize $\mathcal{G}_0 = 2^{X^*}$. At the first step, take all of the unobservable flows coming out of $\mathcal{G}_0$. The collection $\mathcal{G}_0$ is complete so all of these flows are going to a node of $\D^c$ (because we are considering unobservable flows). Call the set of all these nodes $\mathcal{F}_0 \subseteq \D^c$. Since each of these nodes is in $\D^c$, all of their subsets are also in $\D^c$ because $\D^c$ is a lower set.\footnote{Remember that a lower set satisfies the following: $E \in {\mathcal D}^c \implies D \in \D^c$ for all $E \supseteq D$. } In particular, for each $F \in \mathcal{F}_0$ we have $F \setminus X^* \in \D^c$.

Notice also that there is an arc from $\mathcal{G}_0$ to $F \setminus X^*$.\footnote{From $\emptyset$ to $F \setminus X^*$ which is a singleton in the first step. In step $n$, such an arc exists because we may write $F = G\cup x$ for $G \in {\cal G}_n$ and $x \in \tilde{X}$. Notice $F \setminus X^* = (G \cup x) \setminus X^*$. Then by the construction of ${\cal G}_n$ we have $G \setminus X^* \in {\cal G}_n$ and therefore we have the arc $(G \setminus X^*,F \setminus X^*)$ coming out of ${\cal G}_n$} So we conclude that $F \setminus X^* \in \C$  (otherwise it contradicts with the fact that all outflows from ${\mathcal C}$ are observable).  By the completeness of $\C$, we have $\{(F \setminus X^*) \cup E : E \in 2^{X^*}\} \subseteq \C$.

Define $\mathcal{H}_0 = \bigcup_{F \in \F_0}\{(F \setminus X^*) \cup E : E \in 2^{X^*}\}$. Now define $\mathcal{G}_1 = \mathcal{G}_0 \cup \mathcal{H}_0$. At the $n$th step, define $\mathcal{H}_{n-1}$ in the same way as $\mathcal{H}_0$ and let $\mathcal{G}_n = \mathcal{G}_{n-1} \cup \mathcal{H}_{n-1}$. Since $2^X$ is finite, at some step $n$ the set $\mathcal{G}_n$ will have no unobservable outflows. Call this terminal collection  $\mathcal{G}$.

The collection $\mathcal{G}$ has no unobservable outflows and contains $\emptyset$ by its construction. It is straightforward to show that $\mathcal{G}$ has no inflows, which  implies $\mathcal{G}$ is a lower set. To see why, suppose that $(D, D \cup x) \in \mathcal{G}^c \times \mathcal{G}$. Since $D \not \in {\mathcal G}$, by the construction of ${\mathcal G}$, we have $D \setminus X^* \notin \mathcal{G}$. Since $D \cup x \in {\mathcal G}$ we have $(D \cup x) \setminus X^* \in \D^c$  by the construction of ${\mathcal G}$ again. It follows from that $D \setminus X^*$ and all  subsets of $D \setminus X^*$  belong to  $\D^c$.\footnote{This is because $\D^c$ is a lower set.}   Since $\emptyset \in {\mathcal G}$ there must be an arc from ${\mathcal G}$ to a subset $E$ of $D \setminus X^*$, where $E\not \in {\mathcal G}$. Since $E \not \not\in {\mathcal D}$, this means  ${\mathcal G}$ has an unobservable outflow, which is impossible by its construction.

Since (i) all the outflows from $\mathcal{G}$ are observable; (ii) there are no observable inflows to $\mathcal{G}$;  (iii) and $\emptyset \in \mathcal{G}$, therefore we conclude that $\delta_{\rho}(\mathcal{G}) = 0$.

Note that $\C \setminus \mathcal{G} = \C \cap \mathcal{G}^c$ is complete  because $\mathcal{G}$ is a lower set (in particular ${\cal G}^c$ is an upper set and thus is complete) and the intersection of complete sets is complete.  Since $\emptyset \in {\mathcal G}$, $\C \setminus \mathcal{G}$ does not contain $\emptyset$.  It follows from the supposition of the step that $\delta_\rho(\C\setminus \mathcal{G}) \ge 0$.

Since $\delta_\rho(\C) = \delta_\rho(\C \setminus \mathcal{G}) + \delta_\rho(\mathcal{G})$, we have $ \delta_{\rho}(\C) \ge 0$, which means the inequality (\ref{eq:1}) for $\C$.
\end{myproof} 

\vspace{-0.4cm}

\subsection{Disjoint Additivity of $\delta$}

\vspace{-0.1cm}

\begin{lemma}\label{lem:delta-is-additive}
For any $\C \subseteq 2^X$, $\delta_{\rho} (\C) = \sum_{D \in \C} \delta_{\rho}(D)$.
\end{lemma}

\begin{myproof}
It suffices to show $\delta_{\rho} ({\C} \cup {\cal E}) = \delta_{\rho}({\C}) + \delta_{\rho}({\cal E})$ for any disjoint sets ${\C}, {\cal E} \subseteq 2^X.$ If there are no arcs connecting ${\C}$ and ${\cal E}$, then the result is trivial. 
Suppose otherwise. The values of a flow on the connecting arcs will be canceled out in $\delta_{\rho} ({\C}) + \delta_{\rho}({\cal E})$.  Without loss of generality, suppose that there is a connecting arc $(D, D\cup x)$ and $D \in {\C}$ and $D \cup x \in {\cal E}$. Then the value $K(\rho, D\cup x, x)$ of the flow on the arc  is added to $\delta_{\rho}({\cal C})$ and subtracted from $\delta_{\rho}({\E})$. Thus the value is canceled out in $\delta_{\rho}({\C}) + \delta_{\rho}({\cal E})$. In the same way, the value $K(\rho, D\cup x, x)$ of the flow on the connecting arc $(D, D\cup x)$ and $D \in {\cal E}$ and $D \cup x \in {\C}$ will be canceled out.
\end{myproof}
\begin{comment}
\footnote{Algebraic proof: To simplify the notation, for any ${\C}, {\cal E} \subseteq 2^X$, define\\ $r({\C},{\cal E}) = \sum_{(D,x): D\setminus  x \in {\C}, D \in {\cal E}} r^*(D\setminus x, D)$. Then $\delta({\C})= r({\C},{\C}^c)-r({\C}^c,{\C})-1(s \in {\C}\& t \in {\C})+1(s \in {\C}\& t \not\in {\C})$.

It is clear that $1(t\in \C\& s \not \in \C)- 1(s\in \C\& t \not \in \C) + 1(t\in \E\& s \not \in \E) -1(s\in \E\& t \not \in \E)$ $= 1(t\in \C\cup \E\& s \not \in \C\cup \E) - 1(s\in \C\cup \E\& t \not \in \C\cup \E)$. Also, 
\begin{flalign*}
&r(\C, \C^c) - r(\C^c,\C) + r(\E, \E^c) - r(\E^c,\E)&&
\\
=&r(\C,\E) + r(\C , \C^c \setminus \E) - r(\E,\C) + r(\C^c \setminus \E, \C) + r(\E,\C) + r(\E, \E^c \setminus \C) - r(\C,\E) - r(\E^c \setminus \C, \E)&&\\
=& r(\C , \C^c \setminus \E) + r(\E, \E^c \setminus \C) - r(\C^c \setminus \E, \C) - r(\E^c \setminus \C, \E)&&\\
 =& r(\C \cup \E, (\C \cup \E)^c) - r((\C \cup \E)^c, \C \cup \E),&&
\end{flalign*}
where the last equality holds because the two sets are disjoint.}
\end{comment}

\vspace{-0.4cm}

\subsection{Proof of Corollary \ref{cor:essential-is-enough}}

\vspace{-0.1cm}

We prove the corollary by proving the following two lemmas.
 The first lemma shows that checking all test collections belong to $\D$, rather than all complete collections, is enough.  

\begin{lemma} \label{lem:test-is-enough}
If $\delta_{\rho}(\C)\ge 0$ is for  any test collection $\C \subseteq \D$, then $\delta_{\rho}(\hat{\C})\ge 0 $ for any complete collection $\hat{\C}$ such that $\emptyset \not \in \hat{\C}$.
\end{lemma}

%\textcolor{blue}{\fbox{$\emptyset \not \in \hat \C?$}}

This lemma reduces the number of conditions to be checked because any test collection is complete and it allows us to focus only on $\D$. The next lemma shows that we do not have to check $\delta_{\rho} (\C)\ge 0$ for the nonessential test collections.  

\begin{lemma} \label{lem:non-essential}
Let $\C\equiv\{A\cup E \mid E \in \E\}$ be a test collection with $A \subseteq \tilde{X}$ and $\E\subseteq 2^{X^*}$.  Assume that $\C\subseteq \D$.  (i) If $\E=2^{X^*}$, then $\delta_{\rho}(\C) = 0$; (ii) Suppose $K(\rho, D,x)\ge 0$ for all $(D,x) \in \M$.  If $A=\tilde{X}$ or $A=\emptyset$, then $\delta_{\rho} (\C) \ge 0$.
\end{lemma}

%The next lemma is a symmetric to the previous lemma and shows that in the Boolean lattice, the values of deltas are always nonnegative when $\C$ is a the bottom part and its subset.

Combining Proposition \ref{prop:characterization-complete}, Lemmas \ref{lem:test-is-enough} and \ref{lem:non-essential} immediately imply that checking the essential collections belonging to $\D$, rather than all test collections, is enough, which is the statement of the corollary.

\subsubsection{Proof of Lemma \ref{lem:test-is-enough}}

We will prove the statement by the following two steps.

\noindent\textbf{Step 1:} If $\delta_{\rho} (\hat{\C}) \geq 0$ for any test collection $\hat{\C} \subseteq \D$, then $\delta_{\rho} (\C) \geq 0$ for any test collection $\C$ such that $\emptyset \not \in \C$.

\begin{myproof}
Fix a test collection $\C$ such that $\emptyset \not \in \C$. Assume that $\C \not \subseteq \D$. 

\noindent\textbf{Case 1}: Suppose that $\C \cap \D = \emptyset$. By the property of $\D$ (i.e., $D \in \D\  \&\ E \supseteq D \implies E \in \D$), there are no observable inflow into $\C$. That is, if there exists $(E,y)$ such that $E\not \in \C$ and $E\cup y \in \C$, then $E\cup y \not \in \D$, which shows that $\delta_{\rho}(\C)$ does not contain any $-\sum_{(E,y): E \not\in {\C}, E\cup y \in {\C}, E\cup y \in \D, y \in \tilde{X}}         K(\rho, E \cup y, y)$. Moreover, since $\emptyset \not \in \C$, $\delta_{\rho}(\C)$ does not contain $-1\{\emptyset \in \C, X \not \in \C\}$ either. This means that $\delta_{\rho}(\C)$ does not contain any negative terms. Thus $\delta_{\rho}(\C)\ge 0$.

\noindent\textbf{Case 2}: Suppose that $\C \cap \D \neq \emptyset$. Let $\C^*= \C \cap \D$. By the property of $\D$ (i.e., $D \in \D\  \&\ E \supseteq D \implies E \in \D$), $\D$ is complete. Since $\C$ is complete, its union $\C^*$ is also complete. Since $\C^* \subseteq \D$, it follows from our supposition that $\delta_{\rho} (\C^*) \ge 0$. By the disjoint additivity of $\delta_{\rho}$ (Lemma \ref{lem:delta-is-additive}), $\delta_{\rho}(\C)= \delta_{\rho}(\C^*)+ \sum_{D \in \C \setminus \C^*}\delta_{\rho}(\{D\})$. Since for any $D \in \C \setminus \C^*\equiv \C \cap \D^c$, we have $\{D\} \cap \D=\emptyset$. Since $D\neq \emptyset$, by Case 1, we have $\delta_{\rho}(\{D\})\ge 0$. Thus, we have $\delta_{\rho}(\C) \ge 0$.
\end{myproof}

\noindent\textbf{Step 2}: If $\delta_{\rho}(\C)\ge 0$ is for any test collection $\C$ such that $\emptyset\not \in \C$, then $\delta_{\rho}(\hat{\C})\ge 0 $ for any complete collection $\hat{\C}$ such that $\emptyset\not \in \hat{\C}$.

\begin{myproof}Fix a complete collection $\hat{\C}$ such that $\emptyset\not \in \hat{\C}$. Decompose $\hat{\C}$ as follows: for each $A \subseteq \tilde{X}$ write $\C_A = \{D \in \hat{\C} : D \setminus X^* = A\}$. Clearly $\hat{\C} = \bigcup_{A \subseteq \tilde{X}} \C_A$. 
%We want to show that $\C_A$ is a test collection. By definition, $D \setminus X^* = A$ for all $D \in \C_A$ so we only need to show that $\C_A$ satisfies $(*).$ Fix $D \in \C_A$ and $x \in X^*$. We want to show $D \cup  x \in \C_A$. Since $\C$ is a test collection and satisfies $(*)$ in particular, $D \cup  x \in \C$. Also $D \cup  x \setminus X^* = D \setminus X^* = A$ thus $D \cup  x \in \C_A$.
It is easy to see that each $\C_A$ is a test collection and $\emptyset \not \in \C_A$. Thus by the assumption of the step, $\delta_{\rho}(\C_A)\ge 0$. Notice that for $A \neq B$, $\C_A$ and $\C_B$ are disjoint. By Lemma \ref{lem:delta-is-additive}, $\delta_{\rho}(\C)$ can be written as $\delta_{\rho}(\hat{\C}) = \sum_{A\subseteq \tilde{X}} \delta_{\rho}(\C_A)\ge 0$.
\end{myproof}

\subsubsection{Proof of Lemma \ref{lem:non-essential}}

Let $\C\equiv\{A\cup E \mid E \in \E\}$ be a test collection with $A \subseteq \tilde{X}$ and $\E\subseteq 2^{X^*}$. Assume that $\C\subseteq \D$.

\noindent\textbf{Step }1: If $\E=2^{X^*}$, then $\delta_{\rho}(\C) = 0$. 

\begin{myproof}
By the fact that $\E=2^{X^*}$ and $\C \subseteq \D$, all flows into and out of $\C$ are observable.\footnote{That is, (i) if there exists $(D,x)$ such that $D \in \C$ and $(D,x)\not \in \C$, then $D \in \D$; and  (ii) if there exists $(E,y)$ such that $E\not \in \C$ and $E\cup y \in \C$, then $E\cup y \in \D$.}  By the equality of inflows and  outflows (not necessarily nonnegative), it follows that $\delta_{\rho}(\C)$ is zero. 
\end{myproof}

\noindent \textbf{Step } 2: Suppose $K(\rho, D,x)\ge 0$ for all $(D,x) \in \M$.  If $A=\tilde{X}$ or $A=\emptyset$, then $\delta_{\rho} (\C) \ge 0$.

\begin{myproof}
Assume $A=\tilde{X}$. Fix any $D \in \C$ such that $D \neq X$. By the supposition, there is no observable flows coming out from $D$. Since $K(\rho, D,x) \ge 0$ for all $x \in \tilde{X}$, it follows from the definition of $\delta_{\rho}$, $\delta_{\rho} (D) \le 0$  for all $D \neq X$. Remember $\delta_{\rho}(\C)=\sum _{D \in \C} \delta_{\rho}(D)$. Since $\delta_{\rho} (D) \le 0$ for all $D \in \C \setminus  x$, it suffices to prove $\delta_{\rho}(\C)= 0$ where $\C$ is the largest, or $\C=\{\tilde{X} \cup E \mid E \in 2^{X^*}\}$. By Step 1, we have $\delta_{\rho}(\{\tilde{X} \cup E \mid E \in 2^{X^*}\})=0$.

Assume $A=\emptyset$.  If $\emptyset \in \C$, then $\C=2^{X^*}$ by the fact that $\C$ is complete. Thus, all inflows into $\C$ are not observable and all outflows from $\C$ are observable, $\delta_{\rho}(\C)=\sum_{(D,x): D\setminus  x \in \C, D  \not\in \C} K(\rho, D,x) \ge 0$.
\end{myproof}

\vspace{-0.5cm}

\section{Proof of Statement (b) of Theorem \ref{thm:characterization}}

\vspace{-0.2cm}

First, we explain the outline of the proof of statement (b). In the proof, we first obtain the the nonredundancy results assuming $\D=2^X\setminus \emptyset$; then translate the results into the given incomplete datasets.  Fix an essential test  collection $\C^*$. It suffices to show that there exists an incomplete dataset $\rho$ that satisfies all inequalities in (i) as well as all inequalities in (ii) except the one for $\C^*$. 

We provide a preliminary lemma that allows us to convert a flow from $\emptyset$ to $X$ into a complete dataset:

\begin{lemma}\label{lem:flow_to_prob} Let $\D=2^X \setminus \emptyset$. 
If there exists $r\in \Re^{\{(D\setminus x, D) \mid x \in D \in 2^X \}}$ satisfying the following three conditions: (i) $\sum_{x \in X} r(X \setminus x, X)=1$; (ii) for any $D \in \D$ such that $1\le |D|\le |X|-1$, $\sum_{x \in D}r(D\setminus x, D)=\sum_{y \not \in D}r(D, D \cup y)$; (iii) for any $x \in D \in \D$, $\sum_{E:E\supseteq D} r(E\setminus x, E) \ge 0$, then there exists an complete dataset $\rho^* \in \Re^{\{(D,x) \mid x \in D \in 2^X \} }_{+}$ such that $\sum_{x \in D} \rho^*(D,x)=1$ for all $D \in D$ and  $K(\rho^*, D,x)=r(D\setminus x, D)$ for any $(D,x)$ such that $x \in D \in 2^X$.
\end{lemma}

We provide the proof of Lemma \ref{lem:flow_to_prob} in Section \ref{sec:proof_B1} in the online appendix.

For any $\C \subseteq \N$, define $\delta_r(\C)\equiv  \sum_{\substack{(D,x):\\ D \in {\C},  D\cup x \not\in {\C},x\in \tilde{X}}} r(D, D \cup x) -  \sum_{\substack{(E,y):\\ E \not\in {\C}, E\cup y \in {\C},y \in \tilde{X}}}r(E,$ $ E \cup y)
    +
    1\{X \in {\C}, \emptyset \not \in {\C}\}
    -
    1\{\emptyset \in {\C}, X \not \in {\C}\}$.

Given Lemma \ref{lem:flow_to_prob}, we will construct a flow $r\in \Re^{\{(D\setminus x, D) \mid x \in D \in \D \}}$ that satisfies conditions (i)--(iii) in the lemma and the following two conditions: (a) $r(D\setminus x, D)\ge 0$ for all $x \in \tilde{X}$; (b) $\delta_r(\C^*) < 0$ and $\delta_r(\C) \ge  0$ for any essential test collection $\C$ except $\C^*$. (See Lemma \ref{cor:essential-is-minimal2} below for the complete statements.) By using Lemma \ref{lem:flow_to_prob}, we translate the flow $r$ into a complete dataset $\rho^* \in \Re^{\{(D,x) \mid x \in D \in 2^X \} }_{+}$; then convert $\rho^*$ to an incomplete dataset $\rho \in \Re^{\M}_{+}$. (See Claim  below.)  The most difficult part is the construction of the flow $r$. The difficulty comes from the fact that we need to change the value of $\delta_r$ only on one particular essential test collection $\C^*$ but not the others; the values of $\delta_r(\C)$ across test collections $\C$ are interdependent through the conservation law of the network flow; and essential test collections exist across the network. We overcome this difficulty by constructing several flows and combine them into one desirable flow $r$ in an intricate way.

\vspace{-0.4cm}

\subsection{Main proof of statement (b)}\label{sec:ap:essential}

\vspace{-0.1cm}

To prove statement (b) of Theorem \ref{thm:characterization}, we prove the following lemma:

\begin{lemma}\label{cor:essential-is-minimal2} Let $\D=2^X\setminus \emptyset$. \\
(i) For each essential test  collection $\C^*$,  there exists $r\in \Re^{\{(D\setminus x, D) \mid x \in D \in \D \}}$ that satisfies conditions (i)--(iii) in Lemma \ref{lem:flow_to_prob} and the following two conditions: (a) $r(D\setminus x, D)\ge 0$ for all $x \in \tilde{X}$; (b) $\delta_r(\C^*) < 0$ and $\delta_r(\C) \ge  0$ for any essential test collection $\C$ except $\C^*$.\\
(ii) For each $(D,x) \in \M$ such that $1<|D|<|X|$, there  exists $r\in \Re^{\{(D\setminus x, D) \mid x \in D \in \D \}}$ that satisfies conditions (i)--(iii) in Lemma \ref{lem:flow_to_prob} and the following two conditions: (a) $r (D \setminus x,D)<0$;  $r(E\setminus y,E)\ge 0$ for all $(E,y) \in \M$ s.t. and $(E,y)\neq (D,x)$; 
(b)$\delta_r(\C) \ge 0$ for any essential test collection $\C$. 
\end{lemma}

We provide the proof of Lemma \ref{cor:essential-is-minimal2} in section \ref{sec:lemma_b_2} in the online appendix. Lemma \ref{cor:essential-is-minimal2} shows the following claim, which implies statement (b) of Theorem \ref{thm:characterization}.

\noindent \textbf{Claim:} {\it (i) For each essential test collection $\C^* \subseteq \D$, there exists an incomplete dataset $\rho \in \Re^{\M}$ such that (a) $K(\rho,D,x)\ge 0$  for all $(D,x) \in \M$;(b) $\delta_{\rho}(\C^*) < 0$ and $\delta_{\rho}(\C) \ge  0$ for all essential test collection $\C \subseteq \D$ except $\C^*$.\\
(ii) For each $(D,x) \in \M$ such that $1<|D|<|X|$, there exists an incomplete dataset $\rho \in \Re^{\M}$ such that (a) $K(\rho,D,x)<0 $; and $K(\rho,E,y)\ge 0$  for all $(E,y) \in \M\setminus  \{(D,x)\}$; (b) $\delta_{\rho}(\C) \ge  0$ for all essential test collection $\C \subseteq \D$.
}

\noindent \textbf{Proof of Claim:} We first prove statement (i),  fix $\C^* \subseteq \D$. By Lemma \ref{cor:essential-is-minimal2} (ii), there exists $r\in \Re^{\{(D\setminus x, D) \mid x \in D \in 2^X\}}$ that satisfies conditions (i)--(iii) in Lemma \ref{lem:flow_to_prob} and conditions (a) and (b) in Lemma \ref{cor:essential-is-minimal2} (i). By Lemma \ref{lem:flow_to_prob}, there exists a complete dataset $\rho^* \in \Re^{\{(D,x) \mid x \in D \in 2^X \}}_{+}$ such that $\sum_{x \in D} \rho^*(D,x)=1$ for all $D \in 2^X$ and  $K(\rho^*, D,x)=r(D\setminus x, D)$ for any $(D,x)$ such that $x \in D \in 2^X$. Let $\rho$ be the restriction of $\rho^*$ to $\M$. In the following, we will show that $\delta_{\rho}(\C^*) < 0$ and $\delta_{\rho}(\C) \ge  0$ for all essential test collection $\C \subseteq \D$ except $\C^*$.

Let $\delta_{\rho^*}$ be the function  defined by (\ref{eq:delta_o}) with respect to the complete dataset $\rho^*$ with $\D=2^X$ and $X^*=\emptyset$.\footnote{That is, $
    \delta_{\rho^*} (\C)
    =
    \left(
        \sum_{(D,x): D \in {\C},  D\cup x \not\in {\C}}  
        K(\rho^*, D \cup x, x)
        - 
        \sum_{(E,y): E \not\in {\C}, E\cup y \in {\C}}  
        K(\rho^*, E \cup y, y)
    \right)\\
    +
    1\{X \in {\C}, \emptyset \not \in {\C}\}
    -
    1\{\emptyset \in {\C}, X \not \in {\C}\}$.}  Let $\delta_{\rho}$ be the function defined by (\ref{eq:delta_o}) with respect to the incomplete dataset of $\rho$ with given $\D$ and $X^*$. Remember $\delta_r$ is the function defined after Lemma \ref{lem:flow_to_prob}. Note that for any test collection $\C \subseteq \D$, $\delta_r(\C)=\delta_{\rho^*}(\C)= \delta_{\rho}(\C)$, where the first equality holds because $K(\rho^*, D,x)=r(D\setminus x, D)$ and the second equality holds because the value of $\delta$ does not depend on the values of $\rho^*$ and $\rho$ outside of $\M$; and $\rho=\rho^*$ on $\M$. Thus, we have $\delta_{\rho}(\C^*) < 0$ and $\delta_{\rho}(\C) \ge  0$ for all essential test collection $\C \subseteq \D$ except $\C^*$.

Statement (ii) can be proved exactly in the same way using Lemma \ref{cor:essential-is-minimal2} (ii) in stead of (i).

\vspace{-0.5cm}

\section{Proof of Propositions}

\vspace{-0.4cm}

\subsection{Proof of Proposition \ref{prop:outside}}

\vspace{-0.1cm}

By
\cite{falmagne1978representation}, it suffices to show that $K( \hat{\rho},\hat{D},x) \ge 0$ for all $(\hat{D},x)$ such that $x \in \hat{D} \in \hat{{\cal D}}$. Choose $\hat{D} \in \hat{{\cal D}}$. Write $D = \hat{D} \cap \tilde{X}$.\\
\textbf{Case 1:} $x_0 \in \hat{D}$. Then $\hat{D}=D \cup x_0$. By calculation, we have $K(\hat{\rho},D \cup x_0 ,x)=K(\rho, D \cup X^*, x)$ for $x \in D.$
Thus the nonnegativity of $K(\hat{\rho},D \cup x_0 ,x)$  is immediate from condition (i). By calculation, again, we have $K(\hat{\rho},D \cup x_0 ,x_0)
=\sum_{y \notin D \cup X^*}K(\rho,(D \cup y) \cup X^* ,y) - \sum_{x \in D}K(\rho, D \cup X^* ,x)
\ge 0$, where the inequality holds by the inequality (\ref{eq:th1}) in condition (ii) for   the test collection $\C = \{D \cup X^*\}$.\footnote{Note that if a test collection $\C$ is a singleton then $\C=\{A \cup X^*\} \subseteq \D$ for some $D$ such that $\emptyset \subsetneq D \subsetneq \tilde{X}$. The term $\sum_{y \notin D \cup X^*}K(\rho,(D \cup y) \cup X^* ,y)$ is the value of total outflows from $\{D \cup X^*\}$; $\sum_{x \in D}K(\rho, D \cup X^* ,x)$ is the value of observable inflow into the node. Thus the sum is the total outflow minus the observable inflows.}

%$$\sum_{(D,x): D \in {\C},  D\cup x \not\in {\C}}  K(\rho, D \cup x, x)- \sum_{(F,y): F \not\in {\C}, F\cup y \in {\C}, y \in \tilde{X}}  K(\rho, F \cup y, y) = \sum_{x^* \in X^*} K(\rho, D \cup X^*, x^*)$$

\noindent\textbf{Case 2:}  $x_0 \notin \hat{D}$ (i.e., $D \subseteq \tilde{X} $). Then $\hat{D}=D$. By the definition, we have $\hat{\rho}(D,x) = \rho(D,x)$, thus $\sum_{E \supseteq D}K(\rho,E,x) = \sum_{\hat{E} \supseteq D}K(\hat{\rho},\hat{E},x)$, which gives 
\begin{align*}
K(\hat{\rho},D,x) 
&= \sum_{E:E \supseteq D} K(\rho,E,x) - \sum_{\hat{E}:\hat{E} \supsetneq D}K(\hat{\rho},\hat{E},x)\\
&= \sum_{E: E \supseteq D, E\not \supseteq X^*}K(\rho,E,x) - \sum_{E:\tilde{X} \supseteq E\supsetneq D}K(\hat{\rho},E,x),
\end{align*}
 where the second equality holds because $K(\rho, E,x)=K(\hat{\rho},\hat{E},x)$, where $\hat{E} = (E \cap \tilde{X}) \cup x_0$,  for all $E\supseteq X^*$.  Thus for any $D$ such that $D \subseteq \tilde{X}$, we have 
\begin{equation}\label{eq:prop3.4}
K(\hat{\rho},D,x)= \sum_{E: E \supseteq D, E\not \supseteq X^*}K(\rho,E,x) - \sum_{E:\tilde{X} \supseteq E\supsetneq D}K(\hat{\rho},E,x).
\end{equation}

By the assumptions of the proposition,  $K(\rho,E,x) \ge 0$ for all $E \subseteq X$ and $x \in \tilde{X}$ and $K(\hat{\rho}, E \cup x_0, x) \ge 0$ for all $E \subseteq \tilde{X}$ and $x \in E \cup x_0$. We wish to show that $K( \hat{\rho} ,D,x) \ge 0$ for all $D \subseteq \tilde{X}$. We proceed by induction. The base case is $D = \tilde{X}$, for which that claim holds vacuously because $\sum_{E:\tilde{X} \supseteq E\supsetneq D}K(\hat{\rho},E,x)=0$. 

Fix $D \subseteq \tilde{X}$ and $x \in D$ and suppose that $K(\hat{\rho},E,x) \ge 0$ for all $\tilde{X} \supseteq E \supsetneq D$. It suffices to show the result for deterministic $\rho$.  By the induction assumption we see that $\hat{\rho}$ is rationalizable on the upper set domain $\{(E,x): x\in E \in \hat {\D}, E \supsetneq D\}$. Since $\rho$ is deterministic, so is $\hat{\rho}$ and thus there is a a single order $\hat{\succ}$ on $\hat{X}$ that rationalizes $\hat{\rho}$. We see that $K(\hat{\rho},E,x) = 1(E^c \ \hat{\succ} \ x \ \hat{\succ}\  E \setminus \{x\})$ for $E$ on the restricted domain. Thus we see that there is at most one $E' \supsetneq D$ with $K(\hat{\rho},E',x) > 0$. If there is no such $E'$, then the term $\sum_{E:\tilde{X} \supseteq E\supsetneq D}K(\hat{\rho},E,x) = 0$; thus (\ref{eq:prop3.4}) and the supposition of the proposition implies $K(\rho,D ,x) \ge 0$. 

If such $E'$ does exist then $\sum_{E:\tilde{X} \supseteq E\supsetneq D}K(\hat{\rho},E,x)=K(\hat{\rho},E',x) =1$. Also, 
\begin{align*}
1&=K(\hat{\rho},E',x) = \sum_{F:F \supseteq E',F \not\supseteq X^*}K(\rho,F,x) - \sum_{F:\tilde{X} \supseteq F\supsetneq E'}K(\hat{\rho},F,x) \\
&= \sum_{F: F \supseteq E', F \not\supseteq X^*}K(\rho,F,x),
\end{align*}
where the second  equality holds by (\ref{eq:prop3.4}) and the third equality holds by the fact that there is  no $F$ such that  $F \supsetneq E' $ with $K(\hat{\rho},F,x) > 0$.  Therefore $\sum_{E: E \supseteq D, \ E \not\supseteq X^*}K(\rho,E,x)\ge  \sum_{F: F \supseteq E', F \not\supseteq X^*}K(\rho,F,x) = 1$. It follows from (\ref{eq:prop3.4}) and $\sum_{E:\tilde{X} \supseteq E\supsetneq D}K(\hat{\rho},E,x)=1$ that $K(\hat{\rho},D,x) \ge 0$.  Thus, the claim holds for all $D\subseteq \tilde{X}$ by induction.

\vspace{-0.4cm}

\subsection{Proof of Proposition \ref{coro:bound2}}\label{sec:bounds_new}

By M\"{o}bius inversion,  the condition (ii) in Definition \ref{def:gamma_0} can be written as follows: for all $(D,x)\text{ such that }x\in D \in 2^X$, $\mu(\{\succ \in \L \mid D^c  \succ x \succ D \setminus x  ) =K(\rho^*, D,x)$. Thus by condition (\ref{eq:0con}), if $r$ is a solution to (P2) then $\mu(\{\succ \in \L \mid D^c  \succ x \succ D \setminus x  \}) =r(D\setminus x, D)$. Therefore condition (ii) is equivalent to 
\begin{equation}\label{eq:rk}
r(D \setminus x, D) =K(\rho^*, D,x)\text{ for all }(D,x) \text{ such that }x\in D \in 2^X.
\end{equation}
Thus we can rewrite the set $\Gamma$ (i.e., (\ref{eq:boundP1})) as follows:
\vspace{-0.1cm}
\begin{align}\label{eq:boundP1.5}
    \footnotesize
    \left\{
        \rho^*
        \in 
        \Re_+^{\{(D,x) \mid x \in D \in 2^X\}}
        \ \bigg | \ 
        {\small
            \begin{array}{l}
                \text{There exists  a solution $r \in \Re_+^{\{(D\setminus x, D)| x\in D\in 2^X\}}$ to   }\\
                 \text{(P2) satisfying  (\ref{eq:rk}) and } \rho^*=\rho \text{ on }\M.
            \end{array}
        }
    \right\}
    \normalsize
\end{align}
By eliminating observable flows $r$ (i.e., $r(D\setminus x, D)=K(\rho, D,x)$ for all $(D,x) \in \M$) in (P2), it can be verified that the conditions (\ref{eq:0sum1}) and  (\ref{eq:0in=out}) of (P2) are equivalent to (\ref{eq:dmnd=sply}).\footnote{ Condition (\ref{eq:0con}) in (P2) is implied by (\ref{eq:rk}) and $\rho^*=\rho \text{ on }\M$.} 
 Using the M\"{o}bius inversion formula, we also can rewrite (\ref{eq:rk}) into the following:  for all $(D, x)$ such that $x \in D \in 2^X$,  we have 
\begin{equation}\label{eq:mobi}
\rho^*(D, x) 
    = 
    \sum_{E: E \supseteq D}
    r (E \setminus x, E)
    ,
\end{equation}
where $r(E\setminus x, E)=K(\rho, E,x)$ for all \textcolor{black}{$(E,x) \in \M$}. These observations imply that we can rewrite the set (\ref{eq:boundP1.5}) into the following set:
\begin{align}\label{eq:boundP2}
    \footnotesize
    \left\{
        \rho^*
        \in 
        \Re_+^{\{(D,x) \mid x \in D \in 2^X\}}
        \ \bigg | \ 
        {\small
            \begin{array}{l}
             \text{ There exists  $r \in \Re_+^{\{(D\setminus x, D) \mid x\in D\in 2^X\}}$ } \\
            \text{ satisfying (\ref{eq:dmnd=sply})\text{ and }(\ref{eq:mobi})   and } \rho^*=\rho \text{ on }\M\\
            \end{array}
        }
    \right\}.
    \normalsize
\end{align}
It follows that  the  upper bound becomes 
   $\overline \rho (D, x)=  \max_{r\in \Re_+^{\{(D,x) \mid x \in D \in 2^X\}}} 
    \sum_{E: E \supseteq D}
    r (E \setminus x, E)$
 subject to (\ref{eq:dmnd=sply}),  where $r(E\setminus x, E)=K(\rho, E,x)$ for all \textcolor{black}{$(E,x) \in \M$}.  The corresponding result for the lower bound can be obtained by changing $\max$ to $\min$.

\vspace{-0.7cm}

\begin{singlespace}
\bibliographystyle{new_ecma}
\bibliography{discrete_ru}

@article{saito2025aggregate,
  title={Random utility model with aggregated alternatives},
  author={Liao, Yuexin and Saito, Kota and Sandroni, Alec},
  journal={Working paper can be found at https://arxiv.org/abs/2506.00372.},
  year={2025}
}

@article{dean2022better,
  title={A Better Test of Choice Overload},
  author={Dean, Mark and Ravindran, Dilip and Stoye, J{\"o}rg},
  journal={arXiv preprint arXiv:2212.03931},
  year={2022}
}

@article{cerreia2023multinomial,
  title={Multinomial logit processes and preference discovery: inside and outside the black box},
  author={Cerreia-Vioglio, Simone and Maccheroni, Fabio and Marinacci, Massimo and Rustichini, Aldo},
  journal={The Review of Economic Studies},
  volume={90},
  number={3},
  pages={1155--1194},
  year={2023},
  publisher={Oxford University Press US}
}

@article{cerreia2021canon,
  title={A canon of probabilistic rationality},
  author={Cerreia-Vioglio, Simone and Lindberg, Per Olov and Maccheroni, Fabio and Marinacci, Massimo and Rustichini, Aldo},
  journal={Journal of Economic Theory},
  volume={196},
  pages={105289},
  year={2021},
  publisher={Elsevier}
}

@book{luce2005individual,
  title={Individual choice behavior: A theoretical analysis},
  author={Luce, R Duncan},
  year={2005},
  publisher={Courier Corporation}
}

@article{apesteguia2017single,
  title={Single-crossing random utility models},
  author={Apesteguia, Jose and Ballester, Miguel A and Lu, Jay},
  journal={Econometrica},
  volume={85},
  number={2},
  pages={661--674},
  year={2017},
  publisher={Wiley Online Library}
}

@article{petri2023binary,
  title={Binary single-crossing random utility models},
  author={Petri, Henrik},
  journal={Games and Economic Behavior},
  volume={138},
  pages={311--320},
  year={2023},
  publisher={Elsevier}
}

@article{reinelt1993note,
  title={A note on small linear-ordering polytopes},
  author={Reinelt, Gerhard},
  journal={Discrete \& Computational Geometry},
  volume={10},
  pages={67--78},
  year={1993},
  publisher={Springer}
}

@inproceedings{suck2002binary,
  title={From binary choice to complete choice; combinatorics and polytopes},
  author={Suck, Reinhard},
  booktitle={33rd EMPG meeting in Bremen},
  year={2002}
}

@article{mccausland2020testing,
  title={Testing the random utility hypothesis directly},
  author={McCausland, William J and Davis-Stober, Clintin and Marley, Anthony AJ and Park, Sanghyuk and Brown, Nicholas},
  journal={The Economic Journal},
  volume={130},
  number={625},
  pages={183--207},
  year={2020},
  publisher={Oxford University Press}
}

@article{suck2016regular,
  title={Regular choice systems: A general technique to represent them by random variables},
  author={Suck, Reinhard},
  journal={Journal of Mathematical Psychology},
  volume={75},
  pages={110--117},
  year={2016},
  publisher={Elsevier}
}

@article{sprumont2022regular,
  title={Regular random choice and the triangle inequalities},
  author={Sprumont, Yves},
  journal={Journal of Mathematical Psychology},
  volume={110},
  pages={102710},
  year={2022},
  publisher={Elsevier}
}

@article{brady2016menu,
  title={Menu-dependent stochastic feasibility},
  author={Brady, Richard L and Rehbeck, John},
  journal={Econometrica},
  volume={84},
  number={3},
  pages={1203--1223},
  year={2016},
  publisher={Wiley Online Library}
}

@article{chambers2021correlated,
  title={Correlated choice},
  author={Chambers, Christopher P and Masatlioglu, Yusufcan and Turansick, Christopher},
  journal={arXiv preprint arXiv:2103.05084},
  year={2021}
}

@article{turansick2022identification,
  title={Identification in the random utility model},
  author={Turansick, Christopher},
  journal={Journal of Economic Theory},
  volume={203},
  pages={105489},
  year={2022},
  publisher={Elsevier}
}

@book{NetworkFlows,
  title={Network Flows and Monotropic Optimization},
  author={Rockafellar, Tyrrell},
  year={1998},
  publisher={Athena Scientific}
}

@book{linear_ordering_polytope,
  title={The Linear Ordering Problem},
  author={Martí, Rafael and Reinelt, Gerhard},
  year={2011},
  publisher={Springer}
}

@InCollection{mcfad91,
  author={McFadden, D. and Richter, M.K.},
title={{Stochastic rationality and revealed stochastic preference}},
  booktitle = 	 {Preferences, Uncertainty, and Optimality, Essays in Honor of Leo Hurwicz, Westview Press: Boulder, CO},
  pages={161--186},
  year={1990}
}

@article{falmagne1978representation,
  title={A representation theorem for finite random scale systems},
  author={Falmagne, Jean-Claude},
  journal={Journal of Mathematical Psychology},
  volume={18},
  number={1},
  pages={52--72},
  year={1978},
  publisher={Elsevier}
}

@article{fiorini2004short,
  title={A short proof of a theorem of Falmagne},
  author={Fiorini, Samuel},
  journal={Journal of mathematical psychology},
  volume={48},
  number={1},
  pages={80--82},
  year={2004},
  publisher={Elsevier}
}

@article{article,
author = {Lu, Jay and Saito, Kota},
year = {2022},
month = {02},
pages = {},
title = {Mixed Logit and Pure Characteristics Models}
}

@article{doignon2022adjacencies,
  title={Adjacencies on random ordering polytopes and flow polytopes},
  author={Doignon, Jean-Paul and Saito, Kota},
  journal={arXiv preprint arXiv:2207.06925},
  year={2022}
}

@book{ford2015flows,
  title={Flows in networks},
  author={Ford Jr, Lester Randolph and Fulkerson, Delbert Ray},
  volume={56},
  year={2015},
  publisher={Princeton university press}
}

@inproceedings{mcfadden2006revealed,
  title={Revealed stochastic preference: a synthesis},
  author={McFadden, Daniel L},
  booktitle={Rationality and Equilibrium: A Symposium in Honor of Marcel K. Richter},
  pages={1--20},
  year={2006},
  organization={Springer}
}

@article{manski2007partial,
  title={Partial identification of counterfactual choice probabilities},
  author={Manski, Charles F},
  journal={International Economic Review},
  volume={48},
  number={4},
  pages={1393--1410},
  year={2007},
  publisher={Wiley Online Library}
}

@article{orlin1997polynomial,
  title={A polynomial time primal network simplex algorithm for minimum cost flows},
  author={Orlin, James B},
  journal={Mathematical Programming},
  volume={78},
  pages={109--129},
  year={1997},
  publisher={Springer}
}

@article{orlin1993polynomial,
  title={Polynomial dual network simplex algorithms},
  author={Orlin, James B and Plotkin, Serge A and Tardos, {\'E}va},
  journal={Mathematical programming},
  volume={60},
  number={1-3},
  pages={255--276},
  year={1993},
  publisher={Springer}
}

@article{ahuja1988network,
  title={Network flows},
  author={Ahuja, Ravindra K and Magnanti, Thomas L and Orlin, James B},
  year={1988},
  publisher={Cambridge, Mass.: Alfred P. Sloan School of Management, Massachusetts~…}
}

@article{kitamura2018nonparametric,
  title={Nonparametric analysis of random utility models},
  author={Kitamura, Yuichi and Stoye, J{\"o}rg},
  journal={Econometrica},
  volume={86},
  number={6},
  pages={1883--1909},
  year={2018},
  publisher={Wiley Online Library}
}

@article{kitamura2019nonparametric,
  title={Nonparametric counterfactuals in random utility models},
  author={Kitamura, Yuichi and Stoye, J{\"o}rg},
  journal={arXiv preprint arXiv:1902.08350},
  year={2019}
}

@TechReport{turansickgraaphicalmethods,
  author={Christopher Turansick},
  title={{On Graphical Methods in Stochastic Choice}},
  year=2023,
  month=Mar,
  institution={arXiv.org},
  type={Papers},
  url={https://ideas.repec.org/p/arx/papers/2303.14249.html},
  number={2303.14249},
  doi={},
}

@TechReport{chambers_identification,
  author={Christopher P. Chambers and Christopher Turansick},
  title={{The Limits of Identification in Discrete Choice}},
  year=2024,
  month=Mar,
  institution={arXiv.org},
  type={Papers},
  url={http://arxiv.org/abs/2403.13773v1},
}

@article{canay2023user,
  title={A user’s guide for inference in models defined by moment inequalities},
  author={Canay, Ivan A and Illanes, Gast{\'o}n and Velez, Amilcar},
  journal={Journal of Econometrics},
  pages={105558},
  year={2023},
  publisher={Elsevier}
}

@article{andrews2010inference,
  title={Inference for parameters defined by moment inequalities using generalized moment selection},
  author={Andrews, Donald WK and Soares, Gustavo},
  journal={Econometrica},
  volume={78},
  number={1},
  pages={119--157},
  year={2010},
  publisher={Wiley Online Library}
}
\end{singlespace}

\newpage

\section*{**For Online Publication**}

\section*{Online Appendix}

\section{Omitted Proofs}

\subsection{Proof of Lemma \ref{lem:network-flow-feasibility}}\label{sec:lem:network-flow-feasibility}

To prove the lemma we prove the following general result:

\begin{theorem}
Let $T, S \subseteq {\cal N}$ such that $S \cap T =\emptyset$,  $a:S \to \Re_+$, $b : T \to \Re_+$ such that $\sum_{s \in S} a(s)=1=\sum_{t \in T} b(t)$. There exists $r:{\cal A} \to \Re_+$ such that $ \forall s\in S [r(s, {\cal N})-r({\cal N},s)=a(s)]$, $\forall D\in {\cal N} \setminus (S \cup T)    [r(D, {\cal N})-r({\cal N},D)=0]$, $\forall t\in T [r({\cal N},t)-r(t, {\cal N})=b(t)]$, $\forall (D,E) \in {\cal A} [l(D,E) \le r (D,E) \le u(D,E)]$ if and only if the following conditions hold for any $\C \subseteq {\cal N}$:
\begin{equation}\label{eq:c1}
\sum_{(D,E) \in \C \times \C^c}u(D,E)-\sum_{(D,E) \in \C^c \times \C}l(D,E) \ge  \sum_{t \in \C^c \cap T}b(t)-\sum_{s \in \C^c \cap S}a(s).
\end{equation}
\end{theorem}

\begin{proof}\ \\
\textbf{Necessity:}  Suppose a feasible flow $r$ exists.
$
\sum_{t \in \C^c \cap T}b(t)-\sum_{s \in \C^c \cap S}a(s)\le r(\C,\C^c)-r(\C^c,\C)\le \sum_{(D,E) \in \C \times \C^c}u(D,E)-\sum_{(D,E) \in \C^c \times \C}l(D,E).
$

\noindent \textbf{Sufficiency:} Define an extended network with lower bound by ${\cal N}^*={\cal N} \cup\{s^*,t^*\}$ and $\A^*= \A \cup\{(s^*,s) \mid s \in S\} \cup \{(t,t^*) \mid t \in T\}$ and $u^*(s^*,s)=a(s)$, $l^*(s^*,s)=0$ for all  $s \in S$,
$u^*(t, t^*)=b(t), l^*(t, t^*)=0$ for all $t \in T$, $u^*(D,E)= u(D,E), l^*(D,E)=l(D,E)$ for all other arcs.

We first define the residual capacity function $e$ in the augmented network: for any $\C^*  \subseteq \N^*$, $ 
e(\C^*, \N^* \setminus \C^*)= \sum_{(D,E) \in \C^* \times (\N^* \setminus \C^*)}u^*(D,E)- \sum_{(D,E) \in  (\N^* \setminus \C^*) \times \C^*}l^*(D,E).\footnote{We write $\N^* \setminus \C^*$ instead of $(\C^*)^c$ to clarify the underling space is $\N^*$ not $\N$.}$ (Similarly, we define the residual capacity function in the original network as follows: for any $\C  \subseteq \N$ $e(\C, \N \setminus \C)= \sum_{(D,E) \in \C \times (\N \setminus \C)}u^*(D,E)- \sum_{(D,E) \in  (\N \setminus \C) \times \C}l^*(D,E)$.)

%\[
%e(\C, \N \setminus \C)= \sum_{(D,E) \in \C \times \N \setminus \C}u(D,E)- \sum_{(D,E) \in  \N \setminus \C \times \C}l(E,D). 
%\]
%Note that $u^*=u$ and $l^*=l$ on $\N$, $e(\C, \N \setminus \C)= e(\C, \N \setminus \C)$  for any nonempty $\C \subseteq \N$.
%}}

Then we will prove that $(\N^* \setminus \{t^*\}, t^*)$ is a minimum $s^*$-$t^*$ cut.   Let be any $\C^* \subseteq {\cal N}^*$ such that $s^* \in \C^*$ and $t^* \not \in \C^*$. That is, $(\C^*,  \N^* \setminus \C^*)$  be an arbitrary cut separating $s^*$ and $t^*$. Let $\C=\C^* \cap {\cal N}$.  Then by the structure of network, 
\begin{flalign*}
&e(\C^*,  \N^* \setminus \C^*)- e(\N^* \setminus \{t^*\}, t^*) \\
&=e(s^*, \N^* \setminus \C^*)+e(\C, \N^* \setminus \C^*)- e(\N^* \setminus \{t^*\}, t^*)\quad (\because s^* \in \C^*)\\
&=e(s^*, \N^* \setminus \C^*)+e(\C, \N \setminus \C)+e(\C, t^*)- e(\N^* \setminus \{t^*\}, t^*)\quad (\because t^* \in \N^* \setminus \C^*)\\
&=e(s^*,(\N \setminus \C) \cap S)+e(\C, \N \setminus \C) +e(\C \cap T, t^*)- e(T, t^*)\\
&=e(s^*,(\N \setminus \C) \cap S)+e(\C, \N \setminus \C) -e((\N \setminus \C) \cap T, t^*)\\
&=\sum_{s \in (\N \setminus \C) \cap S}a(s)+e(\C, \N \setminus \C) -\sum_{t \in (\N \setminus \C) \cap T}b(t)\\
&=\sum_{s \in (\N \setminus \C) \cap S}a(s)+\left( \sum_{(D,E) \in \C \times (\N \setminus \C)}u(D,E)- \sum_{(D,E) \in  (\N \setminus \C) \times \C}l(D,E)\right)-\sum_{t \in (\N \setminus \C) \cap T}b(t),
\end{flalign*}
which is nonnegative by (\ref{eq:c1}). Thus $e(\C^*, \N^* \setminus \C^*)\ge  e(\N^* \setminus \{t^*\}, t^*)$ for any cut $(\C^*,\N^* \setminus \C^*)$ separating $s^*$ and $t^*$ if and only  (\ref{eq:c1}) holds for any $\C \subseteq {\cal N}$. 

It follows from the maximum-flow theorem with lower bounds (Theorem 6.1 \cite{ahuja1988network}) that (\ref{eq:c1}) implies the existence of a flow $r^*$ from $s^*$ to $t^*$ that saturates all arcs of $(T,t^*)$, that is, $r^*(t,t^*)=b(t)$ for all $t \in T$. Since $\sum_{s \in S}a(s)=1=\sum_{t \in T} b(t)$, we must have $r^*(s^*,s)=a(s)$ for all $s \in S$. These equalities imply that $r^*(S,{\cal N})=1$ and $r^*({\cal N},T)=1$ and $r^*(D, {\cal N})=r^*({\cal N},D)$ for all $D \in \N\setminus (T \cup S)$. Now define $r$ as a restriction of $r^*$ on $({\cal N}, {\cal A})$. Then $r$ satisfies all desired conditions. 
\end{proof}

By the theorem, we obtain the lemma by letting both $T$ and $S$ singletons. 

\subsection{Proof of Lemma \ref{lem:flow_to_prob}}\label{sec:proof_B1}
For any $(D,x)$ such that $x \in D \in 2^X$, define $\rho(D,x)= \sum_{E \supseteq D}r(E\setminus x,E)$.  By (iii), we have $\rho(D,x)\ge 0$ for all $(D,x)$  such that $x \in D \in 2^X$. Fix any $D$ to show $\sum_{x \in D} \rho (D,x)=1$.  Then we have
    $\sum_{x\in D}\rho(D,x) = \sum_{x \in D} \sum_{E \supseteq D} r(E \setminus x, E)
    = \sum_{y\in D}r(D \setminus y , D)+\sum_{x \in D}\sum_{\substack{ E\supseteq D \\|E| \ge |D|+1}}r(E\setminus x, E)
    =\sum_{\substack{E\supseteq D\\ |E| = |D|+1}}r(D,E)+ \sum_{x \in D}\sum_{\substack{E\supseteq D\\ |E| \ge |D|+1}}r(E\setminus x, E)
    =\sum_{\substack{E\supseteq D\\ |E| = |D|+1}}  \sum_{y \in E \cap D^c }r(E\setminus y,E)+\sum_{x\in D}\sum_{\substack{E\supseteq D\\ |E|=|D|+1}}r(E\setminus x, E)+ \sum_{x \in D}\sum_{\substack{E\supseteq D\\ |E| \ge |D|+2}}r(E\setminus x, E)    =\sum_{y \in E}\sum_{\substack{E\supseteq D\\ |E| = |D|+1}}r(E \setminus y, E)+\sum_{x\in D}\sum_{\substack{E\supseteq D\\ |E| \ge |D|+2}}$$r(E\setminus x, E)
    =\sum_{x \in E}\sum_{\substack{E\supseteq D\\ |E| \ge |D|+1}}r(E \setminus x, E)$, where the third equality holds by appling (ii) for the first term;  the fourth equality is obtained by rewriting the first term and dividing the second term into the two terms; and the second to the last equality is obtained by combining the first two terms into one. Note that the last term has the same form as the term in the first equation but in the last term the summation over $E=D$ is deleted. By repeating this, we get $\sum_{x\in D}\rho(D,x)= \sum_{x \in E}\sum_{\substack{E\supseteq D\\ |E| \ge |D|+2}}r(E \setminus x, E)$. Finally we get $\sum_{x\in D}\rho(D,x)=\sum_{\substack{y \in X}}r(X \setminus y, X)$, which is equal to $1$ by (i).

\begin{comment}
\begin{align*}
    \sum_{x\in D}\rho(D,x) &= \sum_{x \in D} \sum_{E \supseteq D} r(E \setminus x, E)\\
    &= \sum_{y\in D}r(D \setminus y , D)+\sum_{x \in D}\sum_{\substack{ E\supseteq D \\|E| \ge |D|+1}}r(E\setminus x, E)\\
    &=\sum_{\substack{E\supseteq D\\ |E| = |D|+1}}r(D,E)+ \sum_{x \in D}\sum_{\substack{E\supseteq D\\ |E| \ge |D|+1}}r(E\setminus x, E)\\
    &(\because  \text{apply (ii) for the first term})\\
    &=\sum_{\substack{E\supseteq D\\ |E| = |D|+1}}  \sum_{y \in E \cap D^c }r(E\setminus y,E)+\sum_{x\in D}\sum_{\substack{E\supseteq D\\ |E|=|D|+1}}r(E\setminus x, E)+ \sum_{x \in D}\sum_{\substack{E\supseteq D\\ |E| \ge |D|+2}}r(E\setminus x, E)\\ 
    &(\because \text{rewrite the first term and divide the second term into two terms})\\
    &=\sum_{y \in E}\sum_{\substack{E\supseteq D\\ |E| = |D|+1}}r(E \setminus y, E)+\sum_{x\in D}\sum_{\substack{E\supseteq D\\ |E| \ge |D|+2}}r(E\setminus x, E)\\
    &(\because \text{combine the first two terms into one})\\
    &=\sum_{x \in E}\sum_{\substack{E\supseteq D\\ |E| \ge |D|+1}}r(E \setminus x, E).
    \end{align*}
\end{comment}

\subsection{Proof of Lemma \ref{cor:essential-is-minimal2}}\label{sec:lemma_b_2}

\subsubsection{Proof of statement (i) in Lemma \ref{cor:essential-is-minimal2}}

Fix an essential test collection $\C^*$.  In the following, we will construct  a flow $r$ from $\emptyset$ to $X$ such that $\delta_r(\C^*)<0$ and $\delta_r(\C)\ge 0$ for any other essential test collection $\C\neq \C^*$.

Let $A^*$ be such that $D \setminus X^* = A^*$ for all $D \in \C^*$. (Such $A^*$ exists because $\C^*$ is a test collection.) Since $\C^*$ is essential, $A^* \neq \emptyset$ and $A^* \neq \tilde{X}$. Let $\hat{\D}\equiv \{D \mid \text{there exists an essential test collection }\C \text{ such that } D \in \C \}$. 

In the following we prove five claims to prove this lemma. 

\begin{claim}  There exists a $\emptyset - X$ directed path $\Pi_1$ avoiding any nodes in $\hat{\D}$.
\end{claim}
\begin{myproof}
We first  construct a directed path $\emptyset - \tilde{X}$ that avoids any node in $\hat{\D}$ by adding each element of $\tilde{X}$ one by one. Note that each node $A$ does not appear in any essential test collection since the only test collection containing $A$ is the  nonessential test collection $\{A \cup E \mid E \in 2^{X^*}\}$, which appears in Lemma \ref{lem:non-essential} (i).

In the same way, we next construct a directed path $\tilde{X} - X$ that avoids any node in $\hat{\D}$ by adding each element of $X^*$ one by one. Note that each such node can be written as $\tilde{X} \cup E$ for some $E \in 2^{X^*}$ and does not appear in any essential test collection since the only test collection containing the set is the  nonessential test collection $\{\tilde{X} \cup E \mid E \in \E \}$, which appears in Lemma \ref{lem:non-essential} (ii).

By combining these two directed paths, we obtain a desirable $\emptyset - X$ directed path avoiding any nodes in $\hat{\D}$.
\end{myproof}

The next claim shows that for each node $D$ in $\C^*$, there is a flow in which the value of $\delta$ is negative on the node $D$ and the values of $\delta$ on the other nodes in $\hat{\D}$ are zero. For simplicity, we introduce a notation: Fix $D, E \subseteq X$ such that $D \subsetneq E$. For each directed path $\Pi$ from $D$ to $E$ in the network defined by (\ref{eq:network_def}), define $r^{\Pi}\in \Re^{\{(F, F\cup x) \mid x \in F \in 2^X\}}$ by 
\begin{eqnarray}\label{def:r_pi}
r^{\Pi}(F, F\cup x)=
\left\{
\begin{array}{lll}
1 \text{ if }(F, F\cup x) \text{ is an arc that belongs to $\Pi$,}\\
0  \text{ otherwise}.
\end{array}
\right.
\end{eqnarray}

\begin{claim}\label{claim:fig1} Fix $\varepsilon \in (0,1]$. For any $D$ in $\C^*$, there exists a flow $r^1_D$  such that $\delta (\{D\})=-\varepsilon$ and  $\delta(\{\hat{D}\})=0$ for any $\hat{D} \in \hat{\D} \setminus D$.  Moreover $r^1_D$ satisfies the three conditions in the Lemma \ref{lem:flow_to_prob}.
\end{claim}

\begin{myproof} Fix $D \in \C^*$. 
Consider 
\bit
\item an $\emptyset - D$ directed path $\Pi_2$ containing going through the node $D \setminus A^*$,
\item an $A^*-X$ directed  path $\Pi_3$ which avoid any nodes in $\hat{\D}$ (Such a path exists because we can take the union of any $A^* - \tilde{X}$ directed path and any $\tilde{X} - X$ directed path as in Claim A.3.),
\item The directed path $\Pi_4$ from $A^*$ to $D$ which follows the same order as $\Pi_3$.
\eit

Remember the definition (\ref{def:r_pi}). Fix $\varepsilon>0$ and define $r^1_D\equiv(1-\varepsilon)r^{\Pi_1} + \varepsilon r^{\Pi_2} + \varepsilon r^{\Pi_3}-\varepsilon  r^{\Pi_4}$. Note that $r^1_D$ satisfies the three conditions in Lemma \ref{lem:flow_to_prob}. To confirm the condition (iii) is satisfied it suffices to show that all negative flows are cancelled in the sum $\sum_{E: E \supseteq D} r(E \setminus x, E)$. (All of the negative flows are in $\Pi_4$ and are canceled by some flow in $\Pi_3$ because $\Pi_4$ follows the same order as $\Pi_3$.)

Note also that in the flow,  $\delta_{r^1_D} (\{A^*\})=\delta_{r^1_D} (\{D \setminus A^*\})=\varepsilon$ and $\delta_{r^1_D} (\{\tilde{X}\})=-1$.  To see $\delta_{r^1_D} (\{D\})=-\varepsilon$ note that an arc going into $D$ exists and is observable because $A^*$ is not empty and consists of observable alternatives.

Moreover, for all other $\hat{D} \in \hat{\D}$,  $\delta_{r^1_D}(\{\hat{D}\})=0$. (To see this note that $\delta$ is non-zero only when observable inflows are not equal to observable outflows.\footnote{In the figure, this occurs when a dotted line becomes a solid line or vice-versa in the diagram.} ) 

Since $A^*, \tilde{X}, D \setminus A^*\notin \hat{\D}$ by Lemma \ref{lem:non-essential}, we have $\delta(\{\hat{D}\})=0$ for any $\hat{D} \in \hat{\D} \setminus D$. This completes the proof of the claim. 
\end{myproof}

The next claim shows that for each node $D$ in $\C^*$, there is a flow from $\emptyset$ to $X$ in which the value of $\delta$ is positive on the node $D$ and the values of $\delta$ on the other nodes in $\hat{\D}$ is zero. 

\begin{figure}[h]
\begin{centering}
  \begin{tikzpicture}[scale=.4, transform shape]
      \tikzset{vertex/.style = {shape=circle,draw,thick,minimum size=5em}}
    \tikzset{-to/.style = {thick,arrows={-to}}}

    \node[vertex] (X) at (-2,12.5) {\large $X$};
    \node[vertex] (Xminus) at (-2,8) {\large $\tilde{X}$};
    \node[vertex] (A) at (0,3) {\large $A^*$};
    \node[vertex] (D) at (2,6) {\large $D$};
        \node[vertex] (DminusA) at (4,3) {\large $D \setminus A^*$};
    \node[vertex] (emptyset) at (4,-1) {\Large $\emptyset$};

    \draw[-to,dotted] (emptyset) -- (DminusA) node[midway,right] {\large $+\varepsilon$};
    \draw[-to] (DminusA) -- (D) node[midway,right] {\large $+\varepsilon$};
    \draw[-to,dotted] (A) -- (D) node[midway, below right] {\large $-\varepsilon$};
    \draw[-to] (A) -- (Xminus) node[midway,right] {\large $+\varepsilon$};
    \draw[-to,dotted] (Xminus) -- (X) node[midway,right] {\large $+\varepsilon$};
  \end{tikzpicture}
\quad\quad\quad\quad
  \begin{tikzpicture}[scale=.35, transform shape]
    \tikzset{vertex/.style = {shape=circle,draw,thick,minimum size=5.5em}}
    \tikzset{-to/.style = {thick,arrows={-to}}}

    \node[vertex] (X) at (0,16) {\large $X$};
    \node[vertex] (A) at (0,4) {\large $A$};
    \node[vertex] (D) at (0,8) {\large $D$};    \node[vertex] (Dplus) at (0,12) {\large $D \cup \tilde{X}$};
    \node[vertex] (emptyset) at (0,0) {\Large $\emptyset$};

    \draw[-to] (emptyset) -- (A) node[midway,right] {\large $+\varepsilon$};
    \draw[-to,dotted] (A) -- (D) node[midway,right] {\large $+\varepsilon$};
    \draw[-to] (D) -- (Dplus) node[midway,right] {\large $+\varepsilon$};
    \draw[-to,dotted] (Dplus) -- (X) node[midway,right] {\large $+\varepsilon$};
    \end{tikzpicture}
    
\caption{Flow $r^1_D$ in Claim \ref{claim:fig1} (left);\ Flow $r^2_D$ in Claim \ref{claim:fig2} (right)}
\end{centering}
\begin{footnotesize}
Note: Given an incomplete dataset $\rho \in \Re^{\M}_+$,  solid arrows correspond to observable flows and dotted arrows correspond to unobservalbe flows.
\end{footnotesize}
   \end{figure}

\begin{claim}\label{claim:fig2} Fix $\varepsilon \in (0,1]$. For any $D$ in $\C^*$, there exists a flow $r^2_D$ such that   $\delta_{r^2_D} (\{D\})=\varepsilon$ and   $\delta_{r^2_D}(\{\hat{D}\})= 0$ for all $\hat{D} \in \hat{\D} \setminus D$. Moreover the flow $r^2_D$ satisfies the all conditions in Lemma \ref{lem:flow_to_prob}.
\end{claim}

\begin{myproof}
Choose any directed path from $\emptyset-X$  that goes through nodes $A, D,$ and $\tilde{X} \cup D$. We denote the path by $\Pi_5$. Define $r^2_D\equiv (1-\varepsilon)r^{\Pi_1} + \varepsilon r^{\Pi_5}$. Note that  $r^2_D$ satisfies the all conditions in Lemma \ref{lem:flow_to_prob}.  

In the flow, $\delta_{r^2_D} (\{A^*\})=\delta_{r^2_D}(\tilde{X} \cup D)=-\varepsilon$ and $\delta_{r^2_D}(\{D\})=\varepsilon$. 

Moreover, for all other $\hat{D} \in \hat{\D}$,  $\delta_{r^2_D}(\{\hat{D}\})=0$ by the same reaon as the previous claim.  Since  $A^*, \tilde{X} \cup D\not \in \hat{\D}$, we have $\delta_{r^2_D}(\{\hat{D}\})= 0$ for all $\hat{D} \in \hat{\D} \setminus D$. This completes the proof of the claim.
\end{myproof}

\begin{claim}
 Fix $\varepsilon \in (0,1]$. There exists a flow $\hat{r}$ such that (i) $-1/(2|\C^*|)=\delta_{\hat{r}}(\C^*) \le \delta_{\hat{r}}(\C)$ for any essential test collection $\C$; (ii) If $\C \not\supseteq \C^*$ then $\delta(\C)\ge 0$.
\end{claim}

\begin{proof}
Define $r^3 \equiv \sum_{D \in \C^*}\frac{1}{|\C^*|} r^1(D)$.Then,  $\delta_{r^3}(\{D\})=-\frac{\varepsilon}{|\C^*|}$ for each $D \in \C^*$; moreover, for all  $\hat{D} \in \hat{\D} \setminus \C^* $,  $\delta_{r^3}(\{\hat{D}\})=0$.  Define $r^4\equiv \frac{1}{|\C^*|} r^{\Pi_1}+\frac{|\C^*| - 1}{|\C^*|}r^2(A^*\cup X^*)$. Then, $\delta_{r^4}(\{A^* \cup X^*\})=\frac{|\C^*| - 1}{|\C^*|}\varepsilon$; moreover, for all  $\hat{D} \in \hat{\D}\setminus \C^*$,  $\delta_{r^4}(\{\hat{D}\})=0$.\footnote{Note that $A^*\cup X^* \in \C^*$.} Define $\hat{r}=1/2 r^3+1/2 r^4$. In $\hat{r}$, we have $\delta_{\hat{r}}(\C^*) = \sum_{D \in \C^*} \delta_{\hat{r}}(\{D\})=\frac{1}{2}\big( -1+\frac{|\C^*| - 1}{|\C^*|} \big)\varepsilon= -\frac{1}{2|\C^*|}\varepsilon$. 

\textbf{Step 1:} $\delta_{\hat{r}}(\C^*) \le \delta_{\hat{r}}(\C)$ for any essential test collection $\C$. 

\begin{myproof}
Fix any essential test collection $\C$. We consider the following two cases. 

{\bf Case} 1: There exists $D \in \C$ such that $D \setminus X^* \neq A^*$ . (In fact, in this case, by the definition of essential test collection, $D \setminus X^* \neq A^*$ for all $D \in \C$.) Then, $\C \cap \C^* =\emptyset$. Since $\hat{D} \in \hat{\D}\setminus \C^*$,  $\delta_{\hat{r}}(\{\hat{D}\})=0$, we have $\delta_{\hat{r}}(\C) = 0 \ge \delta (\C^*)$. 

{\bf Case} 2: $D \setminus X^* =A^*$ for all $D \in \C$. Since $\C$ is complete, $\C$ contains $A^* \cup X^*$. Since $\C^*$ contains all $D \in \hat{\D}$ such that $\delta_{\hat{r}}(\{D\}) < 0$ it is clear that $\delta_{\hat{r}}(\C) \ge \delta_{\hat{r}}(\C^*)$.
\end{myproof}

\textbf{Step 2:} If $\C \not\supseteq \C^*$ then $\delta(\C)\ge 0$.

\begin{myproof}
Suppose that $\C \not\supseteq \C^*$. Then there exists $D^* \in \C^*$ such that $D^* \notin \C$. By Step 1,  $\delta_{\hat{r}}(\C \cup \{D^*\}) \ge \delta(\C^*)= -\frac{1}{2|\C^*|}\varepsilon$. Also by definition of $\hat{r}$,  $\delta_{\hat{r}}(\{D^*\}) = -\frac{1}{2|\C^*|}\varepsilon$ so $\delta_{\hat{r}}(\C)=\delta_{\hat{r}}(\C \cup \{D^*\})-\delta_{\hat{r}}(\{D^*\})  \ge 0$. 
\end{myproof}
\vspace{-0.2cm}
\end{proof}
\vspace{-0.2cm}

We finally prove the statement of the lemma:

\begin{claim}
There exists a flow $r^*$ from $\emptyset$ to $X$ such that $\delta_{r^*}(\C^*)<0$ and $\delta_{r^*}(\C)\ge 0$ for any other essential test collection $\C\neq \C^*$.
\end{claim}

\begin{myproof}
For each essential test collection $\C$ such that $\C^* \not\supseteq \C$, choose $D_{\C}  \in \C \setminus \C^*$.  Let $\F$ be the collection of such $D_{\C}$. (Since the number of test collections is finite, $\F$ is a finite collection.) Define $r^*\equiv \alpha \hat{r}+(1-\alpha) \sum_{D_{\C} \in {\cal F}}\frac{1}{|\F|} r^2_{D_{\C}}$. Then for any essential test collection $\C$, $\delta_{r^*}(\C)=\alpha \delta_{\hat{r}}(\C)+(1-\alpha) \sum_{D_{\C} \in {\cal F}}\frac{1}{|\F|} \delta_{r^2_{D_{\C}}}(\C)$. Since $\delta_{r^2_{D_{\C}}}(D_{\C})>0$, there exists $\alpha$ small enough such that for any essential test collection $\C$ such that $\C^* \not\supseteq \C$, we have $\delta_{r^*}(\C)\ge 0$.

Note that $\delta_{r^*}(\C^*)= -\frac{\alpha}{2|\C^*|}\varepsilon$. Note also that $r^2_{D_{\C}}$ does not decrease values of $\delta$ for any test collection.  Thus by statement (ii) of the previous claim, we have if $\C \not\supseteq \C^*$ then $\delta_{r^*}(\C)\ge 0$. It follows that  $\delta_{r^*}(\C^*)<0$ and $\delta_{r^*}(\C)\ge 0$ for any other essential test collection $\C\neq \C^*$.
\end{myproof}

\subsubsection{Proof of statement (ii) in Lemma \ref{cor:essential-is-minimal2}}\label{section:statementB2}

Choose any arc $(D \setminus x,D)$ with $x \in \tilde{X}$, $D \setminus x \neq \emptyset$, $D \neq X$. Let $\hat{\D}\equiv \{D \mid \text{there exists an essential test collection }\C \text{ such that } D \in \C \}$. 
 We will consider two cases:

\textbf{Case 1:} $D^c \cap \tilde{X}\neq \emptyset$. Let $\Pi_1$ be a $\emptyset$ to $X$ dipath which avoids any nodes in $\hat{\D}$. (Such a dipath exists by Claim A.3 in Section \ref{sec:ap:essential}.) Let $\Pi_2$ be a dipath from $\emptyset$ to $D$ which passes through $D \cap \tilde{X}$. Let $\Pi_3$ be a $D \setminus x$ to $X$ dipath which passes through $D  \cup \tilde{X}$ but not $D$. Such a dipath exists because $D^c \cap \tilde{X}\neq \emptyset$ implies that there exists an observable alternative $y \in D^c \cap \tilde{X}$ and there exists an arc $(D \setminus x, D \cup y \setminus x)$. Fix $\varepsilon>0$ and define $r^*\equiv(1 - \varepsilon)r^{\Pi_1}+ \varepsilon r^{\Pi_2} - \varepsilon r^{(D \setminus x,D)} + \varepsilon r^{\Pi_3}$. By definition, $r^*(D\setminus x, D)<0$ and $r^*(E\setminus y, E) \ge 0$ for any $(E,y)$ such that $y \in \tilde{X}$ and $(D,x)\neq (E,y)$. Moreover, for any essential test collection $\C$, $\delta_{r^*}(\C)\ge 0$.  To see this notice that the flow $r^{\Pi_1}$ does not change any value of $\delta_r (\C)$ for any essential test collection. By the definition of $r^*$, we have $\delta_{r^*}(\{D \setminus x\})=0$ and  $\delta_{r^*}(\{D \})\ge 0$.\footnote{$\delta_{r^*}(\{D \})$ is either $0$ or $\varepsilon$.} For all other nodes $E$, $\delta_{r^*}(\{E \})=0$.  Thus, we have $\delta_{r^*}(\C)\ge 0$ for any essential test collection $\C$.

\begin{figure}[h]

\begin{center}
  \begin{tikzpicture}[scale=.4, transform shape]
      \tikzset{vertex/.style = {shape=circle,draw,thick,minimum size=5em}}
    \tikzset{-to/.style = {thick,arrows={-to}}}

    \node[vertex] (X) at (-2,12.5) {\large $X$};
    \node[vertex] (Xminus) at (-2,8) {$D \cup \tilde{X}$};
    \node[vertex] (A) at (0,3) {\large $D \setminus x$};
    \node[vertex] (D) at (2,6) {\large $D$};
        \node[vertex] (DminusA) at (4,3) {\large $D \cap \tilde{X}$};
    \node[vertex] (emptyset) at (4,-1) {\Large $\emptyset$};

    \draw[-to] (emptyset) -- (DminusA) node[midway,right] {\large $+\varepsilon$};
    \draw[-to,dotted] (DminusA) -- (D) node[midway,right] {\large $+\varepsilon$};
    \draw[-to] (A) -- (D) node[midway, below right] {\large $-\varepsilon$};
    \draw[-to] (A) -- (Xminus) node[midway,right] {\large $+\varepsilon$};
    \draw[-to,dotted] (Xminus) -- (X) node[midway,right] {\large $+\varepsilon$};
  \end{tikzpicture}
\quad\quad\quad\quad
  \begin{tikzpicture}[scale=.4, transform shape]
      \tikzset{vertex/.style = {shape=circle,draw,thick,minimum size=5em}}
    \tikzset{-to/.style = {thick,arrows={-to}}}

    \node[vertex] (X) at (-2,12.5) {\large $X$};
    \node[vertex] (Xminus) at (-2,8) {\small $(D \setminus x) \cup X^*$};
    \node[vertex] (A) at (0,3) {\large $D \setminus x$};
    \node[vertex] (D) at (2,6) {\large $D$};
        \node[vertex] (DminusA) at (4,3) {\large $D \cap \tilde{X}$};
    \node[vertex] (emptyset) at (4,-1) {\Large $\emptyset$};
  
    \draw[-to] (emptyset) -- (DminusA) node[midway,right] {\large $+\varepsilon$};
    \draw[-to,dotted] (DminusA) -- (D) node[midway,right] {\large $+\varepsilon$};
    \draw[-to] (A) -- (D) node[midway, below right] {\large $-\varepsilon$};
    \draw[-to,dotted] (A) -- (Xminus) node[midway,right] {\large $+\varepsilon$};
    \draw[-to] (Xminus) -- (X) node[midway,right] {\large $+\varepsilon$};
    \end{tikzpicture}
    
\caption{Flow $\varepsilon r^{\Pi_2} - \varepsilon r^{(D \setminus x,D)} + \varepsilon r^{\Pi_3}$ in Case 1 (left); Flow $\varepsilon r^{\Pi_2} - \varepsilon r^{(D \setminus x,D)} + \varepsilon r^{\Pi_4}$  in Case 2  (right) of section \ref{section:statementB2}}
\end{center}
\end{figure}

\textbf{
Case 2:} $D^c \cap \tilde{X} = \emptyset$. This means that $D$ contains all elements in $X^*$. Notice $(D \setminus x) \cap X^* \neq X^*$ since otherwise $D = X$. So let $\Pi_4$ be a $D \setminus x$ to $X$ dipath that passes through $(D \setminus x) \cup X^*$. (Note that the last arc  is observable arc $(X\setminus x, X)$, where $x \in \tilde{X}$.) Define $r^*\equiv (1 - \varepsilon)r^{\Pi_1}+ \varepsilon r^{\Pi_2} - \varepsilon r^{(D \setminus x,D)} + \varepsilon r^{\Pi_4}$. By definition, $r^*(D\setminus x, D)<0$ and $r^*(E\setminus y, E) \ge 0$ for any $(E,y)$ such that $y \in \tilde{X}$ and $(D,x)\neq (E,y)$. Moreover, for any essential test collection $\C$, $\delta_{r^*}(\C)\ge 0$.  To see this notice (i)
$\delta_{r^*}(\{D \setminus x\}) = -\varepsilon$ but $\delta_{r^*}(\{(D \setminus x )\cup X^*\}) = \varepsilon$; (ii)   any test collection $\C$ containing $D \setminus x$  contains $(D \setminus x) \cup X^*$. (i) and (ii) implies that the negative value of $\delta_{r^*}(\{D \setminus x\})$ is cancelled by the positive value of $\delta_{r^*}((D \setminus x) \cup X^*)$. For all other nodes $E$, $\delta_{r^*}(\{E \})=0$. Thus, we have $\delta_{r^*}(\C)\ge 0$ for any essential test collection $\C$.

\section{Supplemental Contents}

\subsection{Meaning of condition (ii) in Theorem \ref{thm:characterization}}\label{section:mean_new}
Suppose that an incomplete dataset $\rho$ is RU-rationalizable by $\mu$. To understand the meaning of condition (ii) in Theorem \ref{thm:characterization}, fix  a test collection $\C=\{A \cup E \mid E \in \E\}$ and define $\hat{\mathcal{C}} = \{C \in \C \mid C \setminus x^*\not \in \C \text{ for some }x^*\in X^* \}$.
\footnote{In a Boolean lattice that we will explain later, the subcollection $\hat{\C}$ can be interpreted as bottom parts of $\C$.} 
Also, for each $C \in \hat{\C}$, let $U_C= \{x^* \in X^* \mid C \setminus x^* \notin \C\}$.\footnote{In a Boolean lattice, $U_C$ is the set of unobservable alternatives $x^*$ where $(C \setminus x^*, C)$ is an arc from outside of $\C$ going into $\C$.} We can show that the left hand side of (\ref{eq:th1}) is 
\begin{align}\label{eq:mean}
\sum_{C \in \hat{\C}} \mu \left( \succ \in \L \mid C^c \succ C \text{ and } \max_C \succ \in U_C\right),
\end{align}
where $\max_C \succ$ denotes the best element in $C$ with respect to $\succ$.
In particular, when the test collection $\C$ is a singleton of the form $\{D\}$ where $X^* \subseteq D$,  (\ref{eq:mean}) simplifies to $\mu(\succ \in \L| \exists x^* \in X^*\ s.t. \  D^c \succ x^* \succ D \setminus x^*)$.

Equation (\ref{eq:mean}) can be derived as follows. By the inflow-outflow equality, the total flow out of $\C$ minus the total flow into $\C$ is zero. Note that since $\C$ is an essential test collection, there are no unobservable flows out of $\C$. Thus the left hand side of (\ref{eq:th1}) is only missing the unobservable flows into $\C$. That is,

\begin{align*}
\Bigg(\sum_{(D,x): D \in {\C},  D\cup x \not\in {\C}}  K(\rho, D \cup x, x)&- \sum_{(F,y): F \not\in {\C}, F\cup y \in {\C}, y \in \tilde{X}}  K(\rho, F \cup y, y)\Bigg)\\
&- \sum_{(F,z): F \not\in {\C}, F\cup z \in {\C}, z \in X^*}  K(\rho, F \cup z, z) = 0. 
\end{align*}

It follows that the left hand side of (\ref{eq:th1}) equals $$\sum_{(F,z): F \not\in {\C}, F\cup z \in {\C}, z \in X^*}  K(\rho, F \cup z, z).$$ Applying equation (\ref{eq:meaning_BM}) yields (\ref{eq:mean}).

\subsection{Random utility polytope}\label{sec:online_polytope}

In this section, we provide a geometric intuition for the set of RU rational stochastic choice functions. Let $\M$ be the set of pairs of $(D,x)$ such that $\rho(D,x)$ is observable.

\begin{remark}
For each ranking $\succ \in \L$ and $(D,x) \in \M$, define 
\begin{eqnarray}\label{df:rho_u}
\rho^{\succ}(D,x)=
\left\{
\begin{array}{llll}
1 &\text{ if }x \succ y \text{ for all }y \in D \setminus x;\\
0 &\text{ otherwise}.
\end{array}
\right.
\end{eqnarray}
The stochastic choice function $\rho^{\succ}$ gives probability one to the best alternative $x$ in a choice set $D$ according to the ranking $\succ$.  
The set of RU-rationalizable datasets is a polytope, that is, $\co \{\rho^{\succ} \mid \succ \in \L\}$, where $\co$ denotes the convex hull. 
\end{remark}
\vspace{-0.3cm}

%\textcolor{blue}{\fbox{def of co.}}

\begin{figure}[h]
\begin{center}
\begin{tikzpicture}[mystyle/.style={draw,shape=circle,fill=black, inner sep=0pt, minimum size=4pt,  label={[anchor=center, label distance=2mm](90+360/\ngon*(#1-1)):#1}}]
\def\ngon{6}
\node[draw, regular polygon,regular polygon sides=\ngon,minimum size=3cm] (p) {};
\foreach\x in {1,...,\ngon}{
    \node[mystyle=] (p\x) at (p.corner \x){};
}

\node[anchor=west]  at (p.corner 1) {$\rho^{\pi_1}$};
\node[anchor=east]  at (p.corner 2) {$\rho^{\pi_2}$};
\node[anchor=east]  at (p.corner 3) {$\rho^{\pi_3}$};
\node[anchor=east]  at (p.corner 4) {$\rho^{\pi_4}$};
\node[anchor=west]  at (p.corner 5) {$\rho^{\pi_5}$};
\node[anchor=west]  at (p.corner 6) {$\rho^{\pi_6}$};

\draw[fill=yellow!20!white,opacity=0.5] (p.corner 1)-- (p.corner 2)--(p.corner 3)--(p.corner 4)--(p.corner 5)--(p.corner 6)--(p.corner 1);

 	\draw [thick, blue] (p.corner 1) -- (p.corner 6);
 	\draw [thick, blue] (0,2.4)--(p.corner 1) ;
    \draw [thick, blue] (p.corner 6) -- (2,-0.8);

 	\draw [thick, red] (-0.1,2)--(p.corner 1);
 	\draw [thick, red] (p.corner 1) --(2, 0.55);
  
 	\draw [thick, red] (-0.1,1.5)--(p.corner 1);
 	\draw [thick, red] (p.corner 1) --(2, 1.05);

 	\draw [thick, red] (-0.1,1.75)--(p.corner 1);
 	\draw [thick, red] (p.corner 1) --(2, 0.8);

\end{tikzpicture}
\end{center}
\vspace{-0.3cm}
\caption{Random utility polytope}\label{fig:lem2} 
\end{figure}
\vspace{-0.2cm}

The hexagons in Figure \ref{fig:lem2} illustrates the polytope.\footnote{Although the geometric intuition is useful, it is important to notice that the figure oversimplifies the reality since the number (i.e., $|X|!$) of vertices and the dimension of a random utility function can be very large. To see why the dimension of a random utility function can be very large,  notice that it assigns a number for each pair of $(D,x) \in \M$.}  The inequality conditions in Theorem \ref{thm:characterization} consist of the facet defining   inequalities of the polytope, which correspond to the blue hyperplanes. On the other hand, as we will explain in the next section,  McFadden and Richter's approach would contain non-facet defining  inequalities, which correspond to the red hyperplanes.

\subsection{Generalization  \cite{mcfad91}}\label{sec:mc}

\cite{mcfad91} provide a characterization of the random utility model under menu unobservability. 
Unlike the characterization of \cite{falmagne1978representation}, the characterization of \cite{mcfad91} holds when the frequencies of all alternatives are observed on an arbitrary set of observable menus. However, their characterization fails when some alternatives are unobservable.
Their characterization is obtained as the dual of the existence of a rationalizing random utility model through the Farkas's lemma. This dual condition, unlike our main theorem, involves an infinite number of inequalities. In this section we first show how the characterization in \cite{mcfad91} is insufficient in our setup. We then generalize the characterization of \cite{mcfad91} to arbitrary domains, including those with unobservable alternatives.

Let $\M$ be the set of pairs $(D,x)$ such that $x \in D \in 2^X$ and $\rho(D,x)$ is well defined (i.e., $\rho(D,x)$ is observable to the analyst).  In the following, we call $\M$ the set of observable pairs.  We write the set of datasets as $\P(\M)$. In \cite{mcfad91}, it is assumed that all alternatives are observable. That is, if $(D,x) \in \M$ then $(D,y)\in \M$ for all $y \in D$. Equivalently, $\M\coloneqq  \{(D,x)\in \D \times X \mid x \in D \}$ for some $\D \subseteq 2^X \setminus \emptyset$.\footnote{The formal definition of $\P(\M)$ can be provided  as follows: 
$\P(\M)=\{\rho \in \Re_+^{\M} \mid\text{  For all } D \in \D \text{ (i) if }(D,x) \in \M\text{ for any } x \in D, \text{ then }  \sum_{x \in D}\rho(D,x)=1; \text{ (ii) if }(D,x) \not \in \M \text{ for some }x \in D, \text{ then } \sum_{x \in D}\rho(D,x)\le 1 \}$.} In the main body of the paper, we assumed $\M\equiv  \{(D,x)\in \D \times \tilde{X} \mid x \in D \}$.\footnote{In \cite{falmagne1978representation}, $\M$ is assumed to be $\{(D,x)  \mid x \in D\in 2^X\}$.} In the following section we will consider arbitrary $\M \subseteq \{(D,x) \mid x \in D \in 2^X\}$ which generalizes both the setup of \cite{mcfad91} and our setup.

\cite{mcfad91} characterize random utility for the no unobservable alternative case with the following axiom sometimes called  the Axiom of Revealed Stochastic Preference (ARSP).

\begin{definition}
    A stochastic choice function satisfies the Mcfadden-Richter axiom if for any sequence finite $(D_i,x_i)_{i=1}^k \in \M^k$ for all $k$, \[
\max_{\succ\in \L(X)} \sum_i \rho^\succ(D_i,x_i) \ge \sum_i \rho(D_i,x_i).
\]
\end{definition}
However, this is not sufficient to characterize random utility in our setup. Consider the following example:

\begin{remark}
Let $X = \{a,b,c,d\}$ and $X^* = \{c,d\}$ and $\M = \{(X,a ), (X \setminus b,a)\}$. Let $\rho$ be such that $\rho(X,a ) = 0$ and $\rho(X \setminus b,a ) = 1$. This is obviously not RU-rational as it violates monotonicity (in particular condition (i) is violated). However, fix $\succ_0 \in \L(X)$ be such that $a \succ_0 b \succ_0 c \succ_0 d$. That is, $\rho^{\succ_0}(X,a ) = \rho^{\succ_0}(X \setminus b,a ) = 1$. Then since $ \rho^{\succ_0}(D,x) \ge  \rho(D,x) $ for all $(D,x)\in \M$,  we observe that 
\[
\sum_i \rho^{\succ_0}(D_i,x_i) \ge \sum_i \rho(D_i,x_i).
\]
and thus since $\max_{\succ\in \L(X)} \sum_i \rho^\succ(D_i,x_i) \ge  \sum_i \rho^{\succ_0}(D_i,x_i) $, $\rho$ satisfies the McFadden-Richter Axiom but not RU rationalizability. 
\end{remark}

We provide the following characterization that works for arbitrary $\M \subseteq \{(D,x)\in 2^X \times X\mid x \in D \}$.

\begin{definition}\label{def:gen_mcfad}
    A stochastic choice function $\rho$ satisfies the Generalized McFadden-Richter axiom if for any finite sequences $(D_i^+,x_i^+)_{i=1}^k \in \M^k$ and $(D_j^-,x_j^-)_{i=1}^l \in \M^l$ for all $k$ and $l$,\[
\max_{\succ\in \L(X)} \sum_i \rho^\succ(D_i^+,x_i^+)- \sum_j \rho^\succ(D_j^-,x_j^-) \ge \sum_i \rho(D_i^+,x_i^+)-\sum_j \rho(D_j^-,x_j^-).
\]
\end{definition}

\begin{theorem} Suppose $\M \subseteq \{(D,x)\in 2^X \times X\mid x \in D \}$. A stochastic choice function $\rho$ defined on $\M$ is RU-rationalizable if and only if it satisfies the Generalized McFadden-Richter axiom.\footnote{Another way to write a McFadden-Richter type axiom for our setup is to only consider positive sequences but allow terms of the form $\rho(D,a_D) = 1- \sum_{x \in D \cap \tilde{X}}\rho(D,x)$. The interpretation is that $\rho(D,a_D)$ is the total frequency of unobservable alternatives in $D$. This approach also leads to adding the negative terms. However, this approach leads to a weaker characterization of RUM as it ignoring possible restrictions between $\rho(D,a_D)$ and $\rho(E,a_E)$ for different choice sets.}

\end{theorem}

\begin{proof}
First suppose that there exists sequences $(D_i^+,x_i^+)_{i=1}^k \in \M^k$ and $(D_j^-,x_j^-)_{i=1}^l \in \M^l$ for all $k$ and $l$ such that
$\max_{\succ\in \L(X)} \sum_i \rho^\succ(D_i^+,x_i^+)- \sum_j \rho^\succ(D_j^-,x_j^-) < \sum_i \rho(D_i^+,x_i^+)-\sum_j \rho(D_j^-,x_j^-)$. Then for any $\rho^\mu \in \mathcal{P}_r = \operatorname{co}\left( \{\rho^{\succ} \mid \succ \in \mathcal{L}\}\right)$ it follows that $\sum_i \rho^\mu(D_i^+,x_i^+)- \sum_j \rho^\mu(D_j^-,x_j^-) < \sum_i \rho(D_i^+,x_i^+)-\sum_j \rho(D_j^-,x_j^-)$. We conclude that $\rho \notin \mathcal{P}_r$.

To show the other direction, suppose that $\rho \notin \mathcal{P}_r$. Since $\mathcal{P}_r$ is compact and convex then by the separating hyperplane theorem there exists $\nu: \M \rightarrow \mathbb{R}$  such that $\nu \cdot \rho^\mu <  \nu \cdot \rho$ for all $\rho^\mu \in \mathcal{P}_r$. In particular, 
\begin{equation}\label{nu:separating}
\max_{\succ\in \L(X)} \sum_{(D,x) \in \M} \nu(D,x)\rho^\succ(D,x) < \sum_{(D,x) \in \M} \nu(D,x)\rho(D,x).
\end{equation}
Now by density of the rationals and since the maximum is over a finite set, $\nu$ can be taken to be rational valued. Then by multiplying by a large positive integer, $\nu$ can also be taken to be integer valued and still satisfy the inequality. Finally, we obtain the result by letting $(D,x)$ appear $\nu(D,x)$ times in $(D_i^+,x_i^+)_{i=1}^k \in \M^k$ if it is positive and $\nu(D,x)$ times in $(D_i^-,x_i^-)_{j=1}^l \in \M^l$ if it is negative.
\end{proof}
This characterization is similar to the classical Mcfadden-Richter axiom, but it allows negative signs in the sum.  If the dataset satisfies the following condition, then we can focus on positive sequences: For each $D \in \D$, we can observe $\rho(D,x)$ for all $x \in D$. Thus, 
\begin{equation}\label{eq:one_mr}
\sum_{x \in D \text{ s.t.} (D,x) \in \M}\rho(D,x)=1.
\end{equation}

\begin{theorem}
Let $\M \subseteq \{(D,x)\in 2^X \times X\mid x \in D \}$. Suppose that if  $(D,x) \in \M$, then $(D,y) \in \M$ for all $y \in D$. A stochastic choice function $\rho$ defined on $\M$ is RU-rationalizable if and only if it satisfies the McFadden-Richter axiom.
\end{theorem}

The necessity of the proof does not change. In the sufficiency part of proof above, we showed the existence of $\nu$ satisfying (\ref{nu:separating}).
Let $s= -\min_{(D,x) \in \M} \nu(D,x)$. Then define $v^*$ by $v^*(D,x)= v(D,x)+ s$ for all $(D,x) \in \M$. Note that  $\nu^*$ is a nonnegative vector and for any stochastic choice function for each $D \in \D$, $
\sum_{x \in D\ s.t.\ (D,x) \in \M}\nu^*(D,x)\rho(D,x)-\Big(\sum_{x \in D\ s.t.\ (D,x) \in \M}\nu(D,x)\rho(D,x) \Big)= s$, where the last equality holds by (\ref{eq:one_mr}). Thus, (\ref{nu:separating}) holds with $\nu^*$ in the place of $\nu$. The rest of the proof is the same.

One drawback that this characterization shares with \cite{mcfad91} but not the characterization in Theorem \ref{thm:characterization} is that it has an infinite number of inquealities to test, some of them being redundant. In the following we show that these redundancies are inherent to the approach.

\subsection{Redundancy in the \cite{mcfad91} Approach}
The conditions of \cite{mcfad91} and our generalization involve an infinite number of sequences and some of the conditions are redundant. 
%These features result from the fact that their characterization is obtained as a dual of the existence of rationalizing random utility model through the Farkas's lemma; and the dual condition, in general, involves an infinite number of inequalities. 
In this section, we further clarify  the relationship between  our approach and the approach taken by \cite{mcfad91}. The main message is that the approach by \cite{mcfad91} contain redundancy in an essential way, unlike the BM polynomials. 
%Thus, one can improve the results obtained by \cite{mcfad91} by removing some redundancy; however it would be difficult to remove all redundancy from the approach.
\footnote{In other words, if one removes all redundancy from the results by \cite{mcfad91}, such results should reduce to \cite{falmagne1978representation} for the case of complete datasets and our results for the case of incomplete datasets (in our sense).}    

Notice in Definition \ref{def:gen_mcfad}, the same $(D,x)\in \M$ may appear arbitrarily many times in the sequences $(D_i^+,x_i^+)_{i=1}^k \in \M^k$ and $(D_j^-,x_j^-)_{i=1}^l \in \M^l$ . Thus, the number of  sequences to be tested in the Generalized McFadden-Richter Axiom is infinite, although there are finitely many pairs $(D,x)\in \M$. \cite{mcfad91} discuss the difficulty of  providing an upper bound on the number of allowable repetitions needed for their axiom to fully characterize RUM. They prove that sequences containing repetitions must be tested in general. In the following section, we show how Theorem \ref{thm:characterization} can provide an upper bound on the number of required repetitions. We then show that while limiting the number of repetitions does make the number of inequalities to test finite, the number of inequalities to test is much larger than our independent conditions obtained in Theorem \ref{thm:characterization} and therefore contains a large amount of redundancy.

We first show the following remark

\begin{remark} 
The inequality conditions (i) of Theorem \ref{thm:characterization} can be written as  the Generalized McFadden-Richter Axiom with {\it no repetitions}. 
\end{remark}

For each $(E,y)$ such that  $y \in E$, define $K_{(E,y)} \in \Re^{\M}$ by 
\begin{eqnarray*}\label{eq1}
K_{(E, y)}(D,x)=
\left\{
\begin{array}{llll}
-1  &\text{ if }y=x \text{ and } D\supseteq E \text{ and }|D\setminus E| \text{ is even},\\
+1 &\text{ if }y=x \text{ and } D\supseteq E \text{ and }|D\setminus E| \text{ is odd},\\
0  &\text{ otherwise}.
\end{array}
\right.
\end{eqnarray*}
 For any $\rho \in \P$, we have $K_{(E,y)} \cdot \rho = -K(\rho, E,y)$. For each  $(E,y)$, define sequences $(D_i^+,x_i^+)_{i=1}^k \in \M^k$ and $(D_j^-,x_j^-)_{i=1}^l \in \M^l$ so that each $(D, x)$ appears exactly once in $(D_i^+,x_i^+)_{i=1}^k$ if $K_{(E,y)} (D,x) = 1$ and exactly once in $(D_j^-,x_j^-)_{i=1}^l \in \M^l$ if $K_{(E,y)} (D,x) = -1$. Thus $\sum_i \rho(D_i^+,x_i^+)- \sum_j \rho(D_j^-,x_j^-) = -K(\rho,E,y)$ for all $\rho$.
 
 Notice then that \[\max_{\succ\in \L(X)} \sum_i \rho^\succ(D_i^+,x_i^+)- \sum_j \rho^\succ(D_j^-,x_j^-)  = \max_{\succ\in \L(X)} \sum_{(D,x) \in \M} -K(\rho^\succ,E,y) = 0,
 \]
 the Generalized McFadden-Richter Axiom for this sequence is equivalent to $K(\rho,E,y) \ge 0$. In particular, in the case of complete data, it is sufficient to only consider sequences without repetition.

We can do something similar with condition (ii) in Theorem \ref{thm:characterization}. 

\begin{remark} 
The inequality conditions (ii) of Theorem \ref{thm:characterization} are also implied by the Generalized McFadden-Richter Axiom. 
\end{remark}

For each essential test collection $\C$ define $\nu_\C (D,x)$ as the coefficient on $\rho(D,x)$ in the polynomial below: 
\begin{align}\label{eq:delta_oo}
           \delta'_{\rho}(\C)\equiv \sum_{\substack{(D,x): D \in {\C},  D\cup x \not\in {\C},\\ (D\cup x,x)  \in \M}}  
        K(\rho, D \cup x, x)
        - 
        \sum_{\substack{(E,y): E \not\in {\C}, E\cup y \in {\C},\\ (E\cup y, y)  \in \M}}  
        K(\rho, E \cup y, y).
\end{align}
Note that (\ref{eq:delta_oo}) is equivalent to  the left hand-side of condition (ii) of Theorem \ref{thm:characterization}. The value of (\ref{eq:delta_oo}) coincided with $\delta_\rho(\C)$ defined in (\ref{eq:delta_o}) when $\C$ is an essential test collection.\footnote{When $\C$ is an essential test collection, we  have $X \not \in \C$ and $\emptyset \not \in \C$.} 

Define sequences $(D_i^+,x_i^+)_{i=1}^k \in \M^k$ and $(D_j^-,x_j^-)_{i=1}^l \in \M^l$ so that each $(D, x)$ appears exactly $-\nu_\C (D,x)$ in $(D_i^+,x_i^+)_{i=1}^k$ if $\nu_\C (D,x) <0$ and exactly $\nu_\C (D,x)$ many times in $(D_j^-,x_j^-)_{i=1}^l \in \M^l$ if $\nu_\C (D,x)>0$. Then $\sum_i \rho(D_i^+,x_i^+)- \sum_j \rho(D_j^-,x_j^-) = -\delta'_\rho(\C)$ for all $\rho$. As shown in the proof of Theorem \ref{thm:characterization}, $\delta'_{\rho^\succ}(\C) \ge 0$ for all $\succ$; and furthermore there exists $\succ$ such that $\delta'_{\rho^\succ}(\C) = 0$ by the fact that $\delta'_{\rho}(\C) = 0$ is a facet defining inequality and the vertices of  the polytope consist of $\rho^{\succ}$s. Thus
\[\max_{\succ\in \L(X)} \sum_i \rho^\succ(D_i^+,x_i^+)- \sum_j \rho^\succ(D_j^-,x_j^-)  = \max_{\succ\in \L(X)} \sum_{(D,x) \in \M} -\delta'_{\rho^\succ}(\C) = 0\]
and therefore the Generalized McFadden-Richter Axiom for this sequence is equivalent to condition (ii) for the essential test collection $\C$.

\begin{remark}
We have thus shown our conditions can be written as the Generalized McFadden Richter Axiom (by selecting sequences in a very particular way); however the converse is {\it not} true. While our conditions are non-redundant, all of the sequences in the Generalized McFadden Richter Axiom apart from those corresponding to our conditions are redundant.
\end{remark}

See Figure \ref{fig:lem2} for the illustration. Our conditions contain only the blue facet defining hyperplanes, while the Generalized McFadden Richter Axiom contain all non-facet defining red hyperplanes.

To get an idea of how much redundancy there is in the McFadden-Richter axiom (as well as its generalization), consider the following counting argument. Suppose $\D = 2^X \setminus \emptyset$. For finite sequences in $\M$ with no repetitions, each element of $\M$ either appears as a positive, negative, or does not appear in the sequence. Thus (up to reordering) there are $3^{|\M|}$ inequalities without repetitions. This is at least $2^{2^{|X|} -1}$ as the lower bound of $|\M|$ is $2^{|X|} -1$. 
Note that this is a lower bound for the number of sequences {\it without } repetitions. Since repetitions are necessary when the datasets are incomplete, this lower bound is very small one.

We now obtain an upper bound for the number of inequalities in condition (ii) of Theorem \ref{thm:characterization}. The number of inequalities in condition (i) of Theorem \ref{thm:characterization} is $|\M|$, which is bounded by $2^{|X|}\times |\tilde{X}|$. To get an upper bound for the number of inequalities in condition (ii) of Theorem \ref{thm:characterization}, notice that the number of essential test collections is less than $2^{|\tilde{X}|}2^{2^{|X^*|}} =2^{2^{|X^*|}+|\tilde{X}|}$ since there are less than $2^{|\tilde{X}|}$ options for $A$ and less than $2^{2^{|X^*|}}$ options for $\E$. Thus, the upper bound for the number of inequalities in condition (ii) is $2^{2^{|X^*|}+|\tilde{X}|}+2^{|X|}|\tilde{X}|$. We have 
\[
2^{2^{|X^*|}+|\tilde{X}|}+2^{|X|}|\tilde{X}|= 2^{\,2^{|X^{*}|}+|\tilde X|}\Bigl(1+|\tilde X|\;2^{\,|X^{*}|-2^{|X^{*}|}}\Bigr)
\]
as $|X|=|\tilde{X}|+|X^*|$. For $|X^{*}|\ge 2$ we have $
|X^{*}|-2^{|X^{*}|}\le -2$, thus $2^{\,|X^{*}|-2^{|X^{*}|}}\le \frac14$.
Hence
\[
1+|\tilde X|\,2^{\,|X^{*}|-2^{|X^{*}|}}
   \;\le\;
   1+\frac{|\tilde X|}{4}
   \;\le\;
   2^{\,|\tilde X|-1},
\]
where the last inequality holds because $|\tilde X|\ge 2$. Thus an upper bound is 
\[
2^{2^{|X^*|}+|\tilde{X}|}+2^{|X|}|\tilde{X}|= 2^{\,2^{|X^{*}|}+|\tilde X|}\Bigl(1+|\tilde X|\;2^{\,|X^{*}|-2^{|X^{*}|}}\Bigr)\le 2^{\,2^{|X^{*}|}+2|\tilde X|-1}
\]

Therefore the ratio is at least
\[
\dfrac{2^{2^{|X|}-1}}{2^{2^{|X^*|}+2|\tilde{X}|-1}} = 2^{2^{|X|}-2^{|X^*|}-2|\tilde{X}|}\ge2^{2^{|X|-2}}
\]
for $|\tilde{X}| \ge 2$ and $|X^*| \ge 2$. Summarizing the above argument, we have the following: 

\begin{remark}
\[
\dfrac{\#(\text{inequalities of Generalized MR axiom without repetitions})}{\#(\text{inequalities of Theorem \ref{thm:characterization}})} \ge  2^{2^{|X|-2}}
\]
for $|\tilde{X}| \ge 2$ and $|X^*| \ge 2$. This illustrates that even when restricting attention to non-repeating sequences, the number of inequalities required for the Generalized MR axiom is substantially larger than the number of essential test collections identified in Theorem~\ref{thm:characterization}.
\end{remark}

The remaining question is how many repetitions are required to be sufficient in characterizing RUM.  By using Theorem \ref{thm:characterization}, we will obtain a lower bound for the number of repetitions. By the previous observation, this is a lower bound on the number of times that any given $\rho(D,x)$ appears in condition (ii) of Theorem \ref{thm:characterization}. We first construct a test collection. Take arbitrary $A = \tilde{X} \setminus b$ for some $b \in \tilde{X}$. Now take $\E \subseteq 2^{X^*}$ that contains all subsets of $X^*$ with at least $|X^*|/2$ elements. Note then that $\{A \cup E | E \in \E\}$ is an essential test collection.

Now take $a \in A$. We wish to count how many time $\rho(X,a)$ appears in the inequality corresponding to the test collection. First, notice $K(\rho,A \cup E,a)$ only appears as an inflow in the inequality. Now, the sign of $\rho(X,a)$ alternates in $K(\rho,A \cup E,a)$ with respect to the size of $E$. Thus it appears (up to a change of sign) $\sum_{E \in \E} (-1)^{|E|} = \sum^{m/2}_{k=0} (-1)^k{m\choose k}$ times. This equals  $(-1)^{m/2} {m-1 \choose m/2}$ where $m = |X^*|$. Now, it is well-known that ${n \choose k} \ge (n/k)^k$ for all $n$ and $k$, Thus,
\[
{|X^*|-1 \choose |X^*|/2} \ge \frac{1}{|X^*|} {|X^*| \choose |X^*|/2} \ge \frac{2^{|X^*|/2}}{|X^*|}.
\]

%Now, if we allow $m$ repetitions then each $(D,x) \in \M$ may appear up to $m$ times as a positive, up to $m$ times as a negative, or not appear in the sequence. Thus we must test $(2m+1)^{|\M|}$ sequences. We conclude that we must test at least $(\frac{2^{|X^*|/2}}{|X^*|})^{2^{|X|} -1}$ sequences, most of which are redundant.

Thus, we have the following:

\begin{remark}
Let $\M=\{(D,x) \in 2^X \times \tilde{X}| x \in D \}$. Then, in order for the Generalized McFadden-Richter axiom to be sufficient for RU-rationalizability, we must test the axiom for sequences containing at least $\frac{2^{|X^*|/2}}{|X^*|}$ repetitions.
\end{remark}

Remark E.10 and E.11 together demonstrate that the number of inequalities in Theorem \ref{thm:characterization} is significantly smaller than the number of inequities required in McFadden-Richter approach. 

\subsection{Simplification of bounds of unobservable choice probabilities when $\D = 2^X\setminus \emptyset$}\label{sec:online_bounds}

In this section, we assume that $\D = 2^X\setminus \emptyset$, we provide further simplification of bounds of unosbservavle choice frequencies.

\begin{corollary}
Let $\D = 2^X\setminus \emptyset$. 
For $(D, x) \not \in \M$, the upper bound is obtained by 
\begin{align} \label{eq:lp-full}
   \overline \rho (D, x)=  \max_{\{r (D \setminus x, D)\}_{(D, x) \not\in \M}}
         \sum_{A^\prime: A \subseteq A^\prime \subseteq \tilde{X}}
     \sum_{E^\prime: E \subseteq E^\prime \subseteq X^*}
     r (A^\prime \cup E^\prime \setminus x, A^\prime \cup E^\prime)
\end{align}
subject to
\begin{align} \label{eq:const-full}
    \sum_{y \in E^\prime} r (A^\prime \cup E^\prime \setminus y, A^\prime \cup E^\prime)
    -
    \sum_{y \in X^* \setminus E^\prime} r (A^\prime \cup E^\prime, A^\prime \cup E^\prime \cup y)
    =
    \delta_\rho (A^\prime \cup E^\prime)
\end{align}
for all $A^\prime \subseteq \tilde{X}$ and $E^\prime \subseteq X^*.$
The lower bound $\underline \rho (D, x) $ solves a similar problem with a min replacing the max.
\end{corollary}
\begin{remark} \mbox{}
\bit
\item The form of (\ref{eq:const-full}) implies that for $A^\prime, A^{\prime\prime} \subseteq \tilde{X},$ $E^\prime, E^{\prime\prime} \subseteq X^*,$ and $y \in E^\prime, z \in E^{\prime \prime},$ if $A^\prime \neq A^{\prime\prime},$ then variables $r (A^\prime \cup E^\prime \setminus y, A^\prime \cup E^\prime)$ and $r (A^{\prime\prime} \cup E^{\prime\prime} \setminus z, A^{\prime\prime} \cup E^{\prime\prime})$ are independent; either of them does not restrict the other via the constraints (\ref{eq:const-full}).
\item This means that each constraint can be considered separately.
\item Therefore, we can optimize the inner sum of (\ref{eq:lp-full}) separately.
% This form of the problem decomposes the large linear program (\ref{eq:lp-full}) into smaller problems because the constraints (\ref{eq:const-full}) for different $A^\prime$ can be considered independently.
That is, the maximum value of the problem is equivalent to the sum of the maximum values of the following problems for all $A^\prime$ such that $A \subseteq A^\prime \subseteq \tilde{X}:$
\begin{align} \label{eq:lp-smaller}
    \max_{\{r (D \setminus x, D)\}_{(D, x) \not \in \M}}
    \sum_{E^\prime: E \subseteq E^\prime \subseteq X^*}
    r (A^\prime \cup E^\prime \setminus x, A^\prime \cup E^\prime)
\end{align}
subject to (\ref{eq:const-full}) for all $E^\prime \subseteq X^*.$
\item A large linear program (\ref{eq:lp-full}) is now decomposed into smaller problems (\ref{eq:lp-smaller}), which improves the computational efficiency, especially when $A$ is large.
\eit
\end{remark}

\subsection{Implications to statistical testing of rationality}\label{sec:online_polytope2}

Theorem \ref{thm:characterization} establishes that following two representations of the set of random utility polytope are equivalent: $
   co.\{\rho^\succ \mid \succ \in \L\}$ and
$$
    \left\{
        \rho \in \Re_+^\M 
        \mid 
        \rho \text{ satisfies the inequalities in (i) and (ii) of Theorem \ref{thm:characterization}}
    \right\}
    .
$$
The first representation characterizes an RU-rationalizable dataset as the convex hull of a finite set of points.
In convex geometry, this is known as a $V$-representation. The second representation expresses the polytope via its facet defining inequalities, commonly referred to as an $H$-representation.

For testing whether a given incomplete dataset is RU-rationalizable, the $H$-representation is generally more practical, as it requires only checking for a single violated inequality to reject rationalizability. 
In contrast, the $V$-representation involves searching the typically high-dimensional space $\Delta (\L)$ for a supporting weight, making it computationally intensive.

A key challenge with the $H$-representation is that a closed form expression remains unknown for arbitrary patterns of missing data, whereas the $V$-representation follows directly from the definition of rationalizability. 
Consequently, empirical studies on testing RU-rationalizability, such as \cite{kitamura2018nonparametric} and \cite{dean2022better}, rely on the $V$-representation, trading computational cost for broader applicability.

Our contribution in Theorem \ref{thm:characterization} is to explicitly enumerate all facet defining inequalities for a specific structure of data incompleteness. 
Once the $H$-representation is obtained, rationality testing reduces to verifying inequality constraints. 
Statistical inference can then proceed using standard techniques for moment inequalities (see \cite{andrews2010inference}, \cite{canay2023user}), though assessing the performance of such methods is beyond the scope of this paper.

\begin{comment}
\subsection{Outside option approach and welfare loss}

Taking the outside option approach can be severely problematic especially when a policy maker implements one of the options in $X$ based on the choice dataset.
Consider a situation where a half of the population has utility $u$ such that $(u_a, u_b, u_c) = (0, 10, -100)$ and the other half has $v$ such that $(v_a, v_b, v_c) = (0, -100, 10).$
This describes a group of people who think the second best option is acceptable and strongly dislike the worst case.
If the policy maker adopts the outside option approach, $a$ would never be implemented, since $\hat \rho (\{a,x_0\},a) = 0.$
However, implementing either of $b$ or $c$ would make a half of the population mad.
In this case, the utilitarian sum is $.5 \cdot 10 + .5 \cdot (-100) = -45,$ which is much worse than the welfare obtained by implementing $a$.
\end{comment}

\newpage

\subsection{Block-Marschak Polynomial Values for the Motivating Example in Section \ref{subsection:ex}}
In this section, we provide a table that gives all of the observable Block-Marschak polynomial values for $\rho$ of the example in Table $\ref{tale:rho}$ of Section \ref{subsection:ex} for $\ep = 0$.\\

{\centering $K(\rho,D,x):$
\[\begin{array}{c|ccccc}
\tikz{\node[below left, inner sep=1pt] (def) {$D$};%
      \node[above right,inner sep=1pt] (abc) {$x$};%
      \draw (def.north west|-abc.north west) -- (def.south east-|abc.south east);}
                & a & b & (c) & (d) \\
                \hline
                \{a\} & 0 & - & - & -\\
                \{b\} & - & 0 & - & -\\
                \{a,b\} & 0 & 1/3 & - & -\\
                \{a,c\} & 1/6 & - & ? & -\\
                \{a,d\} & 1/6 & - & - & ?\\
                \{b,c\} & - & 1/6 & ? & -\\
                \{b,d\} & - & 1/6 & - & ?\\
                \{c,d\} & - & - & ? & ?\\
                \{a,b,c\} & 1/6 & 0 & ? & -\\
                \{a,b,d\} & 1/6 & 0 & - & ?\\
                \{a,c,d\} & 1/6 & - & ? & ?\\
                \{b,c,d\} & - & 1/6 & ? & ?\\
            \{a,b,c,d\} & 1/6 & 1/6 & ? & ?
\end{array}\]
}

\subsection{Details of calibrated dataset}\label{sec:calib}
Recall that $\L$ is the set of linear orders on $X = \{0, 1, 2, 3, 4\}.$
For a probability distribution $\mu$ over $\L,$ let $\rho (D, x \mid \mu)$ be the choice probability of $x$ out of $D$ induced by $\mu,$ that is,
$$
    \rho (D, x \mid \mu)
    \coloneqq
    \mu \left(
        \left\{
            \succ \in \L
            \mid
            x \succ y
            \text{ for all }
            y \in D \setminus x
        \right\}
    \right)
    .
$$

Given a complete dataset $\rho,$ we shall find an RU-rationalizable dataset that resembles $\rho.$
We solve the following problem:
$$
    \hat \mu
    \in
    \argmin_{\mu \in \Delta (\L)}
    \sum_{x \in D \in \D}
    \left(
        \rho (D, x \mid \mu)
        - 
        \rho (D, x)
    \right)^2
    ,
$$
and we use the calibrated choice data $(\rho (D, x \mid \hat \mu))_{x \in D \in \D}$ in the analysis in Section \ref{sec:app}. Table \ref{tab:calibrated-choice-data} summarizes the calibrated probabilities for the lottery dataset used in the analysis of Section \ref{sec:app}.

\begin{table}[h]
\centering
\caption{Calibrated choice data $\rho$}
\label{tab:calibrated-choice-data}
\begin{tabular}{c|ccccc}
\tikz{\node[below left, inner sep=1pt] (def) {$D$};%
      \node[above right,inner sep=1pt] (abc) {$x$};%
      \draw (def.north west|-abc.north west) -- (def.south east-|abc.south east);}
                & 0        & 1        & 2        & 3        & 4        \\
                \hline\\
\{0, 1\}         & 0.434996 & 0.565004 & -        & -        & -        \\
\{0, 2\}          & 0.171289 & -        & 0.828711 & -        & -        \\
\{0, 3\}          & 0.119875 & -        & -        & 0.880125 & -        \\
\{0, 4\}          & 0.108795 & -        & -        & -        & 0.891205 \\
\{1, 2\}          & -        & 0.444959 & 0.555041 & -        & -        \\
\{1, 3\}          & -        & 0.203263 & -        & 0.796737 & -        \\
\{1, 4\}          & -        & 0.179475 & -        & -        & 0.820525 \\
\{2, 3\}          & -        & -        & 0.475196 & 0.524804 & -        \\
\{2, 4\}          & -        & -        & 0.307597 & -        & 0.692403 \\
\{3, 4\}          & -        & -        & -        & 0.453832 & 0.546168 \\
\{0, 1, 2\}       & 0.162682 & 0.291693 & 0.545626 & -        & -        \\
\{0, 1, 3\}       & 0.119875 & 0.156422 & -        & 0.723703 & -        \\
\{0, 1, 4\}       & 0.091033 & 0.121704 & -        & -        & 0.787263 \\
\{0, 2, 3\}       & 0.064422 & -        & 0.431887 & 0.503691 & -        \\
\{0, 2, 4\}       & 0.066709 & -        & 0.295601 & -        & 0.63769  \\
\{0, 3, 4\}       & 0.053548 & -        & -        & 0.453832 & 0.49262  \\
\{1, 2, 3\}       & -        & 0.198698 & 0.289444 & 0.511858 & -        \\
\{1, 2, 4\}       & -        & 0.147176 & 0.216616 & -        & 0.636208 \\
\{1, 3, 4\}       & -        & 0.102122 & -        & 0.395562 & 0.502315 \\
\{2, 3, 4\}       & -        & -        & 0.27787  & 0.251982 & 0.470148 \\
\{0, 1, 2, 3\}    & 0.064422 & 0.151858 & 0.280029 & 0.503691 & -        \\
\{0, 1, 2, 4\}    & 0.066709 & 0.106637 & 0.215865 & -        & 0.610789 \\
\{0, 1, 3, 4\}    & 0.053548 & 0.08145  & -        & 0.395562 & 0.469439 \\
\{0, 2, 3, 4\}    & 0.046842 & -        & 0.265874 & 0.251982 & 0.435302 \\
\{1, 2, 3, 4\}    & -        & 0.098786 & 0.192909 & 0.251982 & 0.456323 \\
\{0, 1, 2, 3, 4\} & 0.046842 & 0.078113 & 0.192158 & 0.251982 & 0.430904
\end{tabular}
\end{table}

\end{document}